\documentclass[a4paper,showpacs,notitlepage,floatfix,nofootinbib,superscriptaddress,prd]
{revtex4-2}
\usepackage{graphicx,color}
\usepackage{amsfonts}
\usepackage{amsmath}
\usepackage{hyperref}
\usepackage{float}
\usepackage{amssymb}
\usepackage{epstopdf}

\begin{document}

\title{Higher-order dissipative anisotropic
magnetohydrodynamics from the Boltzmann-Vlasov equation}
\author{Etele Moln\'ar}
\affiliation{Incubator of Scientiﬁc Excellence--Centre for Simulations of Superdense Fluids,\\
	University of Wroc{\l}aw, pl. M. Borna 9, PL-50204 Wroc{\l}aw, Poland} 
\affiliation{Institut f\"ur Theoretische Physik,
	Johann Wolfgang Goethe--Universit\"at,
	Max-von-Laue-Str.\ 1, D--60438 Frankfurt am Main, Germany}

\author{Dirk H.\ Rischke}
\affiliation{Institut f\"ur Theoretische Physik,
	Johann Wolfgang Goethe--Universit\"at,
	Max-von-Laue-Str.\ 1, D--60438 Frankfurt am Main, Germany} 
\affiliation{Helmholtz Research Academy Hesse for FAIR, Campus Riedberg,\\
	Max-von-Laue-Str.~12, D-60438 Frankfurt am Main, Germany}


\begin{abstract}
We apply the method of moments to the relativistic Boltzmann-Vlasov equation and derive the equations of motion for the irreducible moments of arbitrary tensor-rank of the invariant single-particle distribution function. 
We study two cases, in the first of which the moments are taken to be irreducible with respect to the little group associated with the time-like fluid four-velocity, while in the second case they are assumed to be also irreducible with respect to a space-like four-vector orthogonal to the fluid four-velocity, which breaks the spatial isotropy to a rotational symmetry in the plane transverse to this vector.
A systematic truncation and closure of the general moment equations leads, in the first case, to a theory of relativistic higher-order dissipative resistive magnetohydrodynamics.
In the second case, we obtain a novel theory of dissipative resistive anisotropic magnetohydrodynamics, where the momentum anisotropy is in principle independent from that introduced by the external magnetic field.
\end{abstract}



\maketitle

\section{Introduction}
\label{Introduction}

Relativistic fluid dynamics describes the space-time evolution of a wide variety of systems 
and various phenomena in astrophysics, cosmology, and heavy-ion physics.
For realistic fluids the effects of dissipation and the interaction of charged particles with electromagnetic fields requires a self-consistent theory of dissipative and resistive magnetohydrodynamics. 
The framework which describes the dynamics of relativistic fluids coupled to electromagnetic 
fields is the theory of relativistic magnetohydrodynamics (MHD)~\cite%
{Alfven_1942,Edward_Harris_1957,Bekenstein_Oron_1978,Hernandez:2017mch,Hattori:2022hyo}.

A resistive and dissipative relativistic MHD framework can be derived from the Boltzmann-Vlasov equation using the Chapman-Enskog method~\cite{Panda:2020zhr,Panda:2021pvq} or Grad's method of moments~\cite{Denicol:2018rbw,Denicol:2019iyh}. 
The so-called ideal-fluid and ideal-MHD limit of these moment equations requires that the system 
is in local thermodynamical equilibrium, with a single-particle distribution function $f_{0\mathbf{k}}$, which is isotropic in momentum space in the local rest frame.
Assuming that the deviations from local equilibrium $\delta f_{\mathbf{k}}$ are small, 
i.e., $|\delta f_{\mathbf{k}}/f_{0\mathbf{k}}|\ll 1$, the equations of motion of second-order resistive dissipative MHD can be obtained using the well-known 14 dynamical-moment approximation of Israel and Stewart~\cite{Israel:1979wp}.  
In second-order fluid dynamics the dissipative quantities relax on finite timescales to their corresponding Navier-Stokes values, i.e., to the values given by the first-order theory. 
The relaxation equations for the dissipative quantities resolve the problems of acausality and instability of first-order theories   \cite{Israel:1979wp,Hiscock:1983zz,Hiscock:1985zz,Hiscock:1987zz,Pu:2009fj,Rocha:2023ilf}.
However, in the case of a rapid longitudinal expansion as in the early stage of a heavy-ion collision, the momentum-space anisotropy becomes dominant and the near-equilibrium expansion 
of the distribution function is no longer appropriate.
To overcome such limitations we introduce an anisotropic distribution function, 
$\hat{f}_{0 \mathbf{k}}$, which comprises a new parameter characterizing the strength 
of the anisotropy in momentum space as well as a space-like four-vector $l^\mu$ in 
the direction of the momentum-space anisotropy.
The corresponding moment equations of the Boltzmann equation lead to the framework of 
leading-order anisotropic fluid dynamics~\cite%
{Barz:1987pq,Kampfer:1990qg,Florkowski:2010cf,Ryblewski:2010bs,Ryblewski:2012rr,Martinez:2010sc,Martinez:2010sd,Martinez:2012tu,Bazow:2013ifa,Bazow:2015cha,Tinti:2013vba,Tinti:2015xwa,Molnar:2016vvu,Molnar:2016gwq,Alqahtani:2017mhy}. 
Furthermore, similar to the near-equilibrium expansion of the single-particle distribution function, one can now use $\hat{f}_{0\mathbf{k}}$ as the starting point of a series expansion. 
Including these additional corrections $\delta \hat{f}_{\mathbf{k}}$  improves the leading-order anisotropic framework and leads to a more complete theory of anisotropic dissipative fluid dynamics derived from the Boltzmann equation~\cite{Molnar:2016vvu}. 

The purpose of this paper is to apply the method of moments to the Boltzmann-Vlasov equation, 
in order to obtain the equations of motion for the irreducible moments of the single-particle 
distribution function for arbitrary tensor-rank, both in the case where the moments are irreducible 
with respect to the little group associated with the fluid four-velocity, as well as in the case where 
the moments are irreducible with respect to the little group associated with the fluid four-velocity $u^\mu$ \textit{and} the anisotropy four-vector $l^\mu$ orthogonal to $u^\mu$.

In the first case, the spatial isotropy, i.e., the symmetry with respect to spatial rotations is preserved, wherefore we will refer to the irreducible moments simply as \textit{isotropic irreducible moments} (IIMs).
In the second case, the spatial isotropy is broken to a rotational symmetry around the axis given by $l^\mu$, i.e., symmetry under rotations in the plane orthogonal to this vector, wherefore we will denote the irreducible moments simply as \textit{anisotropic irreducible moments} (AIMs).
Our results extend previously obtained ones by including interactions with external electromagnetic fields. 
For instance, the equations of motion for the IIMs of arbitrary tensor-rank without electromagnetic 
fields were obtained recently by de Brito and Denicol~\cite{deBrito:2024vhm}.
In the current paper, we extend that work by the coupling of the IIMs to electromagnetic fields.
The additional terms generalize the equations of motion for rank-$\ell$ IIMs to electrically 
conducting fluids. 
Furthermore, we also derive the general equations of motion for rank-$\ell$ AIMs for electrically conducting fluids.
A systematic truncation of these general moment equations leads to novel theories of higher-order resistive and dissipative MHD as well as to dissipative and resistive anisotropic MHD, extending 
earlier dissipative fluid-dynamical and magnetohydrodynamical theories~\cite{Denicol:2018rbw,Denicol:2019iyh,Panda:2020zhr,Panda:2021pvq}.

This paper is organized as follows. 
The notation, conventions, and definitions are presented in Sect.~\ref{sec:definitions}.
In Sect.~\ref{sec:iso_moments} we introduce various rank-$\ell$ IIMs of the single-particle 
distribution function and discuss their properties in the context of the particle four-current 
and the energy-momentum tensor. 
In Sect.~\ref{sec:iso_expansion_local} the expansion of the distribution function around local 
equilibrium $f_{0\mathbf{k}}$ in terms of IIMs is presented. 
In Sect.~\ref{sec:aniso_moments} we introduce various AIMs of the single-particle distribution 
function and decompose the conserved quantities using these.
In Sect.~\ref{sec:aniso_expansion_local}, the expansion of the distribution function around  $\hat{f}_{0\mathbf{k}}$ in terms of AIMs is discussed.

The Boltzmann-Vlasov equation and the fundamental equations of motion of MHD are introduced in Sect.~\ref{sec:Boltzmann_Vlasov}. 
The general equation of motion for IIMs of arbitrary tensor-rank is obtained in Sect.~\ref{sec:iso_moment_equations}. 
Then the systematic truncation of the moment equations is discussed in Sect.~\ref{sec:iso_fluid_dynamics}. 
Similarly as in the previous sections, the equation of motion for AIMs of arbitrary tensor-rank is presented in Sect.~\ref{sec:aniso_moment_equations}, followed by a discussion of the truncation of 
the corresponding moment equations in Sect.~\ref{sec:aniso_fluid_dynamics}. 
The conclusions are given in Sect.~\ref{sec:conclusions}, while additional
definitions, computations, and important relations are relegated to the Appendices.

\subsection{Notation and definitions}
\label{sec:definitions}

In this paper we adopt natural Heaviside-Lorentz units: $\hbar=c=\varepsilon_{0}=\mu _{0}=k_{B}=1$, and work in flat space-time, where the metric tensor is $g^{\mu \nu }=\text{diag}(1,-1,-1,-1)$. 
The rank-four antisymmetric, or Levi-Civit\`{a}, symbol is defined as 
$\varepsilon ^{\mu \nu \alpha \beta}=+1$, 
if $(\mu \nu \alpha \beta)$ is an even permutation of (0123), 
$\varepsilon ^{\mu \nu \alpha \beta}=-1$, 
if $(\mu \nu \alpha \beta)$ is an odd permutation of (0123), and zero otherwise, 
$\varepsilon ^{\mu \nu \alpha \beta }=0$.

The contravariant fluid four-velocity is denoted by 
$u^{\mu }\left(t,\mathbf{x}\right)=\gamma \left( 1,\mathbf{v}\right)^{T}$, with 
$\gamma =(1-\mathbf{v}^{2})^{-1/2}$.
It is a time-like normalized vector, $u^{\mu }u_{\mu }\equiv c^{2}=1$.
In the local rest (LR) frame of the fluid, $u_{LR}^{\mu }=\left( 1,\mathbf{0}\right)^{T}$ 
by definition. 
The four-vector in the direction of the momentum anisotropy is denoted by 
$l^{\mu }\left( t,\mathbf{x}\right) $. 
It is chosen to be orthogonal to the fluid four-velocity, $u^{\mu }l_{\mu }=0$. 
It is a space-like normalized vector, $l^{\mu }l_{\mu }=-1$. 
In the context of heavy-ion collisions the anisotropy four-vector is usually chosen to point into the direction of the beam axis, and hence in the LR frame $l_{LR}^{\mu}=(0,0,0,1)^T$. 
The four-vector in the direction of an external magnetic field is denoted by $b^{\mu}$. 
It is also orthogonal to $u^\mu$, i.e., $b^{\mu} u_{\mu} = 0$, and it is a space-like normalized vector, $b^{\mu }b_{\mu }=-1$.  
In general, $l^\mu \neq b^{\mu}$.

The symmetric rank-two projection operator orthogonal to $u^{\mu }$ is defined as~\cite{deGroot,Cercignani_book,Denicol_Rischke_book} 
\begin{equation}
\Delta^{\mu \nu } \equiv g^{\mu \nu }-u^{\mu }u^{\nu }\,,  \label{Delta_munu}
\end{equation}%
such that $\Delta ^{\mu \nu }u_{\nu }=0$ and $\Delta ^{\mu \nu }g_{\mu \nu}=3$. 
The generalization of the $\Delta ^{\mu \nu }$ projection operator to a new rank-two projection 
operator that is symmetric and orthogonal to both $u^{\mu }$ and $l^{\mu }$ is 
defined as~\cite{Gedalin_1991,Gedalin_1995,Huang:2011dc,Molnar:2016vvu,Denicol_Rischke_book} 
\begin{equation}
\Xi ^{\mu \nu }\equiv g^{\mu \nu }-u^{\mu }u^{\nu }+l^{\mu }l^{\nu }
=\Delta^{\mu \nu }+l^{\mu }l^{\nu }\,,  \label{Xi_munu}
\end{equation}%
where $\Xi^{\mu \nu }u_{\nu }=\Xi^{\mu \nu }l_{\nu }=0$ and $\Xi ^{\mu \nu}g_{\mu \nu }=2$.

Using the elementary rank-two projection operators~(\ref{Delta_munu}) we can construct 
rank-$2n$ symmetric and traceless projection operators that are orthogonal to $u^{\mu }$. 
The definition of these irreducible projection operators is~\cite{deGroot,Denicol:2012cn,Denicol_Rischke_book}, 
\begin{equation}
\Delta _{\nu _{1}\cdots \nu _{n}}^{\mu _{1}\cdots \mu _{n}}\equiv
\sum_{q=0}^{\lfloor n/2\rfloor}\frac{C(n,q)}{\mathcal{N}_{nq}}\sum_{\mathcal{P}_{\mu }^{n}%
\mathcal{P}_{\nu }^{n}}\Delta ^{\mu _{1}\mu _{2}}\cdots \Delta ^{\mu
_{2q-1}\mu _{2q}}\Delta _{\nu _{1}\nu _{2}}\cdots \Delta _{\nu _{2q-1}\nu
_{2q}}\Delta _{\nu _{2q+1}}^{\mu _{2q+1}}\cdots \Delta _{\nu _{n}}^{\mu_{n}}\,,  
\label{Delta_irred_proj}
\end{equation}%
where ${\lfloor n/2\rfloor}$ denotes the largest integer less than or equal to $n/2$, while the 
second sum runs over all distinct permutation of $\mu$- and $\nu$-type indices and the coefficients $C(n,q)$ and $\mathcal{N}_{nq}$ are defined in Eqs.~(\ref{Ccoeff}) and (\ref{N_nq}), respectively.

We can also construct rank-$2n$ symmetric and traceless projector operators that are orthogonal to both $u^{\mu }$ and $l^{\mu }$. 
These irreducible projection operators are formed from the elementary rank-two projection operators~(\ref{Xi_munu}) and are defined similarly to Eq.~(\ref{Delta_irred_proj}), 
\begin{equation}
\Xi _{\nu _{1}\cdots \nu _{n}}^{\mu _{1}\cdots \mu _{n}}=\sum_{q=0}^{\lfloor n/2\rfloor}%
\frac{\hat{C}(n,q)}{\mathcal{N}_{nq}}\sum_{\mathcal{P}_{\mu }^{n}\mathcal{P}_{\nu }^{n}}
\Xi ^{\mu _{1}\mu _{2}}\cdots \Xi ^{\mu _{2q-1}\mu _{2q}}\Xi_{\nu _{1}\nu _{2}}\cdots 
\Xi _{\nu _{2q-1}\nu _{2q}}\Xi _{\nu _{2q+1}}^{\mu_{2q+1}}\cdots \Xi _{\nu _{n}}^{\mu _{n}}\, ,  \label{Xi_irred_proj}
\end{equation}%
where $\hat{C}(n,q)$ is defined in Eq.~(\ref{Chatcoeff}).

The particle four-momentum $k^{\mu }=\left( k^{0},\mathbf{k}\right) ^{T}$ is normalized to the squared rest mass, $k^{\mu }k_{\mu }=m_{0}^{2}$. 
Using the elementary projection operators (\ref{Delta_munu}) and~(\ref{Xi_munu}), the four-momentum is decomposed as 
\begin{equation}
k^{\mu }\equiv E_{\mathbf{k}u}u^{\mu }+k^{\left\langle \mu \right\rangle}
=E_{\mathbf{k}u}u^{\mu }+E_{\mathbf{k}l}l^{\mu }+k^{\left\{ \mu \right\}}\,,  
\label{k_mu_decomposition}
\end{equation}%
where the energy of the particle in the LR frame, $E_{\mathbf{k}u}$, and the momentum of the particle in the direction of the anisotropy, $E_{\mathbf{k}l}$, are defined through 
\begin{equation}
E_{\mathbf{k}u}\equiv k^{\mu }u_{\mu }\,,\quad 
E_{\mathbf{k}l}\equiv -k^{\mu}l_{\mu }\,.
\end{equation}%
The four-momentum orthogonal to the four-velocity $u^\mu$ is denoted by 
$k^{\left\langle\mu \right\rangle}$, while the four-momentum orthogonal to both 
$u^{\mu }$ and $l^{\mu }$ is denoted by $k^{\left\{ \mu \right\} }$,
\begin{equation}
k^{\left\langle \mu \right\rangle }\equiv \Delta ^{\mu \nu }k_{\nu }\,, \quad
k^{\left\{ \mu \right\} }\equiv \Xi ^{\mu \nu }k_{\nu }\,,
\end{equation}%
such that $k^{\left\langle \mu \right\rangle }=E_{\mathbf{k}l}l^{\mu}+k^{\left\{ \mu \right\} }$ and $k^{\left\langle \mu \right\rangle }u_{\mu}=k^{\left\{ \mu \right\} }u_{\mu }
=k^{\left\{ \mu \right\} }l_{\mu }=0$.

Recalling the projection operators~(\ref{Delta_munu}) and (\ref{Delta_irred_proj}) we define the 
following useful projections of the particle four-momentum 
\begin{align}
k^{\left\langle \mu_{1} \right\rangle }\cdots k^{\left\langle \mu_{n} \right\rangle } 
&\equiv  \Delta_{\nu_{1}}^{\mu_{1}}\cdots 
\Delta_{\nu_{n}}^{\mu_{n}} k^{\nu_{1}}\cdots k^{\nu_{n}} \,, \\
k^{\left\langle \mu_{1}\right. }\cdots k^{\left. \mu_{n}\right\rangle }
&\equiv \Delta_{\nu_{1} \cdots \nu_{n}}^{\mu_{1}\cdots \mu_{n}} 
k^{\nu_{1}}\cdots k^{\nu_{n}} \,. \label{k_mu1_k_mun}
\end{align}%
These reducible and irreducible projections obey the following relations, see Ref.~\cite{deBrito:2024vhm} or Appendix~\ref{Projection_properties} for a detailed derivation, 
\begin{align}
k^{\left\langle \mu _{1}\right. }\cdots k^{\left. \mu _{n}\right\rangle} &= 
k^{\left\langle \mu _{1}\right\rangle }\cdots k^{\left\langle \mu_{n}\right\rangle } 
+\sum_{q=1}^{\lfloor n/2\rfloor}C\left( n,q\right) \left( \Delta_{\alpha \beta } 
k^{\alpha }k^{\beta }\right) ^{q}\Delta ^{\left( \mu_{1}\right. \mu _{2}}\cdots 
\Delta ^{\mu _{2q-1}\mu _{2q}}k^{\left\langle\mu _{2q+1}\right\rangle }\cdots 
k^{\left. \left\langle \mu_{n}\right\rangle \right) }\, ,
\label{Main1_iso}  \\ 
k^{\left\langle \mu _{1}\right. }\cdots k^{\left. \mu _{n}\right\rangle}
k^{\left\langle \mu _{n+1}\right\rangle } &= k^{\left\langle \mu _{1}\right.}
\cdots k^{\left. \mu _{n+1}\right\rangle } 
+\frac{n}{2n+1}\left( \Delta_{\alpha \beta }k^{\alpha }k^{\beta }\right) 
\Delta _{\lambda _{1}\cdots\lambda _{n}}^{\mu _{1}\cdots \mu _{n}}
\Delta ^{\lambda _{n}\mu_{n+1}}k^{\left\langle \lambda _{1}\right. }\cdots 
k^{\left. \lambda_{n-1}\right\rangle }\, .  
\label{Main2_iso}
\end{align}%
Analogously, using Eqs.~(\ref{Xi_munu}) and (\ref{Xi_irred_proj}) we define the following anisotropic tensor projections, 
\begin{align}
k^{\left\{ \mu _{1}\right\} }\cdots k^{\left\{ \mu _{n}\right\} } &\equiv
\Xi_{\nu _{1}}^{\mu _{1}}\cdots \Xi _{\nu _{n}}^{\mu _{n}}k^{\nu _{1}}\cdots
k^{\nu _{n}}\, , \\
k^{\left\{ \mu _{1}\right. }\cdots k^{\left. \mu _{n}\right\} } &\equiv 
\Xi _{\nu_{1}\cdots \nu _{n}}^{\mu _{1}\cdots \mu _{n}}k^{\nu _{1}}\cdots k^{\nu_{n}}\, .
\end{align}%
These projections are related via the following equations, see Appendix~\ref{Projection_properties} 
for more details, 
\begin{align}
k^{\left\{ \mu _{1}\right. }\cdots k^{\left. \mu _{n}\right\} }
&= k^{\left\{\mu _{1}\right\} }\cdots k^{\left\{ \mu _{n}\right\} }
+\sum_{q=1}^{\lfloor n/2\rfloor}\hat{C}\left( n,q\right) 
\left( \Xi _{\alpha \beta }k^{\alpha }k^{\beta}\right) ^{q} 
\Xi ^{\left( \mu _{1}\right. \mu _{2}}\cdots \Xi ^{\mu_{2q-1}\mu _{2q}} 
k^{\left\{ \mu _{2q+1}\right\} }\cdots k^{\left. \left\{\mu _{n}\right\} \right) }\, ,  \label{Main1_aniso} \\
k^{\left\{ \mu _{1}\right. }\cdots k^{\left. \mu _{n}\right\} }k^{\left\{\mu _{n+1}\right\} } 
&=k^{\left\{ \mu _{1}\right. }\cdots k^{\left. \mu_{n+1}\right\} }
+\frac{1}{2}\left( \Xi _{\alpha \beta }k^{\alpha }k^{\beta}\right) 
\Xi _{\lambda _{1}\cdots \lambda_{n}}^{\mu _{1}\cdots \mu _{n}}\Xi^{\lambda _{n}\mu _{n+1}} 
k^{\left\{ \lambda _{1}\right. }\cdots k^{\left.\lambda _{n-1}\right\} }\, .  \label{Main2_aniso}
\end{align}

\section{Reducible and irreducible moments}
\label{sec:iso_moments}

The momentum-space integral of the single-particle distribution function $f_{\mathbf{k}}\equiv f_{\mathbf{k}}(x^{\mu },k^{\mu })$, multiplied with a monomial constructed of the particle four-momentum and its LR frame energy, defines the \textit{reducible} rank-$\ell$ tensor moment
\begin{equation}
M_{r}^{\mu _{1}\cdots \mu _{\ell }}\equiv 
\int \mathrm{d}K E_{\mathbf{k}u}^{r}k^{\mu _{1}}\cdots k^{\mu _{\ell }}f_{\mathbf{k}}\, ,
\label{def_M_r_mu1_mun}
\end{equation}%
where the subscript $r$ reflects the power of $E_{\mathbf{k}u}=k^{\mu }u_{\mu }$. 
The Lorentz-invariant integration measure is 
$\mathrm{d}K\equiv g\,\mathrm{d}^{3}\mathbf{k/}\left[ (2\pi )^{3}k^{0}\right] $, where $g$ denotes the degeneracy of the momentum state. 
The \textit{isotropic irreducible} rank-$\ell $ tensor moments (or short, rank-$\ell$ IIMs), which are symmetric, traceless, and orthogonal to $u^\mu$, are denoted by 
\begin{equation}
\mathcal{M}_{r}^{\mu _{1}\cdots \mu _{\ell }}\equiv 
M_{r}^{\left\langle \mu_{1}\cdots \mu _{\ell }\right\rangle }
=\Delta _{\nu _{1}\cdots \nu _{\ell}}^{\mu _{1}\cdots \mu _{\ell }} 
M_{r}^{\nu _{1}\cdots \nu _{\ell}} 
=\int \mathrm{d}K E_{\mathbf{k}u}^{r}k^{\left\langle \mu _{1}\right. }\cdots 
k^{\left. \mu_{\ell}\right\rangle } f_{\mathbf{k}}\, . 
\label{IIMs}
\end{equation}

The microscopic definition of the particle four-current and the energy-momentum tensor are given by the following reducible moments of the single-particle distribution function $f_{\mathbf{k}}$, 
\begin{eqnarray}
N^{\mu } &\equiv & M_{0}^{\mu }=\int \mathrm{d}K k^{\mu }f_{\mathbf{k}}\,,
\label{def_N_mu} \\
T^{\mu \nu } &\equiv & M_{0}^{\mu \nu }=\int \mathrm{d}Kk^{\mu }k^{\nu }f_{\mathbf{k}}\,.  
\label{def_T_munu}
\end{eqnarray}%
The tensor decomposition of $N^{\mu }$ and $T^{\mu \nu }$ with respect to $u^{\mu }$ and 
$\Delta^{\mu \nu }$ reads 
\begin{eqnarray}
N^{\mu } &\equiv &nu^{\mu }+V^{\mu }\, ,  \label{kinetic:N_mu} \\
T^{\mu \nu } &\equiv &eu^{\mu }u^{\nu }-P\Delta ^{\mu \nu }+2W^{\left( \mu
\right. }u^{\left. \nu \right) }+\pi ^{\mu \nu }\, ,  \label{kinetic:T_munu}
\end{eqnarray}%
where all quantities on the right-hand side can be expressed in terms of different projections of 
the particle four-current and the energy-momentum tensor. 
On the other hand, these quantities are also identified as various IIMs of the single-particle distribution function $f_{\mathbf{k}}$. 
The three scalar coefficients in Eqs.~(\ref{kinetic:N_mu}) and (\ref{kinetic:T_munu}), the particle density $n$, the energy density $e$, and the isotropic pressure $P$, are 
\begin{eqnarray}
n &\equiv &\mathcal{M}_{1}=N^{\mu }u_{\mu }\, , \qquad 
e \equiv \mathcal{M}_{2}=T^{\mu \nu }u_{\mu }u_{\nu }\, ,  
\label{kinetic:n_and_e} \\
P &\equiv &-\frac{1}{3}\left( m_{0}^{2}\mathcal{M}_{0}-\mathcal{M}%
_{2}\right) =-\frac{1}{3}T^{\mu \nu }\Delta _{\mu \nu }\, .  
\label{kinetic:P}
\end{eqnarray}%
The particle and energy-momentum diffusion currents $V^{\mu }$ and $W^{\mu }$ are both orthogonal to 
the fluid four-velocity, $V^{\mu }u_{\mu }=W^{\mu}u_{\mu }=0$, and are defined as 
\begin{eqnarray}
V^{\mu } &\equiv &\mathcal{M}_{0}^{\mu }=\Delta _{\nu }^{\mu }N^{\nu }\, ,
\label{kinetic:V_mu} \\
W^{\mu } &\equiv &\mathcal{M}_{1}^{\mu }=\Delta _{\alpha }^{\mu }T^{\alpha\beta }u_{\beta }\, .  
\label{kinetic:W_mu}
\end{eqnarray}%
Finally, the shear-stress tensor is the part of the energy-momentum tensor that is symmetric, 
$\pi ^{\mu \nu }=\pi ^{\nu \mu }$, traceless, $\pi ^{\mu\nu }g_{\mu \nu }=0$, and orthogonal to the four-velocity $\pi ^{\mu \nu}u_{\mu }=0$.
It is defined through 
\begin{equation}
\pi ^{\mu \nu }\equiv \mathcal{M}_{0}^{\mu \nu }=\Delta _{\alpha \beta
}^{\mu \nu }T^{\alpha \beta }\, .  \label{kinetic:pi_munu}
\end{equation}

\subsection{Expansion of the single-particle distribution function around
local equilibrium}
\label{sec:iso_expansion_local}

The state of local thermodynamic equilibrium is specified by the J\"{u}ttner
distribution~\cite{Juttner,Juttner_quantum} 
\begin{equation} 
f_{0\mathbf{k}} \left(\alpha, \beta E_{\mathbf{k}u} \right) 
=\left[ \exp \left( \beta E_{\mathbf{k}u}-\alpha \right) +a\right] ^{-1}\, ,  
\label{f_0k}
\end{equation}%
with $\alpha = \beta \mu$, where $\beta =1/T$ is the inverse temperature and $\mu $ is the chemical potential, while $a=\pm 1$ for fermions/bosons and $a=0$ for classical Boltzmann particles. 
The reducible equilibrium moments of tensor-rank $n$ and of power $r$ in energy $E_{\mathbf{k}u}$ are defined similarly to Eq.~(\ref{def_M_r_mu1_mun}), 
\begin{equation}
\mathcal{I}_{r}^{\mu _{1}\cdots \mu _{n}}\equiv \int \mathrm{d}K E_{\mathbf{k}%
u}^{r}\,k^{\mu _{1}}\cdots k^{\mu _{n}} f_{0\mathbf{k}} 
=\sum_{q=0}^{\lfloor n/2\rfloor }\left( -1\right)
^{q}b_{nq}I_{r+n,q}\Delta ^{\left( \mu _{1}\mu _{2}\right. }\cdots \Delta
^{\mu _{2q-1}\mu _{2q}}u^{\mu _{2q+1}}\cdots u^{\left. \mu _{n}\right) }\,,
\label{I_r_tens}
\end{equation}
where the coefficients $b_{nq}$ are defined in Eq.~(\ref{b_nq}). 
The equilibrium thermodynamic integrals are defined as 
\begin{equation}
I_{nq}\left( \alpha ,\beta \right) \equiv \frac{\left( -1\right) ^{q}}{\left( 2q+1\right) !!} 
\int \mathrm{d}K E_{\mathbf{k}u}^{n-2q}\left( \Delta
^{\alpha \beta }k_{\alpha }k_{\beta }\right) ^{q} f_{0\mathbf{k}} \,.
\label{I_nq}
\end{equation}%
Since the local-equilibrium distribution function is isotropic in momentum space, the projections of the reducible equilibrium moments orthogonal to the fluid four-velocity vanish for any tensor-rank $n\geq 1$,
\begin{equation}
\mathcal{I}_{r}^{\left\langle \mu _{1}\right\rangle \cdots \left\langle \mu_{n}\right\rangle }
=\mathcal{I}_{r}^{\left\langle \mu _{1}\right. \cdots
\left. \mu _{n}\right\rangle }=0\,.  \label{I_r_n_condition}
\end{equation}%
Therefore, the particle-four current and the energy-momentum tensor from Eqs.~(\ref{def_N_mu}), (\ref{def_T_munu}) reduce to the ideal-fluid form,%
\begin{eqnarray}
N_{0}^{\mu } &\equiv &\mathcal{I}_{0}^{\mu }=n_{0}u^{\mu }\,,  
\label{N0_mu} \\
T_{0}^{\mu \nu } &\equiv &\mathcal{I}_{0}^{\mu \nu }
=e_{0}u^{\mu }u^{\nu}-P_{0}\Delta ^{\mu \nu }\,,  \label{T0_munu}
\end{eqnarray}%
where the particle density, the energy density, and the pressure in local thermodynamical equilibrium are 
\begin{eqnarray}
n_{0} &\equiv &\mathcal{I}_{1}=I_{10}\,, \quad e_{0} \equiv \mathcal{I}_{2}=I_{20}\,, \\
P_{0} &\equiv &-\frac{1}{3}\left( m_{0}^{2}\mathcal{I}_{0}-\mathcal{I}_{2}\right) = I_{21}\, .
\end{eqnarray}%
The pressure is defined through an equation of state $P_{0}\equiv P_{0}\left( n_{0},e_{0}\right) $, and hence the temperature $T(e_{0},n_{0})$ and chemical potential $\mu (e_{0},n_{0})$ can be determined.
Therefore the conservation laws of ideal fluid dynamics, $\partial_\mu N^{\mu}_0 = 0$ and 
$\partial_\nu T^{\mu \nu}_0 = 0$, together with an equation of state form a closed system of equations.

Out of equilibrium these intensive thermodynamical quantities are inferred by matching the non-equilibrium particle density and energy density to an arbitrary reference state in local equilibrium. 
These are the so-called Landau matching conditions, 
\begin{eqnarray}
&&\left( N^{\mu }-N_{0}^{\mu }\right) u_{\mu }\equiv n-n_{0}=0\, ,
\label{Landau_matching_iso_n} \\
&&\left( T^{\mu \nu }-T_{0}^{\mu \nu }\right) u_{\mu }u_{\nu }
\equiv e-e_{0}=0\, , \label{Landau_matching_iso_e}
\end{eqnarray}
while the non-equilibrium correction to the pressure defines the bulk viscous pressure
\begin{equation}
\Pi \equiv -\frac{1}{3}\left( T^{\mu \nu }-T_{0}^{\mu \nu }\right) \Delta_{\mu \nu } = P - P_0 \,. 
\label{kinetic:Pi}
\end{equation}

In most cases of interest the single-particle distribution function $f_{\mathbf{k}}$ is not of the local-equilibrium form~(\ref{f_0k}), but if it is sufficiently close to it then we decompose 
\begin{equation}
f_{\mathbf{k}}=f_{0\mathbf{k}}+\delta f_{\mathbf{k}}\,,  \label{f=f0+df_iso}
\end{equation}%
where the deviation $\delta f_{\mathbf{k}}$ from local equilibrium fulfills $|\delta f_{\mathbf{k}}/f_{0\mathbf{k}}|\ll 1$. 
This also means that the IIMs of the single-particle distribution are correspondingly decomposed into equilibrium and non-equilibrium parts, 
\begin{equation}
\mathcal{M}_{r}^{\mu _{1}\cdots \mu _{\ell }}=\mathcal{I}_{r}^{\mu_{1}
\cdots \mu _{\ell }}+\rho _{r}^{\mu _{1}\cdots \mu _{\ell }}\,, \label{M=I+rho}
\end{equation}%
where the non-equilibrium IIMs are 
\begin{equation}
\rho _{r}^{\mu _{1}\cdots \mu _{\ell }}\equiv 
\int \mathrm{d}K E_{\mathbf{k}u}^{r} k^{\left\langle \mu _{1}\right. } 
\cdots k^{\left. \mu _{\ell }\right\rangle }
\delta f_{\mathbf{k}}\,.  \label{kinetic:rho_r}
\end{equation}
The details of the near-equilibrium expansion of $f_{\mathbf{k}}$ are relegated to Appendix~\ref{app:expansion_coefficients}, while the corresponding equations of motions for 
the non-equilibrium IIMs are presented in Appendix~\ref{app:resistiveMHD}.

\section{Reducible and irreducible anisotropic moments}
\label{sec:aniso_moments}

The momentum-space integral of the distribution function $f_{\mathbf{k}}$, multiplied with a monomial constructed of the particle four-momentum, the LR frame energy $E_{\mathbf{k}u}$, as well as its momentum component in the direction of the anisotropy, $E_{\mathbf{k}l}$, forms a \textit{reducible} rank-$\ell$ tensor moment
\begin{equation}
A_{ij}^{\mu _{1}\cdots \mu _{\ell }}\equiv 
\int \mathrm{d}K E_{\mathbf{k}u}^{i}E_{\mathbf{k}l}^{j}k^{\mu _{1}}\cdots k^{\mu
_{\ell }}f_{\mathbf{k}}\, ,  \label{def_A_mu1_mun}
\end{equation}%
where the subscripts $i$ and $j$ define the power of $E_{\mathbf{k}u}=k^{\mu}u_{\mu }$ and $E_{\mathbf{k}l}=-k^{\mu }l_{\mu }$. 
The \textit{anisotropic irreducible} rank-$\ell $ tensor moments (or short, rank-$\ell$ AIMs), which are symmetric, traceless, and orthogonal to both $u^{\mu }$ and $l^{\mu }$, are 
\begin{equation} \label{irred_mom_A}
\mathcal{A}_{ij}^{\mu _{1}\cdots \mu _{\ell }}\equiv A_{ij}^{\left\{ \mu
_{1}\right. \cdots \left. \mu _{\ell }\right\} }=\Xi _{\nu _{1}\cdots \nu
_{\ell }}^{\mu _{1}\cdots \mu _{\ell }}A_{ij}^{\nu _{1}\cdots \nu _{\ell}}
=\int \mathrm{d}KE_{\mathbf{k}u}^{i}E_{\mathbf{k}l}^{j} 
k^{\left\{ \mu _{1}\right. }\cdots k^{\left. \mu _{\ell }\right\} }f_{\mathbf{k}}\, .
\end{equation}%

The particle four-current, $N^{\mu }\equiv A_{00}^{\mu }$, and the energy-momentum tensor, 
$T^{\mu \nu }\equiv A_{00}^{\mu \nu }$, can also be decomposed with respect to 
$u^{\mu }$, $l^{\mu }$, and the projection tensor $\Xi ^{\mu \nu }$, 
see Eqs.~(25) and (26) of Ref.~\cite{Molnar:2016vvu}, 
\begin{eqnarray}
N^{\mu } &=&nu^{\mu }+n_{l}l^{\mu }+V_{\perp }^{\mu }\, ,
\label{kinetic:N_mu_u_l} \\
T^{\mu \nu } &=&eu^{\mu }u^{\nu }+2Mu^{\left( \mu \right. } l^{\left. \nu
\right) }+P_{l}l^{\mu }l^{\nu } - P_{\perp }\Xi ^{\mu \nu } + 2W_{\perp
u}^{\left( \mu \right. }u^{\left. \nu \right) } + 2W_{\perp l}^{\left( \mu
\right. }l^{\left. \nu \right) } + \pi _{\perp }^{\mu \nu }\, .
\label{kinetic:T_munu_u_l}
\end{eqnarray}%
The right-hand sides can either be expressed in terms of different projections of the particle four-current and energy-momentum tensor or with the help of the AIMs 
$\mathcal{A}_{ij}^{\mu_{1}\cdots \mu _{\ell }}$. 
The six scalar moments are:
\begin{eqnarray} 
n &\equiv &\mathcal{A}_{10}=N^{\mu }u_{\mu }\, , \label{kinetic:n} \qquad \quad
e \equiv \mathcal{A}_{20}=T^{\mu \nu }u_{\mu }u_{\nu }\, , \\
n_{l} &\equiv &\mathcal{A}_{01}=-N^{\mu }l_{\mu }\, ,  \label{kinetic:n_l} \qquad
M \equiv \mathcal{A}_{11}=-T^{\mu \nu }u_{\mu }l_{\nu }\, ,
\label{kinetic:M} \\
P_{l} &\equiv &\mathcal{A}_{02}=T^{\mu \nu }l_{\mu }l_{\nu }\, , \qquad 
P_{\perp } \equiv -\frac{1}{2}\left( m_{0}^{2}\mathcal{A}_{00}-\mathcal{A}_{20}
+\mathcal{A}_{02}\right) =-\frac{1}{2}T^{\mu \nu }\Xi _{\mu \nu }\, .
\label{kinetic:P_t}
\end{eqnarray}%
Here, as in Eq.~(\ref{kinetic:n_and_e}) the particle density and energy density are denoted by $n$ 
and $e$, respectively. 
The component of the particle diffusion current in the $l^{\mu }$ direction is denoted by $n_{l}$, 
while the projection of the energy-momentum tensor in both $u^{\mu }$ and $l^{\nu }$ directions 
is denoted by $M$. 
The pressure in the direction of the anisotropy is $P_{l}$, while the pressure in the direction 
transverse to it is denoted by $P_{\perp }$. 
These are related to the isotropic pressure as $P=\frac{1}{3}\left( P_{l}+2P_{\perp }\right)$.
Furthermore,
\begin{eqnarray}
V_{\perp }^{\mu } &\equiv &\mathcal{A}_{00}^{\mu }=\Xi _{\nu }^{\mu }N^{\nu}\,,  
\label{kinetic:Vt_mu} \\
W_{\perp u}^{\mu } &\equiv &\mathcal{A}_{10}^{\mu }=\Xi _{\alpha }^{\mu}
T^{\alpha \beta }u_{\beta }\,,  \label{kinetic:Wu_mu} \\
W_{\perp l}^{\mu } &\equiv &\mathcal{A}_{01}^{\mu }=-\Xi _{\alpha }^{\mu}
T^{\alpha \beta }l_{\beta }\,,  \label{kinetic:Wl_mu} \\
\pi _{\perp }^{\mu \nu } &\equiv &\mathcal{A}_{00}^{\mu \nu }
=\Xi _{\alpha\beta }^{\mu \nu }T^{\alpha \beta }\,,  \label{kinetic:pit_munu}
\end{eqnarray}%
where the particle diffusion current orthogonal to both four-vectors is denoted by $V_{\perp }^{\mu }$. 
The mixed projections of the energy-momentum tensor in either $u^\mu$ or $l^\mu$ direction and 
orthogonal to both are denoted by $W_{\perp u}^{\mu }$ or $W_{\perp l}^{\mu }$, respectively. 
The diffusion currents from Eqs.~(\ref{kinetic:V_mu}) and (\ref{kinetic:W_mu}) are split according 
to the direction defined by $l^{\mu }$ and the direction perpendicular to it, 
$V^{\mu } = n_{l}l^{\mu }+V_{\perp }^{\mu }$ and $W^{\mu } = Ml^{\mu }+W_{\perp u}^{\mu }$. 
The transverse shear-stress tensor $\pi _{\perp }^{\mu \nu }$ is the part of the energy-momentum 
tensor that is symmetric, $\pi _{\perp }^{\mu \nu }=\pi_{\perp }^{\nu \mu }$, traceless, 
$\pi _{\perp }^{\mu \nu }g_{\mu \nu }=0$, and orthogonal to both four-vectors, 
$\pi _{\perp }^{\mu \nu }u_{\mu }=\pi_{\perp }^{\mu \nu }l_{\mu }=0$.

\subsection{Expansion of the single-particle distribution function around a
local anisotropic state}
\label{sec:aniso_expansion_local}

The reference distribution function for a state with momentum anisotropy is denoted by $\hat{f}_{0\mathbf{k}}\left( \hat{\alpha},\hat{\beta}_{u} E_{\mathbf{k}u},\hat{\beta}_{l} E_{\mathbf{k}l}\right)$, where $\hat{\alpha}$, $\hat{\beta}_{u}$ and $\hat{\beta}_{l}$ are parameters.
Here $\hat{\beta}_{u}$ and $\hat{\beta}_{l}$ allow to distinguish particle momenta 
parallel and perpendicular to some direction, i.e., the anisotropy, and thus break the isotropy of the local-equilibrium distribution function.
In principle, the interpretation is that one has an additional inverse "temperature", $\hat{\beta}_l = 1/T_l$, in the direction of the anisotropy.
Such anisotropic distribution functions are well known in plasma physics, where the momentum anisotropy 
is induced by the presence of an external magnetic field, e.g., the bi-Maxwellian distribution function \cite{Yoon:2007,Moseev:2019,Larroche:2021}.
On the other hand, the presence of an external magnetic field is not always necessary, 
since in a high-temperature quark-gluon plasma as created in the initial stage of ultrarelativistic 
heavy-ion collisions the momentum-space distribution of gluons is highly anisotropic.
This is represented by a spheroidal distribution function like the one suggested by Romatschke and Strickland~\cite{Romatschke:2003ms} in the LR frame.
The spheroidal shape incorporates a rescaling of the momentum in the direction of the anisotropy, hence 
$\hat{\beta}_u = \beta $ and $\hat{\beta}_l = \beta \sqrt\xi$, where $\xi$ denotes the so-called anisotropy parameter. 
Thus, in the LR frame this is either a prolate or an oblate spheroid for $\xi <0$ or $\xi >0$, respectively, and a spherically symmetric local-equilibrium distribution for $\xi=0$.

Therefore, the anisotropic reference distribution is chosen such that in the limit of vanishing momentum anisotropy $\hat{\beta}_l \rightarrow 0$, we have $\hat{\alpha} \rightarrow \alpha$ and 
$\hat{\beta}_u \rightarrow \beta$, hence
the local-equilibrium distribution~(\ref{f_0k}) is recovered,
\begin{equation}
	\lim_{\hat{\beta}_{l}\rightarrow 0}\hat{f}_{0\mathbf{k}}\left( \hat{\alpha},%
	\hat{\beta}_{u}E_{\mathbf{k}u}, \hat{\beta}_{l}E_{\mathbf{k}l}\right) 
	=f_{0\mathbf{k}}\left( \alpha,\beta E_{\mathbf{k}u}\right) \,.
	\label{hat_f->f_0}
\end{equation}

In complete analogy to the reducible equilibrium moments~(\ref{I_r_tens}), the reducible moments of tensor-rank $n$ of the anisotropic distribution function $\hat{f}_{0\mathbf{k}}$ are defined as 
(see Eq.~(60) in Ref.~\cite{Molnar:2016vvu}), 
\begin{align}
\hat{\mathcal{I}}_{ij}^{\mu _{1}\cdots \mu _{n}} 
& \equiv \int \mathrm{d}K E_{\mathbf{k}u}^{i}E_{\mathbf{k}l}^{j}k^{\mu _{1}}\cdots k^{\mu_{n}} 
\hat{f}_{0\mathbf{k}}  \notag \\
& =\sum_{q=0}^{\lfloor n/2\rfloor }\sum_{r=0}^{n-2q}\left( -1\right)
^{q}b_{nrq}\hat{I}_{i+j+n,j+r,q}\Xi ^{\left( \mu _{1}\mu _{2}\right. }\cdots
\Xi ^{\mu _{2q-1}\mu _{2q}}l^{\mu _{2q+1}}\cdots l^{\mu _{2q+r}}u^{\mu
_{2q+r+1}}\cdots u^{\left. \mu _{n}\right) }\, ,  \label{I_ij_tens}
\end{align}%
where the coefficient $b_{nrq}$ is defined in Eq.~(\ref{b_nrq}). 
The anisotropic thermodynamic integrals $\hat{I}_{nrq}$ are defined as 
\begin{equation}
\hat{I}_{nrq}\left( \hat{\alpha},\hat{\beta}_{u},\hat{\beta}_{l}\right) =%
\frac{\left( -1\right) ^{q}}{\left( 2q\right) !!} \int \mathrm{d}K 
E_{\mathbf{k}u}^{n-r-2q}E_{\mathbf{k}l}^{r}\left( \Xi ^{\mu \nu }k_{\mu }k_{\nu }\right)^{q} 
\hat{f}_{0\mathbf{k}}\, ,  \label{I_nrq}
\end{equation}%
where $\left( 2q\right) !!=2^{q}q!$ is the double factorial of even numbers.
Also note that similarly to Eq.~(\ref{I_r_n_condition}) for tensor-ranks $n\geq 1$, the orthogonal projections of the anisotropic reducible moments vanish, 
$\hat{\mathcal{I}}_{ij}^{\left\{ \mu _{1}\right\} \cdots \left\{ \mu _{n}\right\}} =\hat{\mathcal{I}}_{ij}^{\left\{ \mu _{1}\right. \cdots \left. \mu_{n}\right\} }=0$.

The tensor decomposition of the anisotropic moments~(\ref{I_ij_tens})
for $n=1, 2$, and $i=j=0$, reads
\begin{eqnarray}
\hat{N}^{\mu } &\equiv &\hat{\mathcal{I}}_{00}^{\mu }=\hat{I}_{100}u^{\mu }+%
\hat{I}_{110}l^{\mu }\, ,  \label{N_0_mu_u_l} \\
\hat{T}^{\mu \nu } &\equiv &\hat{\mathcal{I}}_{00}^{\mu \nu }
=\hat{I}_{200}u^{\mu }u^{\nu }+2\hat{I}_{210}u^{\left( \mu \right. }l^{\left. \nu\right) }
+\hat{I}_{220}l^{\mu }l^{\nu }-\hat{I}_{201}\Xi ^{\mu \nu }\, ,
\label{T_0_munu_u_l}
\end{eqnarray}%
where according to Eqs.~(\ref{kinetic:n})--(\ref{kinetic:P_t}) we obtain 
\begin{eqnarray}
\hat{n} &\equiv& \hat{\mathcal{I}}_{10}=\hat{I}_{100}\, ,  \label{n_aniso}  \qquad
\hat{e} \equiv \hat{\mathcal{I}}_{20}=\hat{I}_{200}\, ,  \label{e_aniso} \\
\hat{n}_{l} &\equiv& \hat{\mathcal{I}}_{01}=\hat{I}_{110} \, ,
\label{n_l_aniso} \qquad
\hat{M} \equiv \hat{\mathcal{I}}_{11}=\hat{I}_{210}\, ,  \label{M_aniso} \\
\hat{P}_{l} &\equiv &\hat{\mathcal{I}}_{02}=\hat{I}_{220}\, ,
\label{P_l_aniso} \qquad
\hat{P}_{\perp } \equiv -\frac{1}{2}\left( m_{0}^{2}\hat{\mathcal{I}}_{00}
-\hat{\mathcal{I}}_{20}+\hat{\mathcal{I}}_{02}\right) =\hat{I}_{201}\, .
\label{P_t_aniso}
\end{eqnarray}%
In an arbitrary anisotropic state the intensive quantities $\hat{\alpha}$ and $\hat{\beta}_{u}$ are obtained from the matching conditions 
\begin{eqnarray}
&&\left( \hat{N}^{\mu }-N_{0}^{\mu }\right) u_{\mu }\equiv \hat{n}-n_{0}=0\, , \\
&&\left( \hat{T}^{\mu \nu }-T_{0}^{\mu \nu }\right) u_{\mu }u_{\nu }\equiv 
\hat{e}-e_{0}=0\, , \label{Landau_matching_aniso}
\end{eqnarray}%
leading to $\mu (\hat{\alpha},\hat{\beta}_{u},\hat{\beta}_{l})$ and $T(\hat{\alpha},\hat{\beta}_{u},\hat{\beta}_{l})$, while the new intensive parameter 
$\hat{\beta}_{l}$ is determined from an additional equation of motion. 
Note that the equations of motion of leading-order anisotropic fluid-dynamics and anisotropic MHD are presented later in Eqs.~(\ref{DI_ij}) and (\ref{DI_ij_mu1}).
The five conservation equations of motion together with an equation of state are used to determine $\hat{\alpha}$, $\hat{\beta_u}$ and the four-velocity, $u^{\mu}$.
Therefore, we need an additional equation of motion to close the conservation equations and to determine the remaining independent parameter $\hat{\beta}_l$.
This equation is provided by a higher-order moment of the Boltzmann-Vlasov equation, but in principle there remains an ambiguity which higher-order moment one chooses, see Ref.~\cite{Molnar:2016gwq} for a more detailed discussion.
Finally, the bulk viscous pressure may be defined according to Eq.~(\ref{kinetic:Pi})
\begin{equation}
\hat{\Pi}(\hat{\alpha},\hat{\beta}_{u},\hat{\beta}_{l}) \equiv 
-\frac{1}{3}\left( \hat{T}^{\mu \nu }-T_{0}^{\mu \nu }\right) \Delta_{\mu \nu } = \hat{P} - P_0 \,.
\end{equation}

Similarly to Eq.~(\ref{f=f0+df_iso}) a complete theory of anisotropic dissipative fluid dynamics 
uses the anisotropic distribution function $\hat{f}_{0\mathbf{k}}$ as the starting point for 
an expansion of the single-particle distribution function, 
\begin{equation}
f_{\mathbf{k}}=\hat{f}_{0\mathbf{k}}+\delta \hat{f}_{\mathbf{k}}\,.
\label{f=f0+df_aniso}
\end{equation}%
The rationale is that, in the case of a strong momentum-space anisotropy, an expansion around $\hat{f}_{0\mathbf{k}}$ instead of around $f_{0\mathbf{k}}$ improves the convergence properties of the expansion, provided $|\delta \hat{f}_{\mathbf{k}}|\ll |\delta f_{\mathbf{k}}|$. 
This also means that the AIMs~(\ref{irred_mom_A}) are decomposed analogously to Eq.~(\ref{M=I+rho}),
\begin{equation}
\mathcal{A}_{ij}^{\mu _{1}\cdots \mu _{\ell }}=\hat{\mathcal{I}}_{ij}^{\mu
_{1}\cdots \mu _{\ell }}+\hat{\rho}_{ij}^{\mu _{1}\cdots \mu _{\ell }}\,,
\end{equation}%
where the AIMs of $\delta \hat{f}_{0\mathbf{k}}$ are 
\begin{equation}
\hat{\rho}_{ij}^{\mu _{1}\cdots \mu _{\ell }}\equiv \int \mathrm{d}KE_{\mathbf{k}u}^{i}
E_{\mathbf{k}l}^{j}k^{\left\langle \mu _{1}\right. }\cdots k^{\left.
\mu _{\ell }\right\rangle }\delta \hat{f}_{\mathbf{k}}\,.
\label{kinetic:rho_ij_hat}
\end{equation}
The expansion of $f_{\mathbf{k}}$ in terms of these AIMs is presented in Appendix~\ref{app:expansion_coefficients}, while the corresponding equations of motions for 
the AIMs are presented in Appendix~\ref{app:aniso_resistiveMHD}.

\section{The Boltzmann-Vlasov equation and the general moment equations}
\label{sec:Boltzmann_Vlasov}

The Boltzmann-Vlasov transport equation describes the space-time evolution of the single-particle distribution function in an external electromagnetic field~\cite{deGroot,Cercignani_book}, 
\begin{equation}
k^{\mu }\partial _{\mu }f_{\mathbf{k}}+\mathbf{q}F^{\mu \nu }k_{\nu }\frac{%
\partial }{\partial k^{\mu }}f_{\mathbf{k}}=C\left[ f\right] \,.
\label{BTE_Fmunu}
\end{equation}%
Here, for the sake of simplicity we assume that the particles have a single charge $\mathbf{q}$, and restrict the collision term $C\left[ f\right] $ to binary elastic scattering processes,
\begin{equation}
C\left[ f\right] \equiv \frac{1}{2}\int \mathrm{d}K^{\prime }\mathrm{d}P\mathrm{d}P^{\prime }\left[ W_{\mathbf{pp}^{\prime }\rightarrow \mathbf{kk}^{\prime }}f_{\mathbf{p}}f_{%
\mathbf{p}^{\prime }}\left( 1-af_{\mathbf{k}}\right) \left( 1-af_{\mathbf{k}%
^{\prime }}\right) -W_{\mathbf{kk}^{\prime }\rightarrow \mathbf{pp}^{\prime
}}f_{\mathbf{k}}f_{\mathbf{k}}\left( 1-af_{\mathbf{p}}\right) 
\left( 1-af_{\mathbf{p}^{\prime }}\right) \right] \,,
\end{equation}%
where the factors $1-af$ represent the corrections from quantum statistics. 
The invariant transition rate $W_{\mathbf{kk}^{\prime }\rightarrow \mathbf{pp}^{\prime }}$ satisfies detailed balance, $W_{\mathbf{kk}^{\prime}\rightarrow \mathbf{pp}^{\prime }}=W_{\mathbf{pp}^{\prime }\rightarrow \mathbf{kk}^{\prime }}$, and it is also symmetric with respect to the exchange of the momenta in binary collisions, $W_{\mathbf{kk}^{\prime}\rightarrow \mathbf{pp}^{\prime }}=W_{\mathbf{k}^{\prime }\mathbf{k}\rightarrow \mathbf{pp}^{\prime }}=W_{\mathbf{kk}^{\prime }\rightarrow \mathbf{p}^{\prime }\mathbf{p}}$.

The electromagnetic field-strength tensor, or Faraday tensor, $F^{\mu \nu }$ is an antisymmetric and therefore traceless rank-two tensor, with six independent components corresponding to the electric field vector $\mathbf{E}$ and the magnetic induction field vector $\mathbf{B}$. 
Without loss of generality it is decomposed with respect to the fluid four-velocity 
as~\cite{Cercignani_book} 
\begin{equation}
F^{\mu \nu }\equiv E^{\mu }u^{\nu }-E^{\nu }u^{\mu }+\varepsilon ^{\mu \nu
\alpha \beta }u_{\alpha }B_{\beta }\,,  \label{F_munu}
\end{equation}%
where the electric and the magnetic four-vectors are 
\begin{equation}
E^{\mu }\equiv F^{\mu \nu }u_{\nu }\,,\quad B^{\mu }\equiv \frac{1}{2}%
\varepsilon ^{\mu \nu \alpha \beta }F_{\alpha \beta }u_{\nu }\,.
\label{E_mu_B_mu}
\end{equation}%
Using the antisymmetry of the Faraday tensor and of the Levi-Civit\`{a} tensor, one can show that 
both $E^{\mu }$ and $B^{\mu }$ are orthogonal to the fluid velocity, $E^{\mu }u_{\mu }=0$ and 
$B^{\mu }u_{\mu }=0$. 
In the LR frame of the fluid, they coincide with the usual electric and magnetic fields, i.e., $E_{LR}^{\mu }=\left( 0,\mathbf{E}\right) ^{T}$ and $B_{LR}^{\mu }=\left( 0,\mathbf{B}\right) ^{T}$, with $E^{i}=F^{i0}$ and $B^{i}=-\frac{1}{2}\varepsilon ^{ijk}F_{jk}$. 
The electric field is a polar vector, while the magnetic induction field is an axial vector dual to $F_{jk}$.
Furthermore, $E^{\mu }E_{\mu }=-E^{2}$ and $B^{\mu }B_{\mu }=-B^{2}$. 
The space-like unit vector in the direction of the magnetic field is 
\begin{equation}
b^{\mu }\equiv \frac{B^{\mu }}{B}\, ,  \label{b_mu}
\end{equation}
such that $b^\mu b_\mu = -1$.
This unit four-vector is also orthogonal to the four-velocity, $b^\mu u_\mu =0$, similarly to $l^{\mu }$, which is pointing in the direction of the momentum anisotropy. 
Hence, when the anisotropy is induced by an external magnetic induction field only, then 
$l^{\mu} \equiv b^{\mu }$. 
This is the case in non-resistive magnetohydrodynamics~\cite{Denicol:2018rbw}, where $E^{\mu} = 0$ such that $\mathbf{E} = -\mathbf{v} \times \mathbf{B}$ and $F^{\mu \nu } \equiv B^{\mu \nu} = \varepsilon ^{\mu\nu\alpha \beta }u_{\alpha }B_{\beta }$, hence the anisotropy is in the direction of the magnetic induction field. 
However, without loss of generality, in the following we will keep the anisotropy four-vector $l^{\mu }$, since there are systems with a spatial anisotropy that is not necessarily restricted to point into the direction of the external magnetic induction field and thus in general $l^{\mu}\neq b^{\mu }$. 
This is the case in non-central heavy-ion collisions, where the (largest component of the) magnetic induction field points in the direction transverse to the reaction plane, while the spatial anisotropy vector points in beam direction, i.e., lies in the reaction plane.

The evolution of the electric and magnetic fields are given by Maxwell's equations,
\begin{eqnarray}
\partial _{\mu }F^{\mu \nu } &=& J^{\mu }\, , \\
\varepsilon ^{\mu \nu \alpha \beta }\partial _{\mu }F_{\alpha \beta } &=& 0\, ,
\end{eqnarray}%
where $J^{\mu }$ is the total electromagnetic four-current. 
The electromagnetic stress-energy tensor for a non-polarizable and non-magnetizable fluid is 
\begin{equation}
T_{\text{em}}^{\mu \nu }\equiv -F^{\mu \lambda }F_{\left. {}\right. \lambda }^{\nu
}+\frac{1}{4}g^{\mu \nu }F^{\alpha \beta }F_{\alpha \beta }\, ,
\end{equation}%
and hence Maxwell's equations equations imply that 
\begin{equation}
\partial _{\nu }T_{\text{em}}^{\mu \nu }=-F^{\mu \lambda }J_{\lambda }\, .
\label{T_em_cons}
\end{equation}

In general, an external charge current $J_{\text{ext}}^{\mu }$ provides an additional source term, such that the total electromagnetic four-current is 
\begin{equation}
J^{\mu }\equiv N_{\mathbf{q}}^{\mu }+J_{\text{ext}}^{\mu }\, ,
\end{equation}%
where $N_{\mathbf{q}}^{\mu }=\mathbf{q}N^{\mu }$ is the electric charge four-current of the fluid 
defined using the particle four-current (\ref{kinetic:N_mu}) or (\ref{kinetic:N_mu_u_l}). 
Therefore, the total energy-momentum tensor of a charged fluid in an external electromagnetic field consists of two parts, 
\begin{equation}
T_{\text{tot}}^{\mu \nu }\equiv T^{\mu \nu }+T_{\text{em}}^{\mu \nu }\, ,
\end{equation}%
where the energy-momentum tensor of the fluid is given in Eq.~(\ref{kinetic:T_munu}), or equivalently 
in Eq.~(\ref{kinetic:T_munu_u_l}).

Finally, the equations of MHD are 
\begin{eqnarray}
\partial _{\mu }N_{\mathbf{q}}^{\mu } &=&0\, ,  \label{N_q_mucons} \\
\partial _{\nu }T_{\text{tot}}^{\mu \nu } &=& -F^{\mu \lambda }J_{\text{ext},\lambda }\, ,  \label{T_munucons}
\end{eqnarray}%
and hence from the latter and from Eq.~(\ref{T_em_cons}) it follows that the energy-momentum tensor 
of the fluid satisfies~\cite{Eckart:1940te} 
\begin{equation}
\partial _{\nu }T^{\mu \nu }=F^{\mu \lambda }N_{\mathbf{q},\lambda }\, .
\label{d_mu_T_munu_f}
\end{equation}%
These equations also follow from the relativistic Boltzmann-Vlasov equation noting that the electric charge as well as the energy and momentum are conserved in microscopic collisions. 
We will derive these equations of motion in terms of the irreducible moments in the next sections.

\subsection{Equation of motion for the rank-$\ell $ IIMs}
\label{sec:iso_moment_equations}

Using the decompositions $\partial _{\mu }\equiv u_{\mu }D+\nabla _{\mu }$ and $k^{\mu }\partial _{\mu }=E_{\mathbf{k}u}D+k^{\left\langle \mu \right\rangle }\nabla _{\mu }$, where $D\equiv u^{\mu }\partial _{\mu }$ is the proper-time derivative and $\nabla ^{\mu }\equiv \Delta ^{\mu \nu}\partial _{\nu }$ is the spatial derivative orthogonal to $u^{\mu }$, the Boltzmann-Vlasov equation~(\ref{BTE_Fmunu}) becomes 
\begin{equation}
Df_{\mathbf{k}}=E_{\mathbf{k}u}^{-1}C\left[ f_{\mathbf{k}}\right] -E_{%
\mathbf{k}u}^{-1}k^{\left\langle \mu \right\rangle }\nabla _{\mu }f_{\mathbf{%
k}}-E_{\mathbf{k}u}^{-1}\mathbf{q}F^{\mu \nu }k_{\nu }\frac{\partial f_{%
\mathbf{k}}}{\partial k^{\mu }}\, .  \label{D_BTE_F_munu}
\end{equation}%
The proper-time derivative of the IIMs are projected with $\Delta _{\nu _{1}\cdots \nu _{\ell }}^{\mu _{1}\cdots \mu _{\ell }}$, in order to ensure that they remain irreducible tensors,
\begin{equation}
D\mathcal{M}_{i}^{\left\langle \mu _{1}\cdots \mu _{\ell }\right\rangle}
\equiv \Delta _{\nu _{1}\cdots \nu _{\ell }}^{\mu _{1}\cdots \mu _{\ell }}
D \mathcal{M}_{i}^{\nu _{1}\cdots \nu _{\ell }} 
=  \Delta _{\nu _{1}\cdots \nu _{\ell }}^{\mu _{1}\cdots \mu _{\ell }} 
D \left( \int \mathrm{d}K  E_{\mathbf{k}u}^{i} 
k^{\left\langle \nu _{1}\right. }\cdots k^{\left. \nu _{\ell}\right\rangle }
f_{\mathbf{k}} \right)\, .  \label{def_D_M_i}
\end{equation}%

Now, inserting Eq.~(\ref{D_BTE_F_munu}) into the above definition, see Appendix~\ref{Appendix:DM_i} for the details of the derivation, we obtain the general equation of motion for the rank-$\ell $ IIMs,
\begin{align}
D\mathcal{M}_{i}^{\left\langle \mu _{1}\cdots \mu _{\ell }\right\rangle } & =%
\mathcal{C}_{i-1}^{\mu _{1}\cdots \mu _{\ell }}+i\mathcal{M}_{i-1}^{\mu
_{1}\cdots \mu _{\ell +1}}Du_{\mu _{\ell +1}}+\frac{1}{3}\theta \left[
m_{0}^{2}\left( i-1\right) \mathcal{M}_{i-2}^{\mu _{1}\cdots \mu _{\ell
}}-\left( i+\ell +2\right) \mathcal{M}_{i}^{\mu _{1}\cdots \mu _{\ell }}%
\right]   \notag \\
& -\Delta _{\nu _{1}\cdots \nu _{\ell }}^{\mu _{1}\cdots \mu _{\ell }}\nabla
_{\nu _{\ell +1}}\mathcal{M}_{i-1}^{\nu _{1}\cdots \nu _{\ell +1}}+\frac{%
\ell }{2\ell +1}\left[ m_{0}^{2}i\mathcal{M}_{i-1}^{\left\langle \mu
_{1}\right. \cdots \mu _{\ell -1}}-\left( i+2\ell +1\right) \mathcal{M}%
_{i+1}^{\left\langle \mu _{1}\right. \cdots \mu _{\ell -1}}\right]
Du^{\left. \mu _{\ell }\right\rangle }  \notag \\
& -\frac{\ell }{2\ell +1}\nabla ^{\left\langle \mu _{1}\right. }\left(
m_{0}^{2}\mathcal{M}_{i-1}^{\mu _{2}\cdots \left. \mu _{\ell }\right\rangle
}-\mathcal{M}_{i+1}^{\mu _{2}\cdots \left. \mu _{\ell }\right\rangle
}\right) +\left( i-1\right) \mathcal{M}_{i-2}^{\mu _{1}\cdots \mu _{\ell
+2}}\sigma _{\mu _{\ell +1}\mu _{\ell +2}}  \notag \\
& +\frac{\ell }{2\ell +3}\left[ m_{0}^{2}\left( 2i-2\right) \mathcal{M}%
_{i-2}^{\lambda \left\langle \mu _{1}\right. \cdots \mu _{\ell -1}}-\left(
2i+2\ell +1\right) \mathcal{M}_{i}^{\lambda \left\langle \mu _{1}\right.
\cdots \mu _{\ell -1}}\right] \sigma _{\lambda }^{\left. \mu _{\ell
}\right\rangle }+\ell \mathcal{M}_{i}^{\lambda \left\langle \mu _{1}\right.
\cdots \mu _{\ell -1}}\omega _{\left. {}\right. \lambda }^{\left. \mu _{\ell
}\right\rangle }  \notag \\
& +\frac{\ell \left( \ell -1\right) }{4\ell ^{2}-1}\left[ m_{0}^{4}\left(
i-1\right) \mathcal{M}_{i-2}^{\left\langle \mu _{1}\right. \cdots \mu _{\ell
-2}}-m_{0}^{2}\left( 2i+2\ell -1\right) \mathcal{M}_{i}^{\left\langle \mu
_{1}\right. \cdots \mu _{\ell -2}}+\left( i+2\ell \right) \mathcal{M}%
_{i+2}^{\left\langle \mu _{1}\right. \cdots \mu _{\ell -2}}\right] \sigma
^{\mu _{\ell -1}\left. \mu _{\ell }\right\rangle }  \notag \\
& -\left( i-1\right) \mathcal{M}_{i-2}^{\left\langle \mu _{1}\right. \cdots
\left. \mu _{\ell }\right\rangle \mu _{\ell +1}}\mathbf{q}E_{\mu _{\ell +1}}-%
\frac{\ell }{2\ell +1}\left[ m_{0}^{2}\left( i-1\right) \mathcal{M}%
_{i-2}^{\left\langle \mu _{1}\right. \cdots \mu _{\ell -1}}-\left( i+2\ell
\right) \mathcal{M}_{i}^{\left\langle \mu _{1}\right. \cdots \mu _{\ell -1}}%
\right] \mathbf{q}E^{\left. \mu _{\ell }\right\rangle }  \notag \\
& +\ell \Delta _{\nu _{1}\cdots \nu _{\ell }}^{\mu _{1}\cdots \mu _{\ell }}%
\mathcal{M}_{i-1}^{\nu _{1}\cdots \nu _{\ell -1}\nu _{\ell +1}}\mathbf{q}%
B^{\nu _{\ell }\nu }g_{\nu \nu _{\ell +1}}\, ,  \label{Main_DM_i_mu1_mun}
\end{align}%
where we denoted the irreducible moments of the collision integral by 
\begin{equation}
\mathcal{C}_{i-1}^{\mu _{1}\cdots \mu _{\ell }}\equiv \int \mathrm{d}K E_{\mathbf{k}u}^{i-1} k^{\left\langle \mu _{1}\right. }\cdots k^{\left. \mu_{\ell }\right\rangle } 
C\left[ f_{\mathbf{k}}\right] \, .
\label{def_C_i_mu1_mun}
\end{equation}%
Note that as a consequence of charge conservation as well as conservation of energy and momentum 
in binary collisions for any distribution function $f_{\mathbf{k}}$, we have 
\begin{equation}
\mathcal{C}_{0}=0\,,\quad \mathcal{C}_{1}=0\,,\quad \mathcal{C}_{0}^{\langle\mu \rangle }=0\, .  \label{C_0_C_1_C_0_mu}
\end{equation}%
Furthermore, in the derivation of Eq.~(\ref{Main_DM_i_mu1_mun}) we repeatedly used the Cauchy-Stokes decomposition 
\begin{equation}
\partial _{\mu }u_{\nu }=u_{\mu }Du_{\nu }+\frac{1}{3}\theta \Delta _{\mu
\nu }+\sigma _{\mu \nu }+\omega _{\mu \nu }\, .
\end{equation}%
where $\theta \equiv \nabla _{\mu }u^{\mu }$ is the expansion scalar, $\sigma^{\mu \nu }\equiv \partial ^{\left\langle \mu \right. }u^{\left. \nu\right\rangle }=\nabla ^{\left( \mu \right. }u^{\left. \nu \right) }-\theta\Delta ^{\mu \nu }/3$ is the shear tensor, and $\omega ^{\mu \nu }\equiv\nabla ^{\left[ \mu \right. }u^{\left. \nu \right] }=\left( \nabla ^{\mu}u^{\nu }-\nabla ^{\nu }u^{\mu }\right) /2$ is the vorticity tensor.
Also note that $\omega^{\mu}_{\   \nu} =- \omega^{\ \mu}_{\nu}$, where $\omega^{\mu}_{\  \nu} = \omega^{\mu \alpha} g_{\alpha \nu}$ and $\omega^{\ \mu}_{\nu} = \omega^{\alpha \mu} g_{\alpha \nu}$.
In the last line of Eq.~(\ref{Main_DM_i_mu1_mun}) we introduced a new tensor $B^{\mu \nu}=\varepsilon ^{\mu \nu \alpha \beta }u_{\alpha }B_{\beta }$ that is normalized to $B^{\mu \nu }B_{\mu \nu }=2B^{2}$ and is orthogonal to both the fluid four-velocity and the magnetic induction field, $B^{\mu \nu }u_{\nu }=0$ and $B^{\mu\nu }B_{\nu }=0$.

The first part of this general moment equation, i.e., the first five lines, derived from the 
Boltzmann equation without the electromagnetic contributions, was obtained recently by 
de Brito and Denicol~\cite{deBrito:2024vhm}, see Eq.~(19) therein. 
The last two lines of Eq.~(\ref{Main_DM_i_mu1_mun}) follow from the last term of Eq.~(\ref{D_BTE_F_munu}), i.e., the Vlasov term, and describe the coupling of the IIMs to the electric and the magnetic fields. 
These additional terms generalize the equations of motion for rank-$\ell$ IIMs to electrically 
conducting fluids. 
Note that the general equations of motion for the reducible rank-$\ell $ tensor moments $M_{r}^{\left\langle \mu _{1}\right\rangle \cdots \left\langle \mu _{\ell }\right\rangle }$ were obtained by Tinti~\textit{et al.}~\cite{Tinti:2018qfb}, see Eq.~(10) and Eq.~(20) therein, corresponding to the equations of motion from the Boltzmann equation with and without the Vlasov term, respectively.

The equation of motion for a given IIM of tensor rank-$\ell$ and energy index $i$ always involves moments with both lower and higher tensor-rank, i.e., $\ell - 2$, $\ell - 1$, $\ell$, $\ell + 1$, and $\ell +2$. 
Furthermore, the energy index of these moments also varies independently from the tensor-rank, i.e., $i-2$, $i-1$, $i$, $i+1$, and $i+2$. 
This leads to an infinite hierarchy of coupled equations of motion that involves moments of arbitrary tensor-rank to obtain the single-particle distribution function as the solution of the Boltzmann-Vlasov equation.

\subsection{Transient fluid dynamics and MHD}
\label{sec:iso_fluid_dynamics}

Relativistic fluid dynamics is an effective theory for the long-wavelength, small-frequency dynamics of macroscopic systems. 
Applying the method of moments, the fluid-dynamical equations of motion are derived from the Boltzmann equation, while the MHD equations are derived from the Boltzmann-Vlasov equation. 
The infinite set of coupled moment equations~(\ref{Main_DM_i_mu1_mun}) is still completely equivalent to the Boltzmann(-Vlasov) equation and independent of the expansion of the single-particle distribution function.
However, in order to derive the fluid-dynamical or MHD equations of motion one needs to expand the single-particle distribution function $f_{\mathbf{k}}$ around a certain specific distribution function, e.g., $f_{0\mathbf{k}}$, and then systematically truncate the system of the moment equations in order to arrive at a closed set of equations.

The charge conservation equation~(\ref{N_q_mucons}) follows from Eq.~(\ref{Main_DM_i_mu1_mun}) for tensor-rank $\ell=0$ and energy index $i=1$, corresponding to the equation of motion for $\mathcal{M}_{1} \equiv n$ with $\mathcal{C}_{0}=0$. 
The evolution of the energy, i.e., Eq.~(\ref{d_mu_T_munu_f}) projected onto $u_\mu $, follows from Eq.~(\ref{Main_DM_i_mu1_mun}) for tensor-rank $\ell=0$ and energy index $i=2$, where $\mathcal{C}_{1}=0$. 
This is precisely the equation of motion for $\mathcal{M}_{2} \equiv e$. 
Finally, the equation for the three-momentum, i.e., Eq.~(\ref{d_mu_T_munu_f}) projected onto the three-space orthogonal to the four-velocity, is obtained for tensor-rank $\ell=1$ and energy index $i=1$ where $\mathcal{C}^{\mu}_{0}=0$. 
This corresponds to the equation for the vector moment $\mathcal{M}^{\mu}_{1} \equiv W^{\mu}$. 
For any given four-velocity, these five equations contain the 14 degrees of freedom that are contained in $N^{\mu} $ and $T^{\mu \nu}$.

In the case of an ideal fluid these five equations represent the equations of motion of non-resistive, non-dissipative MHD that are obtained under the assumption that the fluid is in local thermodynamical equilibrium everywhere in space-time, i.e., $f_{\mathbf{k}} \equiv f_{0\mathbf{k}}$ and therefore $\mathcal{M}_{r}^{\left\langle \mu _{1}\cdots \mu_{\ell }\right\rangle } \equiv \mathcal{I}_{r}^{\left\langle \mu _{1}\cdots \mu _{\ell}\right\rangle }$, while $\delta f_{\mathbf{k}} =0$ and hence $\rho _{r}^{\mu _{1}\cdots \mu _{\ell }} \equiv 0$. 

Using these idealizations the scalar equation of motion follows for tensor rank $\ell =0$ from Eq.~(\ref{Main_DM_i_mu1_mun}) and reads
\begin{equation}
D\mathcal{I}_{i}=\frac{1}{3}\theta \left[ m_{0}^{2}\left( i-1\right) 
\mathcal{I}_{i-2}-\left( i+2\right) \mathcal{I}_{i}\right] \, ,  
\label{DI_i}
\end{equation}%
where all scalar moments of the collision integral of the equilibrium distribution vanish identically, $\mathcal{C}_{i-1}=0$. 
For $\mathcal{I}_{1}=n_{0}$ we obtain the charge conservation equation, and for $\mathcal{I}_{2}=e_{0}$ the energy conservation equation.

The vector equation follows from Eq.~(\ref{Main_DM_i_mu1_mun}) for tensor-rank $\ell =1$,
\begin{align}
0& =\frac{1}{3}\left[ m_{0}^{2}i\mathcal{I}_{i-1}-\left( i+3\right) 
\mathcal{I}_{i+1}\right] Du^{\mu _{1}}-\frac{1}{3}\nabla ^{\mu _{1}} 
\left( m_{0}^{2} \mathcal{I}_{i-1}-\mathcal{I}_{i+1}\right)  \notag \\
& -\left( i-1\right) \mathcal{I}_{i-2}^{\left\langle \mu _{1}\right\rangle
\mu _{2}}\mathbf{q}E_{\mu _{2}}-\frac{1}{3}\left[ m_{0}^{2}\left( i-1\right) 
\mathcal{I}_{i-2}-\left( i+2\right) \mathcal{I}_{i}\right] \mathbf{q}E^{\mu_{1}}\, ,  
\label{DI_i_mu1}
\end{align}%
where all vector moments of the collision integral vanish in equilibrium, $\mathcal{C}_{i-1}^{\mu _{1}}=0$.

However, since the orthogonal projections of the equilibrium moments vanish for any tensor-rank 
$\ell \geq 1$, see Eq.~(\ref{I_r_n_condition}), there are no new or additional equations of motion 
other than the conservation laws. 
Furthermore, noting that $f_{0\mathbf{k}}$ solely depends on five independent variables, see Eq.~(\ref{f_0k}), hence $N^{\mu} \equiv N^{\mu}_0$ and $T^{\mu \nu} \equiv T^{\mu \nu}_0$, see Eqs.~(\ref{N0_mu}) and (\ref{T0_munu}), the equations of ideal fluid dynamics or ideal MHD are 
closed by an equation of state, $P_0=P_0(n_0,e_0)$.

On the other hand, a resistive and dissipative fluid includes non-equilibrium corrections $\delta f_{\mathbf{k}}$, leading to $f_{\mathbf{k}} \equiv f_{0\mathbf{k}}+\delta f_{\mathbf{k}}$ and implying that $\mathcal{M}_{r}^{\mu_{1}\cdots \mu _{\ell }} \equiv \mathcal{I}_{r}^{\mu _{1}\cdots \mu _{\ell }} +\rho _{r}^{\mu _{1}\cdots \mu _{\ell }}$. 
This also means that in order to obtain a closed system of equations of motion, one must truncate the infinite hierarchy of moment equations~(\ref{Main_DM_i_mu1_mun}). 
To this end the general equations of motion of rank $\ell = 0$, $\ell = 1$, and $\ell = 2$ were derived earlier in Ref.~\cite{Denicol:2012cn} from the Boltzmann equation, see Eqs.~(35) -- (37) therein. 
The equations of motion for tensor ranks $\ell=3$ and $\ell=4$ were also obtained earlier by 
de Brito and Denicol, see Eqs.~(26) and~(27) in Ref.~\cite{deBrito:2023tgb}. 
The equations of motion for IIMs up to tensor-rank two and also including the Vlasov term were obtained earlier in Ref.~\cite{Denicol:2019iyh}, and one can easily show that Eq.~(\ref{Main_DM_i_mu1_mun}) is consistent with Eq.~(20) for the scalar, Eq.~(21) for the vector, and Eq.~(22) for the tensor-rank two of Ref.~\cite{Denicol:2019iyh}.
These equations of motion are also derived from Eq.~(\ref{Main_DM_i_mu1_mun}) and are presented in Appendix~\ref{app:resistiveMHD}.

The simplest and most widely used truncation of the relativistic moment equations~(\ref{Main_DM_i_mu1_mun}) is the 14 dynamical-moment approximation of Israel and Stewart~\cite{Israel:1979wp}, which is the proper relativistic generalization of the well-known Grad's 13-moment approximation~\cite{Grad}. 
This method is based on an explicit truncation of the expansion of the distribution function~(\ref{f_iso_expansion}) at tensor-rank two and then using this approximate solution to obtain nine additional equations of motion from the various projections of the third-rank tensor moment $M^{\mu_1 \mu_2 \mu_3}_0$.
These precisely correspond to equations of motion for $\rho_3$, $\rho_2^\mu$, and $\rho_1^{\mu \nu}$, see for example Refs.~\cite{Betz:2010cx,Denicol:2012es}.
Furthermore, the local rest frame of the fluid is usually chosen according to Landau's definition~\cite{Landau_book}, i.e., $\mathcal{M}^{\mu}_{1} \equiv \rho^{\mu}_{1} = W^{\mu} = 0$, meaning that the four-velocity is the timelike eigenvector of the energy-momentum tensor, $u^\mu = T^{\mu\nu} u_\nu/(u_\alpha T^{\alpha \beta} u_\beta)$. 
In case of a single conserved charge one can also define the local rest frame according to 
Eckart~\cite{Eckart:1940te}, i.e., choosing $\mathcal{M}^{\mu}_{0} \equiv \rho^{\mu}_{0} = V^{\mu} = 0$.
The chemical potential and the temperature are inferred from the Landau matching conditions, Eqs.~(\ref{Landau_matching_iso_n}) and (\ref{Landau_matching_iso_e}), i.e., $\rho_1 = 0$ and $\rho_2 = 0$. 

The 14 dynamical-moment approximation can also be applied to truncate the equation of motion~(\ref{Main_DM_i_mu1_mun}) for the IIMs. 
The corresponding equations of motion of second-order fluid dynamics and MHD follow from Eq.~(\ref{Main_DM_i_mu1_mun}) for energy index $i=0$ and tensor-ranks $\ell= 0$, $\ell = 1$, 
and $\ell = 2$, see for example Refs.~\cite{Denicol:2019iyh,Denicol:2012cn}.
To this end, the distribution function is expanded around local thermodynamic equilibrium and the corresponding sums over tensor-rank $\ell$, see Eq.~(\ref{f_iso_expansion}), are truncated at $N_{0}=2$, $N_{1}=1$, and $N_{2}=0$. 
Within this truncation the equations of motion for the primary dissipative fields, the lowest-order IIMs of $\delta f_{\mathbf{k}}$ appearing in $N^{\mu}$ and $T^{\mu \nu}$, i.e., $\rho_{0} = -3 \Pi/m^2_0$, $\rho^{\mu}_{0} = V^{\mu}$, and $\rho^{\mu \nu}_{0} = \pi^{\mu\nu}$ are chosen to be dynamical. 
All remaining IIMs are considered nondynamical moments of the truncation and are approximated in 
various ways, see for example Refs.~\cite{Ambrus:2022vif,Wagner:2023joq} for three slightly different choices.

The natural extension of second-order fluid dynamics beyond the $14$-moment approximation, are higher-order fluid-dynamical theories which include irreducible tensor moments of rank $\ell \ge 3$, see for example Refs.~\cite{deBrito:2023tgb,deBrito:2024vhm}. 
Similarly we also expect that including higher-order dynamical moments from Eq.~(\ref{Main_DM_i_mu1_mun}), e.g., see the equations of motion in Appendix~\ref{app:resistiveMHD}, would also improve on the second-order theories of resistive and dissipative MHD.

\subsection{The equation of motion for rank-$\ell$ AIMs}
\label{sec:aniso_moment_equations}

Analogous to the previous section we here derive the equation of motion for a rank-$\ell$ AIM $\mathcal{A}_{ij}^{\left\{ \mu_{1}\cdots \mu _{\ell }\right\} }$.
Using the decompositions $\partial _{\mu}\equiv u^{\mu }D+l^{\mu }D_{l}+\tilde{\nabla}_{\mu }$ and $k^{\mu }\partial_{\mu }=E_{\mathbf{k}u}D-E_{\mathbf{k}l}D_{l}+k^{\left\{ \mu \right\} }\tilde{\nabla}_{\mu }$, where $D_{l}\equiv -l^{\mu }\partial _{\mu }$ is the spatial derivative in the direction of the momentum anisotropy and $\tilde{\nabla}^{\mu }\equiv \Xi^{\mu \nu }\partial _{\nu }$ is the spatial derivative orthogonal to both $u^{\mu }$ and $l^{\mu }$, the Boltzmann-Vlasov equation (\ref{BTE_Fmunu}) becomes 
\begin{equation}
Df_{\mathbf{k}}=E_{\mathbf{k}u}^{-1}C\left[ f_{\mathbf{k}}\right] -E_{%
\mathbf{k}u}^{-1}k^{\left\{ \mu \right\} }\tilde{\nabla}_{\mu }f_{\mathbf{k}%
}+E_{\mathbf{k}u}^{-1}E_{\mathbf{k}l}D_{l}f_{\mathbf{k}}-E_{\mathbf{k}u}^{-1}%
\mathbf{q}F^{\mu \nu }k_{\nu }\frac{\partial f_{\mathbf{k}}}{\partial k^{\mu}}\,.  
\label{D_BTE_F_munu_aniso}
\end{equation}%
The proper-time derivative of the AIMs is defined analogous to Eq.~(\ref{def_D_M_i}),%
\begin{equation}
D\mathcal{A}_{ij}^{\left\{ \mu _{1}\cdots \mu _{\ell }\right\} }\equiv \Xi
_{\nu _{1}\cdots \nu _{\ell }}^{\mu _{1}\cdots \mu _{\ell }}D\mathcal{A}%
_{ij}^{\nu _{1}\cdots \nu _{\ell }} 
= \Xi _{\nu _{1}\cdots \nu _{\ell }}^{\mu _{1}\cdots \mu _{\ell }} 
D\left( \int \mathrm{d}K E_{\mathbf{k}u}^{i}E_{\mathbf{k}l}^{j} 
k^{\left\{ \nu _{1}\right. }\cdots k^{\left. \nu _{\ell }\right\} } f_{\mathbf{k}} \right) \,,
\end{equation}%
and hence inserting Eq.~(\ref{D_BTE_F_munu_aniso}) into the above definition leads to an equation for the proper-time derivative of a rank-$\ell $ AIM, see Appendix~\ref{Appendix:DA_ij} for more details of the derivation, 
\begin{align}
D\mathcal{A}_{ij}^{\left\{ \mu _{1}\cdots \mu _{\ell }\right\} }& =\mathcal{C%
}_{i-1,j}^{\mu _{1}\cdots \mu _{\ell }}+\left[ i\mathcal{A}_{i-1,j+1}^{\mu
_{1}\cdots \mu _{\ell }}+j\mathcal{A}_{i+1,j-1}^{\mu _{1}\cdots \mu _{\ell }}%
\right] l_{\lambda }Du^{\lambda }-\left[ \left( i-1\right) \mathcal{A}%
_{i-2,j+2}^{\mu _{1}\cdots \mu _{\ell }}+\left( j+1\right) \mathcal{A}%
_{i,j}^{\mu _{1}\cdots \mu _{\ell }}\right] l_{\lambda }D_{l}u^{\lambda } 
\notag \\
& +i\mathcal{A}_{i-1,j}^{\mu _{1}\cdots \mu _{\ell +1}}Du_{\mu _{\ell
+1}}-\left( i-1\right) \mathcal{A}_{i-2,j+1}^{\mu _{1}\cdots \mu _{\ell
+1}}D_{l}u_{\mu _{\ell +1}}-j\mathcal{A}_{i,j-1}^{\mu _{1}\cdots \mu _{\ell
+1}}Dl_{\mu _{\ell +1}}+\left( j+1\right) \mathcal{A}_{i-1,j}^{\mu
_{1}\cdots \mu _{\ell +1}}D_{l}l_{\mu _{\ell +1}}  \notag \\
& +\Xi _{\nu _{1}\cdots \nu _{\ell }}^{\mu _{1}\cdots \mu _{\ell }}\left(
D_{l}\mathcal{A}_{i-1,j+1}^{\nu _{1}\cdots \nu _{\ell }}\right) -\Xi _{\nu
_{1}\cdots \nu _{\ell }}^{\mu _{1}\cdots \mu _{\ell }}\left( \tilde{\nabla}%
_{\nu _{\ell +1}}\mathcal{A}_{i-1,j}^{\nu _{1}\cdots \nu _{\ell +1}}\right) -%
\frac{1}{2}\tilde{\nabla}^{\left\{ \mu _{1}\right. }\left( m_{0}^{2}\mathcal{%
A}_{i-1,j}^{\mu _{2}\cdots \left. \mu _{\ell }\right\} }-\mathcal{A}%
_{i+1,j}^{\mu _{2}\cdots \left. \mu _{\ell }\right\} }+\mathcal{A}%
_{i-1,j+2}^{\mu _{2}\cdots \left. \mu _{\ell }\right\} }\right)   \notag \\
& +\left( i-1\right) \Xi _{\nu _{1}\cdots \nu _{\ell }}^{\mu _{1}\cdots \mu
_{\ell }}\mathcal{A}_{i-2,j+1}^{\nu _{1}\cdots \nu _{\ell +1}}l_{\lambda }%
\tilde{\nabla}_{\nu _{\ell +1}}u^{\lambda }+\frac{ i-1 }{2}\Xi
_{\nu _{1}\cdots \nu _{\ell }}^{\mu _{1}\cdots \mu _{\ell }}\left( m_{0}^{2}%
\mathcal{A}_{i-2,j+1}^{\nu _{1}\cdots \nu _{\ell -1}}-\mathcal{A}%
_{i,j+1}^{\nu _{1}\cdots \nu _{\ell -1}}+\mathcal{A}_{i-2,j+3}^{\nu
_{1}\cdots \nu _{\ell -1}}\right) l_{\lambda }\tilde{\nabla}^{\nu _{\ell
}}u^{\lambda }  \notag \\
& +j\Xi _{\nu _{1}\cdots \nu _{\ell }}^{\mu _{1}\cdots \mu _{\ell }}\mathcal{%
A}_{i,j-1}^{\nu _{1}\cdots \nu _{\ell +1}}l_{\lambda }\tilde{\nabla}_{\nu
_{\ell +1}}u^{\lambda }+\frac{j}{2}\Xi _{\nu _{1}\cdots \nu _{\ell }}^{\mu
_{1}\cdots \mu _{\ell }}\left( m_{0}^{2}\mathcal{A}_{i,j-1}^{\nu _{1}\cdots
\nu _{\ell -1}}-\mathcal{A}_{i+2,j-1}^{\nu _{1}\cdots \nu _{\ell -1}}+%
\mathcal{A}_{i,j+1}^{\nu _{1}\cdots \nu _{\ell -1}}\right) l_{\lambda }%
\tilde{\nabla}^{\nu _{\ell }}u^{\lambda } \notag \\
& +\frac{1}{2}\left[ m_{0}^{2}i\mathcal{A}_{i-1,j}^{\left\{ \mu _{1}\right.
\cdots \mu _{\ell -1}}-\left( i+2\ell \right) \mathcal{A}_{i+1,j}^{\left\{
\mu _{1}\right. \cdots \mu _{\ell -1}}+i\mathcal{A}_{i-1,j+2}^{\left\{ \mu
_{1}\right. \cdots \mu _{\ell -1}}\right] Du^{\left. \mu _{\ell }\right\} } 
\notag \\
& -\frac{1}{2}\left[ m_{0}^{2}\left( i-1\right) \mathcal{A}%
_{i-2,j+1}^{\left\{ \mu _{1}\right. \cdots \mu _{\ell -1}}-\left( i+2\ell
-1\right) \mathcal{A}_{i,j+1}^{\left\{ \mu _{1}\right. \cdots \mu _{\ell
-1}}+\left( i-1\right) \mathcal{A}_{i-2,j+3}^{\left\{ \mu _{1}\right. \cdots
\mu _{\ell -1}}\right] D_{l}u^{\left. \mu _{\ell }\right\} }  \notag \\
& -\frac{1}{2}\left[ m_{0}^{2}j\mathcal{A}_{i,j-1}^{\left\{ \mu _{1}\right.
\cdots \mu _{\ell -1}}-j\mathcal{A}_{i+2,j-1}^{\left\{ \mu _{1}\right.
\cdots \mu _{\ell -1}}+\left( j+2\ell \right) \mathcal{A}_{i,j+1}^{\left\{
\mu _{1}\right. \cdots \mu _{\ell -1}}\right] Dl^{\left. \mu _{\ell
}\right\} }  \notag \\
& +\frac{1}{2}\left[ m_{0}^{2}\left( j+1\right) \mathcal{A}_{i-1,j}^{\left\{
\mu _{1}\right. \cdots \mu _{\ell -1}}-\left( j+1\right) \mathcal{A}%
_{i+1,j}^{\left\{ \mu _{1}\right. \cdots \mu _{\ell -1}}+\left( j+2\ell
+1\right) \mathcal{A}_{i-1,j+2}^{\left\{ \mu _{1}\right. \cdots \mu _{\ell
-1}}\right] D_{l}l^{\left. \mu _{\ell }\right\} }  \notag \\
& +\frac{1}{2}\tilde{\theta}\left[ m_{0}^{2}\left( i-1\right) \mathcal{A}%
_{i-2,j}^{\mu _{1}\cdots \mu _{\ell }}-\left( i+\ell +1\right) \mathcal{A}%
_{i,j}^{\mu _{1}\cdots \mu _{\ell }}+\left( i-1\right) \mathcal{A}%
_{i-2,j+2}^{\mu _{1}\cdots \mu _{\ell }}\right] +\left( i-1\right) \mathcal{A%
}_{i-2,j}^{\mu _{1}\cdots \mu _{\ell +2}}\tilde{\sigma}_{\mu _{\ell +1}\mu
_{\ell +2}}  \notag \\
& +\frac{\ell }{\ell +1}\left[ m_{0}^{2}\left( i-1\right) \mathcal{A}%
_{i-2,j}^{\lambda \left\{ \mu _{1}\right. \cdots \mu _{\ell -1}}-\left(
i+\ell \right) \mathcal{A}_{i,j}^{\lambda \left\{ \mu _{1}\right. \cdots \mu
_{\ell -1}}+\left( i-1\right) \mathcal{A}_{i-2,j+2}^{\lambda \left\{ \mu
_{1}\right. \cdots \mu _{\ell -1}}\right] \tilde{\sigma}_{\lambda }^{\left.
\mu _{\ell }\right\} }+\ell \mathcal{A}_{i,j}^{\lambda \left\{ \mu
_{1}\right. \cdots \mu _{\ell -1}}\tilde{\omega}_{\left. {}\right. \lambda
}^{\left. \mu _{\ell }\right\} }  \notag \\
& +\frac{1}{4}\left[ m_{0}^{4}\left( i-1\right) \mathcal{A}_{i-2,j}^{\left\{
\mu _{1}\right. \cdots \mu _{\ell -2}}-2m_{0}^{2}\left( i+\ell -1\right) 
\mathcal{A}_{i,j}^{\left\{ \mu _{1}\right. \cdots \mu _{\ell -2}}-2\left(
i+\ell -1\right) \mathcal{A}_{i,j+2}^{\left\{ \mu _{1}\right. \cdots \mu
_{\ell -2}}\right] \tilde{\sigma}^{\mu _{\ell -1}\left. \mu _{\ell }\right\}
}  \notag \\
& +\frac{1}{4}\left[ 2m_{0}^{2}\left( i-1\right) \mathcal{A}%
_{i-2,j+2}^{\left\{ \mu _{1}\right. \cdots \mu _{\ell -2}}+\left( i+2\ell
-1\right) \mathcal{A}_{i+2,j}^{\left\{ \mu _{1}\right. \cdots \mu _{\ell
-2}}+\left( i-1\right) \mathcal{A}_{i-2,j+4}^{\left\{ \mu _{1}\right. \cdots
\mu _{\ell -2}}\right] \tilde{\sigma}^{\mu _{\ell -1}\left. \mu _{\ell
}\right\} }  \notag \\
& -\frac{1}{2}\tilde{\theta}_{l}\left[ m_{0}^{2}j\mathcal{A}_{i-1,j-1}^{\mu
_{1}\cdots \mu _{\ell }}-j\mathcal{A}_{i+1,j-1}^{\mu _{1}\cdots \mu _{\ell
}}+\left( j+\ell +2\right) \mathcal{A}_{i-1,j+1}^{\mu _{1}\cdots \mu _{\ell
}}\right] -j\mathcal{A}_{i-1,j-1}^{\mu _{1}\cdots \mu _{\ell +2}}\tilde{%
\sigma}_{l,\mu _{\ell +1}\mu _{\ell +2}}  \notag \\
& -\frac{\ell }{\ell +1}\left[ m_{0}^{2}j\mathcal{A}_{i-1,j-1}^{\lambda
\left\{ \mu _{1}\right. \cdots \mu _{\ell -1}}-j\mathcal{A}%
_{i+1,j-1}^{\lambda \left\{ \mu _{1}\right. \cdots \mu _{\ell -1}}+\left(
j+\ell +1\right) \mathcal{A}_{i-1,j+1}^{\lambda \left\{ \mu _{1}\right.
\cdots \mu _{\ell -1}}\right] \tilde{\sigma}_{l,\lambda }^{\left. \mu _{\ell
}\right\} }+\ell \mathcal{A}_{i-1,j+1}^{\lambda \left\{ \mu _{1}\right.
\cdots \mu _{\ell -1}}\tilde{\omega}_{l,\left. {}\right. \lambda }^{\left.
\mu _{\ell }\right\} }  \notag \\
& -\frac{1}{4}\left[ m_{0}^{4}j\mathcal{A}_{i-1,j-1}^{\left\{ \mu
_{1}\right. \cdots \mu _{\ell -2}}-2m_{0}^{2}j\mathcal{A}_{i+1,j-1}^{\left\{
\mu _{1}\right. \cdots \mu _{\ell -2}}-2\left( j+\ell \right) \mathcal{A}%
_{i+1,j+1}^{\left\{ \mu _{1}\right. \cdots \mu _{\ell -2}}\right] \tilde{%
\sigma}_{l}^{\mu _{\ell -1}\left. \mu _{\ell }\right\} }  \notag \\
& -\frac{1}{4}\left[ 2m_{0}^{2}\left( j+\ell \right) \mathcal{A}%
_{i-1,j+1}^{\left\{ \mu _{1}\right. \cdots \mu _{\ell -2}}+j\mathcal{A}%
_{i+3,j-1}^{\left\{ \mu _{1}\right. \cdots \mu _{\ell -2}}+\left( j+2\ell
\right) \mathcal{A}_{i-1,j+3}^{\left\{ \mu _{1}\right. \cdots \mu _{\ell -2}}%
\right] \tilde{\sigma}_{l}^{\mu _{\ell -1}\left. \mu _{\ell }\right\} } 
\notag \\
& -\left[ \left( i-1\right) \mathcal{A}_{i-2,j+1}^{\mu _{1}\cdots \mu _{\ell
}}+j\mathcal{A}_{i,j-1}^{\mu _{1}\cdots \mu _{\ell }}\right] \mathbf{q}%
E^{\nu }l_{\nu }-\left( i-1\right) \Xi _{\nu _{1}\cdots \nu _{\ell }}^{\mu
_{1}\cdots \mu _{\ell }}\mathcal{A}_{i-2,j}^{\nu _{1}\cdots \nu _{\ell +1}}%
\mathbf{q}E_{\nu _{\ell +1}}  \notag \\
& -\frac{1}{2}\left[ m_{0}^{2}\left( i-1\right) \mathcal{A}_{i-2,j}^{\left\{
\mu _{1}\right. \cdots \mu _{\ell -1}}-\left( i+2\ell -1\right) \mathcal{A}%
_{i,j}^{\left\{ \mu _{1}\right. \cdots \mu _{\ell -1}}+\left( i-1\right) 
\mathcal{A}_{i-2,j+2}^{\left\{ \mu _{1}\right. \cdots \mu _{\ell -1}}\right] 
\mathbf{q}E^{\left. \mu _{\ell }\right\} }  \notag \\
& -j\Xi _{\nu _{1}\cdots \nu _{\ell }}^{\mu _{1}\cdots \mu _{\ell }}\mathcal{%
A}_{i-1,j-1}^{\nu _{1}\cdots \nu _{\ell +1}}\mathbf{q}B^{\mu \nu }l_{\mu
}g_{\nu \nu _{\ell +1}}+\ell \Xi _{\nu _{1}\cdots \nu _{\ell }}^{\mu
_{1}\cdots \mu _{\ell }}\mathcal{A}_{i-1,j}^{\nu _{1}\cdots \nu _{\ell
-1}\nu _{\ell +1}}\mathbf{q}B^{\nu _{\ell }\nu }g_{\nu \nu _{\ell +1}} 
\notag \\
& +\frac{1}{2}\left[ m_{0}^{2}j\mathcal{A}_{i-1,j-1}^{\left\{ \mu
_{1}\right. \cdots \mu _{\ell -1}}-j\mathcal{A}_{i+1,j-1}^{\left\{ \mu
_{1}\right. \cdots \mu _{\ell -1}}+\left( j+2\ell \right) \mathcal{A}%
_{i-1,j+1}^{\left\{ \mu _{1}\right. \cdots \mu _{\ell -1}}\right] \mathbf{q}%
B^{\left. \mu _{\ell }\right\} \nu }l_{\nu }\, .  \label{Main_DA_ij_mu1_mun}
\end{align}
Here, in analogy to Eq.~(\ref{def_C_i_mu1_mun}), we define the anisotropic irreducible collision term, 
\begin{equation}
\mathcal{C}_{i-1,j}^{\mu _{1}\cdots \mu _{\ell }}\equiv \int \mathrm{d}K 
E_{\mathbf{k}u}^{i-1}E_{\mathbf{k}l}^{j}k^{\left\{ \mu _{1}\right. }\cdots
k^{\left. \mu _{\ell }\right\} }C\left[ f_{\mathbf{k}}\right] \,,
\label{def_C_ij_mu1_mun}
\end{equation}%
and hence in binary collisions the conservation of charge, the conservation of energy, the 
conservation of momentum in the direction of the momentum anisotropy, as well as transverse 
to it leads to 
\begin{equation}
\mathcal{C}_{00}=0\,,\quad \mathcal{C}_{10}=0\,,\quad \mathcal{C}_{01}=0\,,\quad 
\mathcal{C}_{00}^{\{\mu \}}=0\,,
\label{C_0_C_1_C_0_mu_aniso}
\end{equation}%
while comparing these with Eq.~(\ref{C_0_C_1_C_0_mu}) we find $\mathcal{C}_{00} = \mathcal{C}_{0}$, $\mathcal{C}_{1} = \mathcal{C}_{10}$, and $\mathcal{C}_{01} = -\mathcal{C}_{0}^{\langle \mu \rangle} 
l_\mu $, and $\mathcal{C}_{00}^{\{\mu \}} = \Xi^\mu_\nu \mathcal{C}_{0}^{\langle \nu \rangle}$. 
Furthermore, in this case the relevant Cauchy-Stokes decompositions are 
\begin{eqnarray}
\partial _{\mu }u_{\nu } &=&u_{\mu }Du_{\nu }+l_{\mu }D_{l}u_{\nu }
+\frac{1}{2}\tilde{\theta}\Xi _{\mu \nu }-l_{\beta }l_{\nu }\tilde{\nabla}_{\mu}u^{\beta }
+\tilde{\sigma}_{\mu \nu }+\tilde{\omega}_{\mu \nu }\, , \\
\partial _{\mu }l_{\nu } &=&u_{\mu }Dl_{\nu }+l_{\mu }D_{l}l_{\nu }+\frac{1}{2}
\tilde{\theta}_{l}\Xi _{\mu \nu }+u_{\beta }u_{\nu }\tilde{\nabla}_{\mu}l^{\beta }
+\tilde{\sigma}_{l,\mu \nu }+\tilde{\omega}_{l,\mu \nu }\, ,
\end{eqnarray}%
where $\tilde{\theta}=\tilde{\nabla}_{\mu }u^{\mu }$ and 
$\tilde{\theta}_{l}=\tilde{\nabla}_{\mu }l^{\mu }$ are expansion scalars. 
The transverse shear tensors are 
$\tilde{\sigma}^{\mu \nu }\equiv \partial ^{\left\{ \mu
\right. }u^{\left. \nu \right\} }=\tilde{\nabla}^{\left( \mu \right.
}u^{\left. \nu \right) }-\frac{1}{2}\tilde{\theta}\Xi ^{\mu \nu }+l_{\beta
}l^{\left( \mu \right. }\tilde{\nabla}^{\left. \nu \right) }u^{\beta }$ and $%
\tilde{\sigma}_{l}^{\mu \nu }\equiv \partial ^{\left\{ \mu \right.
}l^{\left. \nu \right\} }=\tilde{\nabla}^{\left( \mu \right. }l^{\left. \nu
\right) }-\frac{1}{2}\,\tilde{\theta}_{l}\,\Xi ^{\mu \nu }-u_{\beta
}u^{\left( \mu \right. }\tilde{\nabla}^{\left. \nu \right) }l^{\beta }$,
while the vorticity tensors are 
$\tilde{\omega}^{\mu \nu }\equiv \Xi ^{\mu\alpha }\Xi ^{\nu \beta }
\partial _{\left[ \alpha \right. }u_{\left. \beta %
\right] }=\tilde{\nabla}^{\left[ \mu \right. }u^{\left. \nu \right]
}-l_{\beta }l^{\left[ \mu \right. }\tilde{\nabla}^{\left. \nu \right]
}u^{\beta }$ and $\tilde{\omega}_{l}^{\mu \nu }\equiv \Xi ^{\mu \alpha }\Xi
^{\nu \beta }\partial _{\left[ \alpha \right. }l_{\left. \beta \right] }=%
\tilde{\nabla}^{\left[ \mu \right. }l^{\left. \nu \right] }+u_{\beta }u^{%
\left[ \mu \right. }\tilde{\nabla}^{\left. \nu \right] }l^{\beta }$.

The equation of motion for an AIM of given tensor rank-$\ell$, energy index $i$, and index $j$ in the direction of the momentum anisotropy always couple to moments with both lower and higher tensor-rank. 
Furthermore, the indices $i$ and $j$ of these moments also vary independently from the tensor-rank. 
Similarly to Eq.~(\ref{Main_DM_i_mu1_mun}) the resulting equation forms an infinite hierarchy of coupled equations of motion.

The first part of the above equation, i.e., the first seventeen lines, without the electromagnetic contributions is consistent with the equation of motions obtained from the Boltzmann equation in Ref.~\cite{Molnar:2016vvu}.
Namely for tensor-rank $\ell = 0$ for the anisotropic scalar moment, the above general equation 
is exactly the same as Eq.~(110) in Ref.~\cite{Molnar:2016vvu}, while the $\ell=1$ and $\ell=2$ 
cases precisely correspond to the equations for the vector moment and rank-two tensor moment, 
see Eqs.~(111) and~(112) in Ref.~\cite{Molnar:2016vvu}.

The last four lines follow from the Vlasov term, i.e., the last term in Eq.~(\ref{D_BTE_F_munu_aniso}), and represent the coupling of the AIMs to the electric and the magnetic fields. 
Thereby this equation of motion for rank-$\ell$ AIMs describes electrically conducting fluids that have an explicit and additional momentum anisotropy independently from the external magnetic induction field.
This is also apparent from the various terms of the moment equation; some terms directly project the electromagnetic fields in the direction of the anisotropy $l_\nu$ as well as in the directions transverse to it $\Xi^{\mu \nu}$, while the projections with respect to the direction of the magnetic induction field are included through the $B^{\mu\nu}$ tensor.

\subsection{Leading-order anisotropic fluid dynamics and anisotropic MHD}
\label{sec:leading_aniso_fluid_dynamics}

Following the discussions of the previous section here we also summarize the approximations to obtain 
the equations of motion of anisotropic fluid dynamics and anisotropic MHD.
Similar to the assumption $f_{\mathbf{k}} \equiv f_{0\mathbf{k}}$, which leads to ideal fluid dynamics, 
we can derive equations of motion assuming $f_{\mathbf{k}} \equiv \hat{f}_{0\mathbf{k}}$. 
In this way, we derive a fluid-dynamical framework that extends beyond ideal fluid dynamics and ideal MHD, since all irreducible moments include a non-vanishing momentum anisotropy through the anisotropic distribution function $\hat{f}_{0\mathbf{k}} \left( \hat{\alpha}, \hat{\beta}_{u}E_{\mathbf{k}u}, \hat{\beta}_{l}E_{\mathbf{k}l}\right)$.
For instance, $N^{\mu} \equiv \hat{N}^{\mu}$ and $T^{\mu \nu} \equiv \hat{T}^{\mu\nu}$, see Eqs.~(\ref{N_0_mu_u_l}) and (\ref{T_0_munu_u_l}).
Such a fluid-dynamical theory is a representative of the so-called leading-order anisotropic fluid-dynamical framework developed recently in Refs.~\cite{Barz:1987pq,Kampfer:1990qg,Florkowski:2010cf,Ryblewski:2010bs,Ryblewski:2012rr,Martinez:2010sc,
Martinez:2010sd,Martinez:2012tu,Bazow:2013ifa,Bazow:2015cha,Tinti:2013vba,Tinti:2015xwa,
Molnar:2016vvu,Molnar:2016gwq,Alqahtani:2017mhy}.

The equations of motion of leading-order anisotropic fluid dynamics follow directly from Eq.~(\ref{Main_DA_ij_mu1_mun}) under the following replacements: $\mathcal{A}_{ij}^{\mu _{1}\cdots \mu _{\ell }} \equiv \hat{\mathcal{I}}_{ij}^{\mu_{1}\cdots \mu _{\ell }}$, and since $\delta \hat{f}_{\mathbf{k}}=0$ then $\hat{\rho}_{ij}^{\mu _{1}\cdots \mu _{\ell}}=0$.
However, for any tensor-rank $\ell \geq 1$ all irreducible projections of the anisotropic moments $\hat{\mathcal{I}}_{ij}^{\left\{ \mu _{1}\right. \cdots \left. \mu_{\ell}\right\}}$ vanish, $\hat{\mathcal{I}}_{ij}^{\left\{ \mu _{1}\right. \cdots \left. \mu_{\ell}\right\}}=0$, and hence only scalar moments $\hat{\mathcal{I}}_{ij}$ appear in leading-order anisotropic fluid dynamics.

Using these approximations and recalling Eq.~(\ref{Main_DA_ij_mu1_mun}) for tensor-rank $\ell=0$ we obtain
\begin{align}
D\hat{\mathcal{I}}_{ij}& = \mathcal{C}_{i-1,j} + D_{l}\hat{\mathcal{I}}_{i-1,j+1} 
+\left[ i\hat{\mathcal{I}}_{i-1,j+1}+j\hat{\mathcal{I}}_{i+1,j-1}\right] 
l_{\lambda }Du^{\lambda } - 
\left[ \left( i-1\right) \hat{\mathcal{I}}_{i-2,j+2}+\left( j+1\right) \hat{\mathcal{I}}_{i,j}\right] l_{\lambda }D_{l}u^{\lambda }  \notag \\
& +\frac{1}{2}\tilde{\theta}\left[m_{0}^{2}\left( i-1\right) \hat{\mathcal{I}}_{i-2,j} 
-\left( i+1\right) \hat{\mathcal{I}}_{i,j}+\left( i-1\right) \hat{\mathcal{I}}_{i-2,j+2}\right] 
-\frac{1}{2}\tilde{\theta}_{l}\left[ m_{0}^{2}j\hat{\mathcal{I}}_{i-1,j-1}-j\hat{\mathcal{I}}_{i+1,j-1}+\left( j+2\right) \hat{\mathcal{I}}_{i-1,j+1}\right]  \notag \\
& -\left[ \left( i-1\right) \hat{\mathcal{I}}_{i-2,j+1}+j\hat{\mathcal{I}}_{i,j-1}\right] 
\mathbf{q}E^{\nu }l_{\nu } \, .  
\label{DI_ij}
\end{align}%
From this general scalar equation, the charge conservation equation is obtained for $i=1$ and $j=0$, the energy conservation equation for $i=2$ and $j=0$, while the conservation equation for the momentum in the direction of the anisotropy can be obtained for $i=1$ and $j=1$. 
In these cases the corresponding moments of the collision integral vanish, see Eqs.~(\ref{C_0_C_1_C_0_mu_aniso}). 
However for all other cases this is no longer true. 

The anisotropic vector equation is obtained for tensor rank $\ell =1$, 
\begin{align}
0 & =\mathcal{C}_{i-1,j}^{\mu _{1}}-\frac{1}{2}\tilde{\nabla}^{\mu_{1}}\left( m_{0}^{2}\hat{\mathcal{I}}_{i-1,j}-\hat{\mathcal{I}}_{i+1,j} 
+ \hat{\mathcal{I}}_{i-1,j+2}\right)  \notag \\
& +\frac{ i-1 }{2}\left( m_{0}^{2}\hat{\mathcal{I}}_{i-2,j+1}-%
\hat{\mathcal{I}}_{i,j+1}+\hat{\mathcal{I}}_{i-2,j+3}\right) l_{\lambda }%
\tilde{\nabla}^{\mu _{1}}u^{\lambda }+\frac{j}{2}\left( m_{0}^{2}\hat{%
\mathcal{I}}_{i,j-1}-\hat{\mathcal{I}}_{i+2,j-1}+\hat{\mathcal{I}}_{i,j+1}\right) 
l_{\lambda }\tilde{\nabla}^{\mu _{1}}u^{\lambda }  \notag \\
& +\frac{1}{2}\left[ m_{0}^{2}i\hat{\mathcal{I}}_{i-1,j}-\left( i+2\right) 
\hat{\mathcal{I}}_{i+1,j}+i\hat{\mathcal{I}}_{i-1,j+2}\right] 
\Xi _{\lambda}^{\mu _{1}}Du^{\lambda } 
-\frac{1}{2}\left[ m_{0}^{2}j\hat{\mathcal{I}}_{i,j-1} - j\hat{\mathcal{I}}_{i+2,j-1}
+\left( j+2\right) \hat{\mathcal{I}}_{i,j+1}\right] \Xi _{\lambda}^{\mu _{1}}Dl^{\lambda }  \notag \\
&-\frac{1}{2}\left[ m_{0}^{2}\left( i-1\right) \hat{\mathcal{I}}_{i-2,j+1}-\left( i+1\right) \hat{\mathcal{I}}_{i,j+1}+\left(i-1\right) \hat{\mathcal{I}}_{i-2,j+3}\right] 
\Xi _{\lambda }^{\mu_{1}}D_{l}u^{\lambda } \notag \\
&+ \frac{1}{2}\left[ m_{0}^{2}\left( j+1\right) \hat{\mathcal{I}}_{i-1,j}-\left( j+1\right) \hat{\mathcal{I}}_{i+1,j}+\left(j+3\right) \hat{\mathcal{I}}_{i-1,j+2}\right] 
\Xi _{\lambda }^{\mu_{1}}D_{l}l^{\lambda }  \notag \\
& -\frac{1}{2}\left[ m_{0}^{2}\left( i-1\right) \hat{\mathcal{I}}_{i-2,j}-\left( i+1\right) \hat{\mathcal{I}}_{i,j}+\left( i-1\right) \hat{\mathcal{I}}_{i-2,j+2}\right] 
\Xi _{\nu _{1}}^{\mu _{1}}\mathbf{q}E^{\nu_{1}}  \notag \\
&+\frac{1}{2}\left[ m_{0}^{2}j\hat{\mathcal{I}}_{i-1,j-1}-j\hat{\mathcal{I}}_{i+1,j-1} 
+\left( j+2\right) \hat{\mathcal{I}}_{i-1,j+1}\right] \mathbf{q}B^{\mu _{1}\nu }l_{\nu }\, .  
\label{DI_ij_mu1}
\end{align}%
From here the conservation of momentum in the direction transverse to $l^{\mu}$ follows 
for $i=1$ and $j=0$.
All conservation equations, without the electromagnetic contributions, were presented 
in Eqs.~(26)--(29) of Ref.~\cite{Molnar:2016gwq}. 
Also note that these five conservation equations do not determine all independent 
variables entering $\hat{f}_{0\mathbf{k}}$, hence not only an equation of state but an 
additional moment equation must be provided in order to close the conservation equations, 
see also the discussion after Eq.~(\ref{Landau_matching_aniso}) and Ref.~\cite{Molnar:2016gwq}.

These two equations of motion take into account the additional coupling terms to the external electromagnetic field and correspond to the equations of leading-order anisotropic MHD, i.e., extending the equations of leading-order anisotropic fluid dynamics to electrically conducting fluids.
This approach represents a specific class of dissipative and resistive MHD that is based on an explicit local momentum-space anisotropy and an additional independent variable contained in $\hat{f}_{0\mathbf{k}}$ compared to an equilibrium distribution function.

\subsection{Higher-order anisotropic fluid dynamics and anisotropic MHD}
\label{sec:aniso_fluid_dynamics}

Dissipative phenomena can be described by also including the $\delta \hat{f}_{\mathbf{k}}$ correction. 
This improves the leading-order anisotropic framework further as $f_{\mathbf{k}} \equiv \hat{f}_{0\mathbf{k}} + \delta \hat{f}_{\mathbf{k}}$, which then implies that $\mathcal{A}_{ij}^{\mu _{1}\cdots \mu _{\ell }}=\hat{\mathcal{I}}_{ij}^{\mu_{1}\cdots \mu _{\ell }}+\hat{\rho}_{ij}^{\mu _{1}\cdots \mu _{\ell }}$. 
As mentioned before, the corresponding equations of motion for tensor-ranks $\ell = 0$, $\ell = 1$, and $\ell = 2$ were obtained in Ref.~\cite{Molnar:2016vvu} from the Boltzmann equation without electromagnetic fields. 
Therefore the natural magnetohydrodynamical extension of those moment equations follows directly from Eq.~(\ref{Main_DA_ij_mu1_mun}) considering the additional terms coupling to the external electromagnetic fields. These are presented in Appendix~\ref{app:aniso_resistiveMHD}.

The truncation of these equations of motion in the 14 dynamical-moment approximation, i.e., 
including only the lowest-order anisotropic moments appearing in $N^{\mu}$ and $T^{\mu \nu}$, see Eqs.~(\ref{kinetic:N_mu_u_l}, \ref{kinetic:T_munu_u_l}), would lead to the equations of motion of dissipative anisotropic MHD in the 14 dynamical-moment approximation. 
This is similar to the traditional near-equilibrium expansion with $5$+$9$ dynamical moments discussed 
in the previous section but now applied to $6$+$8$ anisotropic moments.
These equations would extend those obtained in Ref.~\cite{Molnar:2016vvu} to electrically conducting fluids. 
Furthermore, going beyond the usual 14-moment approximation by taking into account the equations for higher-rank tensor moments would lead to the framework of higher-order anisotropic fluid dynamics or MHD.

\section{Conclusions and Outlook}
\label{sec:conclusions}

In this paper we have derived the general equations of motion for all isotropic as well as anisotropic irreducible moments of the single-particle distribution function from the Boltzmann-Vlasov equation. 
The equation of motion~(\ref{Main_DM_i_mu1_mun}) extends that obtained in Ref.~\cite{deBrito:2024vhm} for rank-$\ell $ IIMs to electrically conducting fluids. 
Furthermore, the equations of motion Eq.~(\ref{Main_DA_ij_mu1_mun}) for the AIMs generalize those obtained previously in Ref.~\cite{Molnar:2016vvu} to arbitrary tensor-rank and by taking into account the Vlasov term we also include the couplings to external electromagnetic fields. 
The truncation of these equations~(\ref{Main_DM_i_mu1_mun}) in the 14 dynamical-moment approximation leads to the equations of motion of second-order dissipative fluid dynamics and MHD. 
Similarly, the equations of anisotropic fluid dynamics and anisotropic MHD follow from Eq.~(\ref{Main_DA_ij_mu1_mun}).
A suitable truncation beyond the $14$ dynamical-moment approximation of these general equations leads to higher-order anisotropic fluid dynamics and MHD. 
While the present work provides a general derivation of such fluid-dynamical theories, their study and application is left for future work.

\begin{acknowledgments}
	We thank  A.~Dash and P.~Huovinen for reading the manuscript and for constructive comments and useful discussions.
	The authors acknowledge support by the Deutsche Forschungsgemeinschaft (DFG, German Research Foundation) through the CRC-TR 211 ``Strong-interaction matter 	under extreme conditions'' -- Project No. 315477589 -- TRR 211.
	E.M.~was also supported by the program Excellence Initiative--Research University of the University of Wroc{\l}aw of the Ministry of Education and Science. 
	D.H.R.~is supported by the State of Hesse within the Research Cluster ELEMENTS (Project ID 500/10.006).
\end{acknowledgments}

\appendix
\section{Projection operators and properties}
\label{Projection_properties}

The symmetrization of tensor indices is denoted by round brackets around greek indices, 
\begin{equation}
T^{\left( \mu _{1}\cdots \mu _{n}\right) } = \frac{1}{n!}\sum_{\bar{\mathcal{P}}_{\mu }^{n}} 
T^{\mu _{1}\mu _{2}\cdots \mu _{n}}\,,
\end{equation}%
where the sum runs over \textit{all} permutations $\bar{\mathcal{P}}_{\mu }$ of indices. 
The total number of terms in this sum is $n!$, so the sum is normalized to this number.
The antisymmetrization of tensors is denoted by square brackets around indices,
\begin{equation}
T^{\left[ \mu _{1}\cdots \mu _{n}\right] }=\frac{1}{n!}\varepsilon _{\mu_{1}\cdots 
\mu _{n}}\sum_{\bar{\mathcal{P}}_{\mu }^{n}}T^{\mu _{1}\mu _{2}\cdots \mu _{n}}\,,
\end{equation}%
where the pseudotensor $\varepsilon _{\mu _{1}\cdots \mu _{n}}$ is the permutation or 
Levi-Civit\`{a} symbol. 

The symmetrized tensor product of $\Delta $'s and $u$'s is defined by the following expression~\cite{Israel:1979wp}, 
\begin{equation}
\Delta ^{\left( \mu _{1}\mu _{2}\right. }\cdots \Delta ^{\mu _{2q-1}\mu_{2q}}u^{\mu _{2q+1}}\cdots u^{\left. \mu _{n}\right) }\equiv \frac{1}{b_{nq}}\sum_{\mathcal{P}_{\mu }^{n}}\Delta ^{\mu _{1}\mu _{2}}\cdots \Delta ^{\mu_{2q-1}\mu _{2q}}u^{\mu _{2q+1}}\cdots u^{\mu _{n}}\,,
\label{iso_sym_tens_prod}
\end{equation}%
where now the sum runs over all \textit{distinct} permutations $\mathcal{P}_{\mu }^{n}$ of $\mu$-type indices and it is normalized to the number of such terms,
\begin{equation}
b_{nq}\equiv \frac{n!}{2^{q}q!\left( n-2q\right) !}=\frac{n!\left(2q-1\right) !!}{\left( 2q\right)!
\left( n-2q\right) !}\,.  \label{b_nq}
\end{equation}%
This is calculated as follows.
The total number of permutations of $n$ indices is $n!$. 
However, the $q!$ permutations of the $q$ projection operators $\Delta ^{\mu _{i}\mu _{j}}$ among each other and the $(n-2q)!$ permutations of the $n-2q$ factors of $u^{\mu_k}$ among each other do not lead to distinct terms. 
In addition, there are also $2^q$ permutations of the two indices of the $q$ symmetric projection operators that do not lead to distinct terms.
Thus, one needs to divide the total number of permutations by these factors.
Note that the double factorial of odd numbers is identical to 
$\left(2q-1\right) !!=\left( 2q\right) !/\left( 2^{q}q!\right)$.

Similarly to Eq.~(\ref{iso_sym_tens_prod}) the symmetrized tensor product of 
$\Xi $'s, $u$'s, and $l$'s is given by the following expression \cite{Molnar:2016vvu}, 
\begin{equation}
\Xi ^{\left( \mu _{1}\mu _{2}\right. }\cdots \Xi ^{\mu _{2q-1}\mu_{2q}}l^{\mu _{2q+1}}\cdots 
l^{\mu _{2q+r}}u^{\mu _{2q+r+1}}\cdots u^{\left.\mu _{n}\right) }\equiv \frac{1}{b_{nrq}}\sum_{\mathcal{P}_{\mu }^{n}}\Xi^{\mu _{1}\mu _{2}}\cdots 
\Xi ^{\mu _{2q-1}\mu _{2q}}l^{\mu _{2q+1}}\cdots l^{\mu _{2q+r}}u^{\mu _{2q+r+1}}\cdots u^{\mu _{n}}\,,
\label{aniso_sym_tens_prod}
\end{equation}%
where the coefficient counting the distinct terms generalizes Eq.~(\ref{b_nq}) to the presence of the $l$'s and reads 
\begin{equation}
b_{nrq}\equiv \frac{n!}{2^{q}q!r!\left( n-r-2q\right) !} 
=\frac{n!\left(2q-1\right) !!}{\left( 2q\right) !r!\left( n-r-2q\right) !}\,.  
\label{b_nrq}
\end{equation}

The coefficient in Eq.~(\ref{Delta_irred_proj}) is defined as
\begin{equation}
C(n,q)=(-1)^{q}\frac{(n!)^{2}}{(2n)!}\,\frac{(2n-2q)!}{q!(n-q)!(n-2q)!}\,.
\label{Ccoeff}
\end{equation}%
while the coefficient in Eq.~(\ref{Xi_irred_proj}) is defined as (see Ref.~\cite{Molnar:2016vvu} for more details)
\begin{equation}
\hat{C}(n,q)=(-1)^{q}\frac{1}{4^{q}}\,\frac{(n-q)!}{q!(n-2q)!}\frac{n}{n-q}\,.  
\label{Chatcoeff}
\end{equation}
The second sum in Eqs.~(\ref{Delta_irred_proj}), (\ref{Xi_irred_proj}) runs over all distinct permutations $\mathcal{P}_{\mu }^{n}\mathcal{P}_{\nu }^{n}$ of $\mu $- and $\nu $-type indices.
While the \textit{total} number of permutations of these indices is $(n!)^{2}$, in order to get the number of \textit{distinct} permutations one has to divide this by the number of trivial reorderings of the $\Delta ^{\mu_{2q-1}\mu _{2q}}$ and $\Delta _{\nu _{2q-1}\nu _{2q}}$ projectors ($\Xi^{\mu_{2q-1}\mu _{2q}}$ and $\Xi_{\nu _{2q-1}\nu _{2q}}$ projectors, respectively), which is $(q!)^{2} $, as well by the number of trivial permutations of the $\mu $- and $\nu $-type indices on these symmetric projectors, which is $(2^{q})^{2}$. 
Finally, the number of trivial reorderings of the projectors with mixed indices is $(n-2q)!$. 
Therefore the total number of distinct permutations in the second sum is 
\begin{equation}
\mathcal{N}_{nq}\equiv \frac{1}{(n-2q)!}\,\left( \frac{n!}{2^{q}q!}\right)^{2}=b_{nq}\, 
\frac{n!}{2^{q}q!} \,,  \label{N_nq}
\end{equation}%
and the sum is normalized by this number.

The isotropic irreducible projection operators~(\ref{Delta_irred_proj}) are symmetric under the permutation of $\mu$- and $\nu$-type of indices, traceless with respect to contraction of either 
$\mu$- or $\nu$-type of indices, and they are orthogonal to the fluid four-velocity,
\begin{eqnarray}
\Delta _{\nu _{1}\cdots \nu _{n}}^{\mu _{1}\cdots \mu _{n}} 
&=& \Delta_{\left( \nu _{1}\cdots \nu _{n}\right) }^{\left(\mu _{1}\cdots \mu_{n}\right) }\,, \\
\Delta_{\nu _{1}\cdots \nu _{n}}^{\mu _{1}\cdots \mu _{n}}g_{\mu _{i} 
\mu_{j}} &=&\Delta _{\nu _{1}\cdots \nu _{n}}^{\mu _{1}\cdots \mu _{n}} 
g^{\nu_{i}\nu_{j}}=0\,,  \label{tracelessnessDelta}\\
\Delta _{\nu _{1}\cdots \nu _{n}}^{\mu _{1}\cdots \mu _{n}}u_{\mu _{i}}
&=&\Delta _{\nu _{1}\cdots \nu _{n}}^{\mu_{1}\cdots \mu_{n}}u^{\nu_{i}}=0\,.
\end{eqnarray}%
Furthermore one can also show that upon contraction 
\begin{equation}
\Delta _{\nu _{1}\cdots \nu _{n-1}\nu _{n}}^{\mu _{1}\cdots \mu _{n-1}\mu_{n}} 
g_{\mu _{n}}^{\nu _{n}}=\frac{2n+1}{2n-1}
\Delta _{\nu _{1}\cdots \nu _{n-1}}^{\mu _{1}\cdots \mu _{n-1}}\,,
\end{equation}%
and hence upon complete contraction%
\begin{equation}
\Delta _{\nu _{1}\cdots \nu _{n}}^{\mu _{1}\cdots \mu _{n}}g_{\mu _{1}}^{\nu_{1}}\cdots 
g_{\mu _{n}}^{\nu _{n}}=2n+1 \,.
\end{equation}

The anisotropic irreducible projection operators~(\ref{Xi_irred_proj}) are symmetric under the permutation of $\mu$-and $\nu$-type of indices. They are traceless with respect to contraction
with either $\mu$- or $\nu$-type of indices, as well as orthogonal to both
the fluid four-velocity and the anisotropy four-vector, 
\begin{eqnarray}
\Xi_{\nu _{1}\cdots \nu _{n}}^{\mu_{1}\cdots \mu_{n}} 
&=& \Xi_{\left( \nu_{1}\cdots \nu _{n}\right) }^{\left( \mu_{1}\cdots \mu_{n}\right) }\, , \\
\Xi_{\nu_{1}\cdots \nu _{n}}^{\mu_{1}\cdots \mu _{n}}g_{\mu_{i}\mu _{j}}
&=&\Xi_{\nu _{1}\cdots \nu _{n}}^{\mu _{1}\cdots \mu _{n}}g^{\nu _{i}\nu_{j}}=0\, , \\
\Xi_{\nu _{1}\cdots \nu _{n}}^{\mu _{1}\cdots \mu _{n}}u_{\mu _{i}} 
&=&\Xi_{\nu _{1}\cdots \nu _{n}}^{\mu _{1}\cdots \mu _{n}}u^{\nu _{i}}=0\, , \\
\Xi _{\nu _{1}\cdots \nu _{n}}^{\mu _{1}\cdots \mu _{n}}l_{\mu _{i}} 
&=& \Xi_{\nu _{1}\cdots \nu _{n}}^{\mu _{1}\cdots \mu _{n}}l^{\nu _{i}}=0\, .
\end{eqnarray}%
Furthermore, upon contraction of two indices, 
\begin{equation}
\Xi _{\nu _{1}\cdots \nu _{n-1}\nu _{n}}^{\mu _{1}\cdots \mu _{n-1}\mu_{n}} 
g_{\mu _{n}}^{\nu _{n}} = \Xi _{\nu _{1}\cdots \nu _{n-1}}^{\mu_{1}\cdots \mu _{n-1}}\, ,
\end{equation}%
and upon complete contraction 
\begin{equation}
\Xi _{\nu _{1}\cdots \nu _{n}}^{\mu _{1}\cdots \mu _{n}}g_{\mu _{1}}^{\nu_{1}}\cdots 
g_{\mu _{n}}^{\nu _{n}}=2\, .
\end{equation}

Recalling the definition of the isotropic irreducible projection operator from Eq.~(\ref{Delta_irred_proj}) we have 
\begin{align}
\Delta _{\nu _{1}\cdots \nu _{n}}^{\mu _{1}\cdots \mu _{n}}k^{\nu
_{1}}\cdots k^{\nu _{n}} &\equiv k^{\left\langle \mu _{1}\right. }\cdots
k^{\left. \mu _{n}\right\rangle }  \notag \\
&=  \sum_{q=0}^{\lfloor n/2\rfloor}\frac{C\left( n,q\right) }{\mathcal{N}_{nq}}%
\sum_{\mathcal{P}_{\mu }^{n}\mathcal{P}_{\nu }^{n}}\Delta ^{\mu _{1}\mu
_{2}}\cdots \Delta ^{\mu _{2q-1}\mu _{2q}}\Delta _{\nu _{1}\nu _{2}}\cdots
\Delta _{\nu _{2q-1}\nu _{2q}}\Delta _{\nu _{2q+1}}^{\mu _{2q+1}}\cdots
\Delta _{\nu _{n}}^{\mu _{n}} \, k^{\nu _{1}}\cdots k^{\nu _{n}}\, .
\end{align}%
The $q=0$ part of the sum reads
\begin{equation}
 \frac{C\left( n,0\right) }{\mathcal{N}_{n0}}\sum_{\mathcal{P}_{\mu
}^{n}\mathcal{P}_{\nu }^{n}}\Delta _{\nu _{1}}^{\mu _{1}}\cdots \Delta _{\nu
_{n}}^{\mu _{n}}\, k^{\nu _{1}}\cdots k^{\nu _{n}}=  \frac{1%
}{n!}\sum_{\mathcal{P}_{\mu }^{n}\mathcal{P}_{\nu }^{n}}\Delta _{\nu
_{1}}^{\mu _{1}}\cdots \Delta _{\nu _{n}}^{\mu _{n}}\, k^{\nu
_{1}}\cdots k^{\nu _{n}}=k^{\left\langle \mu _{1}\right\rangle }\cdots
k^{\left\langle \mu _{n}\right\rangle }\,,
\end{equation}
where we used the fact that there are only $n!$ terms in the sum, because the permutations of the $\mu$-type indices among themselves do not lead to distinct terms.

Starting from $q\geq 1$ the sums contain $q$ projectors of type $\Delta _{\nu _{i}\nu _{j}}$, and thus a term $\left(\Delta_{\nu_i \nu_j}k^{\nu_i}k^{\nu_j}\right) ^{q}$ will appear in the sum over $\nu $-type indices. 
The total number of terms of this type in the sum over $\mathcal{P}_\nu^n$ is easier calculated by first extending the sum over \textit{distinct} permutations $\mathcal{P}^n_\mu \mathcal{P}^n_\nu$ to the sum over \textit{all} permutations $\bar{\mathcal{P}}^n_\mu \bar{\mathcal{P}}^n_\nu$, which necessitates to multiply the sum with a factor $\mathcal{N}_{nq}/(n!)^2$.
Then, one picks at random the $2q$ $\nu$-type indices that pertain to the $\Delta _{\nu _{i}\nu _{j}}$ projectors out of the total of $n$ $\nu$-type indices, for which there are $ {\scriptsize\left(  \begin{array}{c} n \\ 2q \end{array}  \right) }= \frac{n!}{(2q)! (n-2q)!} $ possibilities.  
Permuting the $2q$ $\nu$-type indices among themselves gives $(2q)!$ identical terms, while permuting the remaining $n-2q$ $\nu$-type indices among themselves gives $(n-2q)!$ identical terms.
Thus, for a given permutation of the $\mu$-type indices we have $(n-2q)! (2q)! \, \frac{n!}{(2q)! (n-2q)!}  = n!$ identical terms of the form
\begin{equation}
\left(\Delta_{\alpha \beta}k^{\alpha}k^{\beta}\right) ^{q} \Delta _{\nu_{2q+1}}^{\mu _{2q+1}}\cdots \Delta _{\nu _{n}}^{\mu _{n}} k^{\nu_{2q+1}}\cdots k^{\nu _{n}}
=\left(\Delta_{\alpha \beta}k^{\alpha}k^{\beta}\right) ^{q}  k^{\left\langle \mu_{2q+1}\right\rangle }\cdots k^{\left\langle \mu _{n}\right\rangle } \, ,
\end{equation}
in the sum over all permutations $\bar{\mathcal{P}}_\nu^n$ (which is obvious, as it corresponds to the total number of terms in this sum).
Inserting this intermediate result we have
\begin{eqnarray}
&&\sum_{q=1}^{\lfloor n/2\rfloor}\frac{C\left( n,q\right) }{\mathcal{N}_{nq}} \frac{\mathcal{N}_{nq}}{(n!)^2} \sum_{\bar{\mathcal{P}}_{\mu }^{n}\bar{\mathcal{P}}_{\nu }^{n}}\Delta ^{\mu _{1}\mu_{2}}\cdots \Delta ^{\mu _{2q-1}\mu _{2q}}\Delta _{\nu _{1}\nu _{2}}\cdots
\Delta _{\nu _{2q-1}\nu _{2q}}\Delta _{\nu _{2q+1}}^{\mu _{2q+1}}\cdots
\Delta _{\nu _{n}}^{\mu _{n}}\; k^{\nu _{1}}\cdots k^{\nu _{n}}  \notag
\\
&=&\sum_{q=1}^{\lfloor n/2\rfloor}C\left( n,q\right)  
\left(\Delta _{\alpha \beta }k^{\alpha}k^{\beta}\right)^{q}\frac{1}{n!}
\sum_{\bar{\mathcal{P}}_{\mu }^{n}}\Delta ^{\mu _{1}\mu _{2}}\cdots
\Delta ^{\mu _{2q-1}\mu _{2q}}k^{\left\langle \mu _{2q+1}\right\rangle
}\cdots k^{\left\langle \mu _{n}\right\rangle }\notag \\
&=& \sum_{q=1}^{\lfloor n/2\rfloor}C\left( n,q\right)   \left(\Delta _{\alpha \beta }k^{\alpha}k^{\beta}\right)^{q}\frac{1}{b_{nq}}
\sum_{\mathcal{P}_{\mu }^{n}}\Delta ^{\mu _{1}\mu _{2}}\cdots
\Delta ^{\mu _{2q-1}\mu _{2q}}k^{\left\langle \mu _{2q+1}\right\rangle
}\cdots k^{\left\langle \mu _{n}\right\rangle } \notag \\
&=& \sum_{q=1}^{\lfloor n/2\rfloor}C\left( n,q\right)  \left(\Delta _{\alpha \beta }k^{\alpha}k^{\beta}\right)^{q}\Delta ^{\left( \mu _{1}\right. \mu_{2}}\cdots 
\Delta ^{\mu _{2q-1}\mu _{2q}}k^{\left\langle \mu_{2q+1}\right\rangle }\cdots 
k^{\left. \left\langle \mu _{n}\right\rangle\right) }\,,
\end{eqnarray}%
where in the next-to-last step we reverted the sum over all permutations of $\mu$-type indices to a sum over all distinct permutations of $\mu$-type indices and in the last step we used Eq.~(\ref{iso_sym_tens_prod}) (with the fluid four-velocity replaced by the four-momentum).
Using these results we obtain
\begin{eqnarray}
k^{\left\langle \mu _{1}\right. }\cdots k^{\left. \mu _{n}\right\rangle }
&=&k^{\left\langle \mu _{1}\right\rangle }\cdots k^{\left\langle \mu_{n}\right\rangle }
+\sum_{q=1}^{\lfloor n/2\rfloor}C\left( n,q\right) \left( \Delta _{\alpha \beta}
k^{\alpha }k^{\beta }\right) ^{q}\Delta ^{\left( \mu _{1}\right. \mu_{2}}
\cdots \Delta ^{\mu _{2q-1}\mu _{2q}}k^{\left\langle \mu_{2q+1}\right\rangle }\cdots 
k^{\left. \left\langle \mu _{n}\right\rangle\right) }\, ,  
\label{iso1_k_mu1_mu_n}
\end{eqnarray}%
which proves Eq.~(\ref{Main1_iso}).

We now use Eq.~(\ref{iso1_k_mu1_mu_n}) for $n \rightarrow n+1$ to compute
\begin{align}
\lefteqn{k^{\left\langle \mu _{1}\right. }\cdots k^{\left. \mu _{n}\right\rangle
}k^{\left\langle \mu _{n+1}\right\rangle } \equiv \Delta _{\nu _{1}\cdots
\nu _{n}}^{\mu _{1}\cdots \mu _{n}}\Delta _{\nu _{n+1}}^{\mu
_{n+1}}k^{\nu _{1} }\cdots k^{ \nu_{n}}k^{ \nu _{n+1} } \equiv \Delta _{\nu _{1}\cdots
\nu _{n}}^{\mu _{1}\cdots \mu _{n}}\Delta _{\nu _{n+1}}^{\mu
_{n+1}}k^{\left\langle \nu _{1}\right\rangle }\cdots k^{\left\langle \nu
_{n}\right\rangle }k^{\left\langle \nu _{n+1}\right\rangle }}\notag \\
&= \Delta _{\nu _{1}\cdots
\nu _{n}}^{\mu _{1}\cdots \mu _{n}}\Delta _{\nu _{n+1}}^{\mu
_{n+1}} \left[ k ^{\left\langle \nu _{1}\right. }\cdots k^{\left. \nu _{n+1}\right\rangle } 
-\sum_{q=1}^{\lfloor (n+1)/2 \rfloor }C\left( n+1,q\right) \left( \Delta _{\alpha \beta}
k^{\alpha }k^{\beta }\right) ^{q}\Delta ^{\left( \nu _{1}\right. \nu_{2}}\cdots 
\Delta ^{\nu _{2q-1}\nu _{2q}}k^{\left\langle \nu_{2q+1}\right\rangle }\cdots 
k^{\left. \left\langle \nu_{n+1}\right\rangle\right) } \right] \notag \\
&= k ^{\left\langle \mu _{1}\right. }\cdots k^{\left. \mu _{n+1}\right\rangle } 
-\Delta _{\nu _{1}\cdots \nu _{n}}^{\mu _{1}\cdots \mu _{n}}\Delta _{\nu
_{n+1}}^{\mu _{n+1}} \, C\left( n+1,1\right) %
\left( \Delta_{\alpha \beta} k^\alpha k^\beta \right) \Delta ^{(\nu _{1} \nu_2}\, k^{\left\langle \nu_{3}\right\rangle }\cdots k^{\left\langle \nu _{n+1}\right\rangle) } \, ,
\label{k_n_k_n+1}
\end{align}%
\newline
where all terms with $q\geq 2$ from Eq.~(\ref{iso1_k_mu1_mu_n}) vanish since they contain at least one other projector $\Delta^{\nu_i \nu_j}$ with indices that contract with two indices of $\Delta_{\nu _{1}\cdots \nu _{n}}^{\mu _{1}\cdots \mu _{n}}$, which then vanishes due to the tracelessness condition~(\ref{tracelessnessDelta}).
The symmetrized tensor product reads explicitly with the definition (\ref{iso_sym_tens_prod})
\begin{align}
\Delta ^{(\nu _{1} \nu_2}\, k^{\left\langle \nu_{3}\right\rangle }\cdots k^{\left\langle \nu _{n+1}\right\rangle) } = \frac{1}{b_{n+1,1}} \sum_{\mathcal{P}_\nu^{n+1}}
\Delta^{\nu _{1} \nu_2}\, k^{\left\langle \nu_{3}\right\rangle }\cdots k^{\left\langle \nu _{n+1}\right\rangle}\,.
\end{align}
When inserting this into Eq.~(\ref{k_n_k_n+1}), we realize that due to the tracelessness condition (\ref{tracelessnessDelta}) of the projector $\Delta _{\nu _{1}\cdots \nu _{n}}^{\mu _{1}\cdots \mu _{n}}$, all terms in the sum over the distinct permutations of the $\nu$-type indices vanish, when the index $\nu_{n+1}$ is not one of the indices of the projector $\Delta^{\nu_i\nu_j}$ in this sum.
If $\nu_{n+1}$ is one of its indices, there are then $n$ distinct terms, namely when the other index of this projector is $\nu_1, \ldots, \nu_n$.
Since the projector $\Delta _{\nu _{1}\cdots \nu _{n}}^{\mu _{1}\cdots \mu _{n}}$ in Eq.~(\ref{k_n_k_n+1}) symmetrizes the result in all $\nu$-type indices anyway, we may without loss of generality take a fixed permutation of the $\nu$-type indices, say $\Delta^{\nu_{n+1} \nu_n} k^{\langle \nu_1 \rangle} \cdots k^{\langle \nu_{n-1} \rangle}$,  and obtain
\begin{align}
k^{\left\langle \mu _{1}\right. }\cdots k^{\left. \mu _{n}\right\rangle
}k^{\left\langle \mu _{n+1}\right\rangle } & = k ^{\left\langle
\mu _{1}\right. }\cdots k^{\left. \mu _{n+1}\right\rangle }  
+\frac{n}{2n+1}\Delta _{\nu _{1}\cdots \nu _{n}}^{\mu _{1}\cdots \mu _{n}}\Delta^{\mu _{n+1} \nu_n} 
\left( \Delta_{\alpha \beta} k^\alpha k^\beta \right) \, k^{\left\langle \nu_{1}\right\rangle }\cdots k^{\left\langle \nu _{n-1}\right\rangle} \,, \label{Eq11_finalproof}
\end{align}
where we have used $C(n+1,1)/b_{n+1,1} = - 1/(2n+1)$.
We now use Eq.~(\ref{iso1_k_mu1_mu_n}) again (for $n \rightarrow n-1$) to replace in the last term
\begin{align}
k^{\left\langle \nu_{1}\right\rangle }\cdots k^{\left\langle \nu _{n-1}\right\rangle} 
& = k^{\left\langle \nu _{1}\right. }\cdots k^{\left. \nu _{n-1}\right\rangle } -
\sum_{q=1}^{\lfloor (n-1)/2 \rfloor} C\left( n-1,q\right) \left( \Delta _{\alpha \beta}
k^{\alpha }k^{\beta }\right) ^{q}\Delta ^{\left( \nu _{1}\right. \nu_{2}}\cdots 
\Delta ^{\nu_{2q-1}\nu _{2q}}k^{\left\langle \nu_{2q+1}\right\rangle }\cdots 
k^{\left. \left\langle \nu _{n-1}\right\rangle\right) }\,.
\end{align}
However, the sum over $q$ vanishes, because all terms contain at least one projector $\Delta^{\nu_i \nu_j}$, where $\nu_i, \nu_j \in \left\{ \nu_1, \ldots, \nu_{n-1} \right\}$, which vanishes due to condition (\ref{tracelessnessDelta}) when multiplied with $\Delta _{\nu _{1}\cdots \nu _{n}}^{\mu _{1}\cdots \mu _{n}}$ in Eq.~(\ref{Eq11_finalproof}).
We thus finally obtain Eq.~(\ref{Main2_iso}).

These results are readily transferred to the $\Xi $-projectors, when using the respective relations for these projectors.
Equation~(\ref{Main1_aniso}) is immediately obtained by replacing the $\Delta$'s in Eq.~(\ref{iso1_k_mu1_mu_n}) by the $\Xi$-projectors,
\begin{eqnarray}
k^{\left\{ \mu _{1}\right. }\cdots k^{\left. \mu _{n}\right\} } &\equiv
&k^{\left\{ \mu _{1}\right\} }\cdots k^{\left\{ \mu _{n}\right\}
}+\sum_{q=1}^{\lfloor n/2\rfloor}\hat{C}\left( n,q\right) \left( \Xi _{\alpha \beta
}k^{\alpha }k^{\beta }\right) ^{q}\Xi ^{(\mu _{1}\mu _{2}}\cdots \Xi ^{\mu _{2q-1}\mu _{2q}}k^{\left\{
\mu _{2q+1}\right\} }\cdots k^{\left\{ \mu _{n}\right\}) }  \, .
\label{aniso1_k_mu1_mu_n}
\end{eqnarray}%
The analogue of Eq.~(\ref{Eq11_finalproof}) is
\begin{align}
k^{\left\{ \mu _{1}\right. }\cdots k^{\left. \mu _{n}\right\} }k^{\left\{\mu _{n+1}\right\} }& 
=k^{\left\{ \mu_{1}\right. }\cdots k^{\left. \mu _{n+1}\right\} }  +\frac{1}{2}\Xi _{\nu_{1}\cdots \nu _{n}}^{\mu _{1}\cdots \mu _{n}}
\Xi ^{\mu_{n+1} \nu_n}\left( \Xi _{\alpha \beta}k^{\alpha}k^{\beta}\right)  k^{\{\nu _{1} \}}\cdots k^{\{\nu _{n-1}\} }\, ,
\end{align}
where the coefficient in front of the last term is different, since $\hat{C}(n+1,1)/b_{n+1,0,1}=-1/(2n)$. 
Using Eq.~(\ref{aniso1_k_mu1_mu_n}) (with $n \rightarrow n-1$), similar arguments as before allow us to replace $k^{\{\nu _{1} \}}\cdots k^{\{\nu _{n-1}\} }$ by $k^{\{\nu _{1}}\cdots k^{\nu _{n-1}\} }$ in the last term, which then proves Eq.~(\ref{Main2_aniso}).

Furthermore, using these results recursively leads to the following relation, 
\begin{align}
k^{\left\langle \mu _{1}\right. }\cdots k^{\left. \mu _{n}\right\rangle }
k^{\left\langle \mu _{n+1}\right\rangle }k^{\left\langle \mu
_{n+2}\right\rangle }& =k^{\left\langle \mu _{1}\right. }\cdots k^{\left.
\mu _{n+2}\right\rangle }+\frac{n+1}{2n+3}\left( \Delta _{\alpha \beta
}k^{\alpha }k^{\beta }\right) \Delta _{\lambda _{1}\cdots \lambda
_{n+1}}^{\mu _{1}\cdots \mu _{n+1}}\Delta ^{\lambda _{n+1}\mu
_{n+2}}k^{\left\langle \lambda _{1}\right. }\cdots k^{\left. \lambda
_{n}\right\rangle }  \notag \\
& +\frac{n}{2n+1}\left( \Delta _{\alpha \beta }k^{\alpha }k^{\beta }\right)
\Delta _{\lambda _{1}\cdots \lambda _{n}}^{\mu _{1}\cdots \mu _{n}}\Delta
^{\lambda _{n}\mu _{n+1}}k^{\left\langle \lambda _{1}\right. }\cdots
k^{\lambda _{n-1}}k^{\left. \mu _{n+2}\right\rangle }  \notag \\
& +\frac{n\left( n-1\right) }{ 4n^{2}-1 }\left( \Delta _{\alpha
\beta }k^{\alpha }k^{\beta }\right) ^{2}\Delta _{\lambda _{1}\cdots \lambda
_{n}}^{\mu _{1}\cdots \mu _{n}}\Delta ^{\lambda _{n}\mu _{n+1}}\Delta
_{\beta _{1}\cdots \beta _{n-1}}^{\lambda _{1}\cdots \lambda _{n-1}}\Delta
^{\beta _{n-1}\mu _{n+2}}k^{\left\langle \beta _{1}\right. }\cdots k^{\left.
\beta _{n-2}\right\rangle }\, .  \label{Main22_iso}
\end{align}%
The anisotropic version of the above relation is, 
\begin{align}
k^{\left\{ \mu _{1}\right. }\cdots k^{\left. \mu _{n}\right\} } k^{\left\{
\mu _{n+1}\right\} }k^{\left\{ \mu _{n+2}\right\} }& =k^{\left\{ \mu
_{1}\right. }\cdots k^{\left. \mu _{n+2}\right\} }+\frac{1}{2}\left( \Xi
_{\alpha \beta }k^{\alpha }k^{\beta }\right) \Xi _{\lambda _{1}\cdots
\lambda _{n+1}}^{\mu _{1}\cdots \mu _{n+1}}\Xi ^{\lambda _{n+1}\mu
_{n+2}}k^{\left\{ \lambda _{1}\right. }\cdots k^{\left. \lambda _{n}\right\}
}  \notag \\
& +\frac{1}{2}\left( \Xi _{\alpha \beta }k^{\alpha }k^{\beta }\right) \Xi
_{\lambda _{1}\cdots \lambda _{n}}^{\mu _{1}\cdots \mu _{n}}\Xi ^{\lambda
_{n}\mu _{n+1}}k^{\left\{ \lambda _{1}\right. }\cdots k^{\lambda
_{n-1}}k^{\left. \mu _{n+2}\right\} }  \notag \\
& +\frac{1}{4}\left( \Xi _{\alpha \beta }k^{\alpha }k^{\beta }\right)
^{2}\Xi _{\lambda _{1}\cdots \lambda _{n}}^{\mu _{1}\cdots \mu _{n}}\Xi
^{\lambda _{n}\mu _{n+1}}\Xi _{\beta _{1}\cdots \beta _{n-1}}^{\lambda
_{1}\cdots \lambda _{n-1}}\Xi ^{\beta _{n-1}\mu _{n+2}}k^{\left\{ \beta
_{1}\right. }\cdots k^{\left. \beta _{n-2}\right\} }\, .  \label{Main22_aniso}
\end{align}%
These relations are used extensively in the derivation of the moment equations.

\section{Properties of reducible and irreducible moments}
\label{app:irreducible_moments}

Using the definition (\ref{def_M_r_mu1_mun}) of the rank-$\ell$ reducible tensor moments in the case when $r=0$ defines the unprojected rank-$\ell $ moments $M_{0}^{\mu _{1}\cdots \mu _{\ell }}$ of the particle four-momenta, and hence%
\begin{equation}
M_{r}^{\mu _{1}\cdots \mu _{\ell }}\equiv u_{\mu _{\ell+1}}\cdots u_{\mu_{\ell+r}} 
\int \mathrm{d}K\, k^{\mu _{1}}\cdots k^{\mu _{\ell +r}}f_{\mathbf{k}}=u_{\mu _{\ell+1}}\cdots 
u_{\mu_{\ell+r}} M_{0}^{\mu _{1}\cdots \mu _{\ell +r}} \, .
\end{equation}
Using the elementary projection operator $\Delta _{\nu }^{\mu }$, or the individually projected moments $k^{\left\langle \mu \right\rangle }=\Delta_{\nu }^{\mu }k^{\nu }$, a new momentum-space integral forming a rank-$\ell$ tensor moment that is orthogonal to the four-velocity is constructed,
\begin{equation}
\mathbf{M}_{r}^{\mu _{1}\cdots \mu _{\ell }}\equiv M_{r}^{\left\langle \mu_{1}\right\rangle 
\cdots \left\langle \mu _{\ell }\right\rangle }=\Delta_{\nu _{1}}^{\mu _{1}}\cdots 
\Delta _{\nu _{\ell }}^{\mu _{\ell }}M_{r}^{\nu_{1}\cdots \nu _{\ell }}
=\int \mathrm{d}KE_{\mathbf{k}u}^{r}k^{\left\langle \mu_{1}\right\rangle }
\cdots k^{\left\langle \mu _{\ell }\right\rangle }f_{\mathbf{k}}\, .  \label{def_M_r_projected}
\end{equation}%
These are projected moments orthogonal to the four-velocity.
In the LR frame they only have spatial components.
Through subsequently inserting Eq.~(\ref{k_mu_decomposition}) into Eq.~(\ref{def_M_r_mu1_mun}) the unprojected moments $M_{r}^{\mu _{1}\cdots \mu _{\ell }}$ may be tensor-decomposed using the orthogonally projected moments $\mathbf{M}_{r}^{\mu _{1}\cdots \mu _{\ell }}$ and the four-velocity as
\begin{equation}
M_{r}^{\mu _{1}\cdots \mu _{\ell }}\equiv \sum_{n=0}^{\ell }\binom{\ell }{n}%
\mathbf{M}_{r+\ell -n}^{\left( \mu _{1}\right. \cdots \mu _{n}}u^{\mu_{n+1}}\cdots 
u^{\left. \mu _{\ell }\right) }=\sum_{n=0}^{\ell }\frac{\ell !}{n!\left( \ell -n\right) !} 
u^{\left( \mu _{1}\right. }\cdots u^{\mu _{n}} 
\mathbf{M}_{r+n}^{\mu _{n+1}\cdots \left. \mu _{\ell }\right) }\, ,
\label{eq:M_vecM_relation}
\end{equation}%
where the binomial coefficient gives the number of
distinct permutations of indices needed for the symmetrization. 

The isotropic irreducible tensors from Eq.~(\ref{k_mu1_k_mun}), i.e., $1,$ $k^{\left\langle \mu _{1}\right\rangle },\ k^{\left\langle \mu _{1}\right. }k^{\left. \mu _{2}\right\rangle },\ldots ,k^{\left\langle \mu _{1}\right. }\cdots k^{\left. \mu _{n}\right\rangle }$, form a complete and orthogonal set and satisfy the following orthogonality conditions for an arbitrary function of energy $\mathrm{F}\left( E_{\mathbf{k}u}\right) $, 
\begin{equation}
\int \mathrm{d}K\, \mathrm{F}\left( E_{\mathbf{k}u}\right) k^{\left\langle \mu _{1}\right. }\cdots k^{\left. \mu _{\ell}\right\rangle }k_{\left\langle \nu_{1}\right. }\cdots 
k_{\left. \nu _{n}\right\rangle }
=\frac{\ell!\ \delta _{\ell n}}{\left( 2\ell+1\right) !!} 
\Delta _{\nu _{1}\cdots \nu _{\ell}}^{\mu _{1}\cdots\mu _{\ell}}
\int \mathrm{d}K\, \mathrm{F}\left( E_{\mathbf{k}u}\right) 
\left( \Delta^{\alpha \beta }k_{\alpha }k_{\beta }\right) ^{\ell}\, .
\label{iso_orthogonality}
\end{equation}%
The IIMs in Eq.~(\ref{IIMs}) are irreducible projections of all previously defined reducible moments,
\begin{equation}
\mathcal{M}_{r}^{\mu _{1}\cdots \mu _{\ell }}\equiv \Delta _{\nu _{1}\cdots
\nu _{\ell }}^{\mu _{1}\cdots \mu _{\ell }}M_{r}^{\nu _{1}\cdots \nu _{\ell}}
\equiv M_{r}^{\left\langle \mu _{1}\right. \cdots \left. \mu _{\ell}\right\rangle }
=\Delta _{\nu _{1}\cdots \nu _{\ell }}^{\mu _{1}\cdots \mu _{\ell }}%
\mathbf{M}_{r}^{\nu _{1}\cdots \nu _{\ell }}\equiv \mathbf{M}_{r}^{\left\langle\mu _{1}\right. 
\cdots \left. \mu _{\ell }\right\rangle}\, ,
\end{equation}%
hence using Eq.~(\ref{Main1_iso}) up to tensor-rank two we find the following useful relations between the orthogonally projected moments and the IIMs, 
\begin{eqnarray}
\mathbf{M}_{r} &=&\mathcal{M}_{r}\, ,\quad \mathbf{M}_{r}^{\mu }=\mathcal{M}_{r}^{\mu }\, , \\
\mathbf{M}_{r}^{\mu \nu } &=&\mathcal{M}_{r}^{\mu \nu } + 
\frac{1}{3}\Delta^{\mu \nu }\left( m_{0}^{2}\mathcal{M}_{r}-\mathcal{M}_{r+2}\right) \, .
\end{eqnarray}

Recalling the definition of the anisotropic reducible moments from Eq.~(\ref{def_A_mu1_mun}) one can show that
\begin{align}
A_{ij}^{\mu _{1}\cdots \mu _{\ell }}& \equiv \left( -1\right) ^{j}u_{\mu_{\ell+1}}\cdots 
u_{\mu _{\ell+i}}l_{\mu _{\ell+i+1}}\cdots l_{\mu _{\ell+i+j}} \int \mathrm{d}K k^{\mu_{1}}\cdots 
k^{\mu _{\ell +i+j}}f_{\mathbf{k}}\notag \\
&=\left( -1\right) ^{j}u_{\mu_{\ell+1}}\cdots u_{\mu _{\ell+i}}l_{\mu _{\ell+i+1}}\cdots 
l_{\mu _{\ell+i+j}} A_{00}^{\mu_{1}\cdots \mu _{\ell +i+j}}\, .
\end{align}%
Here, similarly to Eq.~(\ref{eq:M_vecM_relation}), inserting Eq.~(\ref{k_mu_decomposition}) into Eq.~(\ref{def_A_mu1_mun}) leads to the following relation between the unprojected and orthogonally projected anisotropic moments,
\begin{eqnarray}
A_{ij}^{\mu _{1}\cdots \mu _{\ell }} &\equiv &\sum_{n=0}^{\ell}\sum_{m=0}^{\ell -n}\binom{\ell }{n}\binom{\ell -n}{m}\mathbf{A}_{i+\ell-n-m,j+m}^{\left( \mu _{1}\right. \cdots 
\mu _{n}}l^{\mu _{n+1}}\cdots l^{\mu _{n+m}}u^{\mu _{n+m+1}}\cdots u^{\left. \mu _{\ell }\right) } 
\notag \\
&=&\sum_{n=0}^{\ell }\sum_{m=0}^{\ell -n}\frac{\ell !}{n!m!\left( \ell-n-m\right) !} 
u^{\left( \mu _{1}\right. }\cdots u^{\mu _{n}}l^{\mu_{n+1}}\cdots l^{\mu _{n+m}} 
\mathbf{A}_{i+n,j+m}^{\mu _{n+m+1}\cdots \left.\mu _{\ell }\right) }\, ,  \label{eq:A_vecA_relation}
\end{eqnarray}%
where the binomial coefficients $\binom{\ell }{n}\binom{\ell -n}{m}$ define
the number of distinct permutations of indices.
Furthermore, in analogy to Eq.~(\ref{def_M_r_projected}) the individually projected rank-$\ell$ tensor moments that are orthogonal to both $u^{\mu }$ and $l^{\mu }$ are 
\begin{equation}
\mathbf{A}_{ij}^{\mu _{1}\cdots \mu _{\ell }}\equiv A_{ij}^{\left\{ \mu
_{1}\right\} \cdots \left\{ \mu _{\ell }\right\} }=\Xi _{\nu _{1}}^{\mu
_{1}}\cdots \Xi _{\nu _{\ell }}^{\mu _{\ell }}A_{ij}^{\nu _{1}\cdots \nu_{\ell }}
= \int \mathrm{d}K E_{\mathbf{k}u}^{i}E_{\mathbf{k}l}^{j}k^{\left\{ \mu _{1}\right\} }\cdots
k^{\left\{ \mu _{\ell }\right\} }f_{\mathbf{k}}\, .
\end{equation}%

The anisotropic irreducible tensors $1,$ $k^{\left\{ \mu _{1}\right\} },\ k^{\left\{ \mu_{1}\right. }k^{\left. \mu _{2}\right\} },\ldots ,k^{\left\{ \mu _{1}\right.}\cdots k^{\left. \mu _{n}\right\} }$ also form a complete and orthogonal set. 
Now, similarly to Eq.~(\ref{iso_orthogonality}) but for an arbitrary function of both $E_{\mathbf{k}u}$ and $E_{\mathbf{k}l}$, i.e., $\mathrm{\hat{F}}\left( E_{\mathbf{k}u},E_{\mathbf{k}l}\right)$,  the following orthogonality condition holds, see also Ref.~\cite{Molnar:2016vvu} for details, 
\begin{equation}
\int \mathrm{d}K\, \mathrm{\hat{F}}\left( E_{\mathbf{k}u},E_{\mathbf{k}l}\right)
k^{\left\{ \mu _{1}\right. }\cdots k^{\left. \mu _{\ell}\right\} }k_{\left\{\nu _{1}\right. }
\cdots k_{\left. \nu _{n}\right\} } 
= \frac{\delta _{\ell n}}{2^{\ell}} \, \Xi _{\nu _{1}\cdots \nu _{\ell}}^{\mu _{1}\cdots \mu _{\ell}} 
\int \mathrm{d}K\, \mathrm{\hat{F}}\left( E_{\mathbf{k}u},E_{\mathbf{k}l}\right) 
\left( \Xi ^{\alpha \beta }k_{\alpha }k_{\beta }\right)^{\ell}\, . \label{aniso_orthogonality}
\end{equation}%
The AIMs are irreducible projections of all previously defined reducible moments,
\begin{equation}
\mathcal{A}_{ij}^{\mu _{1}\cdots \mu _{\ell }}\equiv 
\Xi _{\nu _{1}\cdots \nu _{\ell }}^{\mu _{1}\cdots \mu _{\ell }}A_{ij}^{\nu _{1}\cdots \nu _{\ell}}
\equiv A_{ij}^{\left\{ \mu _{1}\right. \cdots \left. \mu _{\ell }\right\} } 
=\Xi _{\nu _{1}\cdots \nu _{\ell }}^{\mu _{1}\cdots \mu _{\ell }} 
\mathbf{A}_{ij}^{\nu _{1}\cdots \nu _{\ell }}\equiv \mathbf{A}_{ij}^{\left\{ \mu _{1}\right.
\cdots \left. \mu _{\ell }\right\} }\, .
\end{equation}
Note that there are relations between the IIMs and the AIMs, which we write down up to tensor-rank 2, 
\begin{align}
\mathcal{M}_{i}& =\mathcal{A}_{i0}\,,\quad \mathcal{M}_{i}^{\mu } 
=\mathcal{A}_{i1}l^{\mu }+\mathcal{A}_{i0}^{\mu }\,, \\
\mathcal{M}_{i}^{\mu \nu }& =\mathcal{A}_{i0}^{\mu \nu } 
+2\mathcal{A}_{i1}^{\left( \mu \right.} l^{\left. \nu \right) } 
+\frac{1}{6}\left(m_{0}^{2}\mathcal{A}_{i0} -\mathcal{A}_{i+2,0}+3\mathcal{A}_{i2}\right)
\left( 2l^{\mu }l^{\nu }+\Xi ^{\mu \nu }\right) \,.
\end{align}%

\section{The expansion of the distribution function}
\label{app:expansion_coefficients}

The non-equilibrium distribution function $f_{\mathbf{k}}$ is expanded around a local-equilibrium state $f_{0\mathbf{k}}$, such that $f_{\mathbf{k}} = f_{0\mathbf{k}} + \delta f_{\mathbf{k}}$, with $\delta f_{\mathbf{k}}\equiv f_{0\mathbf{k}}\left( 1-af_{0\mathbf{k}}\right) \phi _{\mathbf{k}}$.
In turn, $\phi _{\mathbf{k}}$ is expanded in terms of a complete and orthogonal set of isotropic irreducible tensors $\{1, k_{\langle \mu \rangle}, k_{\langle \mu } k_{\nu \rangle } , \ldots \}$, see Eq.~(30) of Ref.~\cite{Denicol:2012cn}, 
\begin{equation}
f_{\mathbf{k}}=f_{0\mathbf{k}}+f_{0\mathbf{k}}\left( 1-af_{0\mathbf{k}}\right) 
\sum_{\ell =0}^{\infty }\sum_{n=0}^{N_{\ell }}\rho _{n}^{\mu_{1}\cdots \mu _{\ell }} 
k_{\left\langle \mu _{1}\right. }\cdots k_{\left.\mu _{\ell }\right\rangle }
\mathcal{H}_{\mathbf{k}n}^{(\ell )}\,.
\label{f_iso_expansion}
\end{equation}%
Here, the coefficient $\rho_n^{\mu_1 \cdots \mu_\ell}$ is the IIM of $\delta f_{\mathbf{k}}$, 
cf. Eq.~(\ref{kinetic:rho_r}),  while $\mathcal{H}_{\mathbf{k}n}^{(\ell )}$ is a polynomial in energy $E_{\mathbf{k}u}$ of order $N_{\ell }$, see Eq.~(\ref{eq:Hfunction_k}).
In principle, the expansion is complete only in the limit $N_{\ell }\rightarrow \infty $.
However, a complete expansion is in most cases practically impossible and hence one must truncate the series at some finite order $N_\ell$ to obtain an approximation for $f_{\mathbf{k}}$.

The moments of negative order $\rho_{-r}^{\mu _{1}\cdots \mu _{\ell }}$ are not explicitly included in the expansion (\ref{f_iso_expansion}) but they can be constructed in terms of those that are included in the expansion and therefore any IIM of $\delta f_{\mathbf{k}}$ of tensor-rank $\ell $ and of arbitrary order $r$ can be expressed as a linear combination of rank-$\ell$ moments with positive order $n\geq 0$, 
\begin{equation}
\rho _{\pm r}^{\mu _{1}\cdots \mu _{\ell }}=\sum_{n=0}^{N_{\ell }}
\rho_{n}^{\mu _{1}\cdots \mu _{\ell }} 
\mathcal{F}_{\mp r,n}^{\left( \ell\right)}\,,  \label{F_rn_useful}
\end{equation}
where the coefficient $\mathcal{F}_{\mp r,n}^{\left( \ell\right)}$ is defined in Eq.~(\ref{F_rn}).
Note that in order to derive the fluid-dynamical equations of motion one also has other choices and approximations for the negative-order moments that have more favorable convergence properties, see for example Ref.~\cite{Wagner:2023joq} for a comparison.

The coefficient $\mathcal{H}_{\mathbf{k}n}^{(\ell )}$ in Eq.~(\ref{F_rn_useful}) is a polynomial in energy $E_{\mathbf{k}u}$ of order $N_{\ell }\rightarrow \infty $, see for example Ref.~\cite{Denicol:2012cn}, 
\begin{equation}
\mathcal{H}_{\mathbf{k}n}^{(\ell )}=\frac{(-1)^{\ell }}{\ell ! J_{2\ell,\ell}} 
\sum_{i=n}^{N_{\ell }}a_{in}^{(\ell )}P_{\mathbf{k}i}^{(\ell )}\,,
\label{eq:Hfunction_k}
\end{equation}%
where  
\begin{equation}
P_{\mathbf{k}n}^{(\ell )}=\sum_{i=0}^{n}a_{ni}^{(\ell )}E_{\mathbf{k}u}^{i}\,.  \label{eq:P_k}
\end{equation}
are polynomials of order $n$ in $E_{\mathbf{k}u}$.
The coefficients $a_{ni}^{(\ell )}$ are independent of $E_{\mathbf{k}u}$ and found from the Gram-Schmidt procedure imposing the following orthogonality condition: 
\begin{equation}
\int \mathrm{d} K\omega ^{(\ell )}P_{\mathbf{k}m}^{(\ell )}P_{\mathbf{k}n}^{(\ell)} =\delta _{mn}\,,  \label{eq:P_ortho}
\end{equation}%
where 
\begin{equation}
\omega ^{(\ell )}=\frac{(-1)^{\ell }}{(2\ell +1)!!}\frac{1}{J_{2\ell ,\ell }}%
(\Delta ^{\alpha \beta }k_{\alpha }k_{\beta })^{\ell }f_{0\mathbf{k}}
\left(1-af_{0\mathbf{k}}\right) \,.  
\label{eq:omega_def}
\end{equation}
The moments of negative order $r < 0$ are constructed in terms of those that are included in the expansion~\cite{Denicol:2012cn}. 
They are obtained by substituting Eq.~(\ref{f_iso_expansion}) into Eq.~(\ref{kinetic:rho_r}) and 
using the orthogonality condition~(\ref{iso_orthogonality}). 
This leads to Eq.~(\ref{F_rn_useful}) with the following coefficient 
\begin{equation}
\mathcal{F}_{\mp rn}^{\left( \ell \right) }=\frac{\ell !}{\left( 2\ell+1\right) !!}
\int \mathrm{d}K\,E_{\mathbf{k}u}^{\pm r}\left( \Delta ^{\alpha\beta } 
k_{\alpha }k_{\beta }\right) ^{\ell }\mathcal{H}_{\mathbf{k}n}^{\left( \ell \right) }
f_{0\mathbf{k}}\left( 1-af_{0\mathbf{k}}\right) \,.
\label{F_rn}
\end{equation}%
Finally, $J_{nq}(\alpha,\beta)$ are the auxiliary thermodynamic integrals defined through the 
derivatives of $I_{nq}$, cf.~Eq.~(\ref{I_nq}), with respect to $\alpha $ and $\beta$, 
\begin{equation}
J_{nq}\left( \alpha ,\beta \right) \equiv 
\left( \frac{\partial I_{nq}}{\partial \alpha }\right)_{\beta }
=-\left( \frac{\partial I_{n-1,q}}{\partial \beta }\right) _{\alpha } 
=\frac{(-1)^{q}}{(2q+1)!!} \int dK E_{\mathbf{k}u}^{n-2q} 
\left( \Delta ^{\alpha \beta }k_{\alpha }k_{\beta}\right)^{q}f_{0\mathbf{k}}
\left( 1-af_{0\mathbf{k}}\right) \, .
\label{J_nq}
\end{equation}

In complete analogy to the expansion around the local-equilibrium distribution $f_{0\mathbf{k}}$, we can also expand $f_{\mathbf{k}}$ around the anisotropic distribution $\hat{f}_{0\mathbf{k}}$, i.e., $f_{\mathbf{k}} = \hat{f}_{0\mathbf{k}} + \delta \hat{f}_{\mathbf{k}}$, where $\delta \hat{f}_{\mathbf{k}}\equiv \hat{f}_{0\mathbf{k}}\left( 1-a\hat{f}_{0\mathbf{k}}\right) \hat{\phi}_{\mathbf{k}}$.
Like $\phi_{\mathbf{k}}$ in the isotropic case, $\hat{\phi}_{\mathbf{k}}$ has an expansion in terms of the complete and orthogonal set of anisotropic irreducible tensors $\{1, k_{\{\mu\}}, k_{\{ \mu } k_{\nu \} }, \ldots \}$, see Eq.~(98) of Ref.~\cite{Molnar:2016vvu}, 
\begin{equation}
f_{\mathbf{k}}=\hat{f}_{0\mathbf{k}}+\hat{f}_{0\mathbf{k}}\left( 1-a\hat{f}_{0\mathbf{k}}\right) \sum_{\ell =0}^{\infty }\sum_{n=0}^{N_{\ell}}\sum_{m=0}^{N_{\ell }-n} 
\hat{\rho}_{nm}^{\mu _{1}\cdots \mu _{\ell}}k_{\left\{ \mu _{1}\right. }\cdots 
k_{\left. \mu _{\ell }\right\} }\hat{\mathcal{H}}_{\mathbf{k}nm}^{(\ell )}\, ,  \label{kinetic:f_full_anisotropic}
\end{equation}
where $\hat{\rho}_{nm}^{\mu_1 \cdots \mu_\ell}$ are the AIMs of $\delta \hat{f}_{\mathbf{k}}$, cf.\ Eq.~(\ref{kinetic:rho_ij_hat}), and $\hat{\mathcal{H}}_{\mathbf{k}nm}^{(\ell )}$ is defined as, see Ref.~\cite{Molnar:2016vvu},
\begin{equation}
\hat{\mathcal{H}}_{\mathbf{k}nm}^{(\ell )}=\frac{(-1)^{\ell }}{\ell ! \hat{J}_{2\ell ,0 ,\ell}} \sum_{i=n}^{N_{\ell }-m}\sum_{j=m}^{N_{\ell}-i}a_{injm}^{(\ell )} 
P_{\mathbf{k}ij}^{\left( \ell \right) }\, ,
\label{kinetic:H_nm}
\end{equation}
where $P_{\mathbf{k}nm}^{\left( \ell \right) }$ are polynomials in both $E_{\mathbf{k}u}$ and $E_{\mathbf{k}l}$, 
\begin{equation}
P_{\mathbf{k}nm}^{\left( \ell\right)}=\sum_{i=0}^{n}\sum_{j=0}^{m}a_{nimj}^{(\ell )} 
E_{\mathbf{k}u}^{i}E_{\mathbf{k}l}^{j}\, .  \label{kinetic:P_nm}
\end{equation}
The coefficients $a_{nimj}^{(\ell )}$ are independent of $E_{\mathbf{k}u}$ and $E_{\mathbf{k}l}$, and are determined by demanding the following orthonormality relation for the polynomials $P_{\mathbf{k}nm}^{\left( \ell \right) }$, 
\begin{equation}
\int \mathrm{d}K\,\hat{\omega}^{\left( \ell \right) }P_{\mathbf{k}nm}^{\left( \ell\right) } P_{\mathbf{k}n^{\prime }m^{\prime }}^{\left( \ell \right)}
=\delta_{nn^{\prime }} \delta _{mm^{\prime }}\, ,
\label{kinetic:orthonormality}
\end{equation}
where the weight is defined as 
\begin{equation}
\hat{\omega}^{\left( \ell \right) }=\frac{(-1)^\ell}{\left( 2\ell \right) !!}
\frac{1}{\hat{J}_{2\ell,0 ,\ell }} \left( \Xi ^{\alpha \beta }k_{\alpha}k_{\beta}\right)^{\ell }
\hat{f}_{0\mathbf{k}} \left( 1-a\hat{f}_{0\mathbf{k}}\right)\, .  
\label{kinetic:weight}
\end{equation}

Furthermore, similar to Eq.~(\ref{F_rn_useful}) any AIM of tensor-rank $\ell$ and of arbitrary order $i,j$ can be expressed as a linear combination of rank-$\ell $ moments with positive order $n,m\geq 0$,
\begin{equation}
\hat{\rho}_{\pm ij}^{\mu _{1}\cdots \mu _{\ell }}
=\sum_{n=0}^{N_{\ell}}\sum_{m=0}^{N_{\ell }-n}\hat{\rho}_{nm}^{\mu _{1}\cdots \mu _{\ell }}
\hat{\mathcal{F}}_{\mp ij,nm}^{\left( \ell \right) }\,.  
\label{F_ijnm_useful}
\end{equation}
Here, the coefficients $\hat{\mathcal{F}}_{\mp ij,nm}^{\left( \ell \right) }$ are constructed similarly to $\mathcal{F}_{\mp r,n}^{\left( \ell \right) }$, cf.~ Eq.~(\ref{F_rn}).
Namely, substituting Eq.~(\ref{kinetic:f_full_anisotropic}) into Eq.~(\ref{kinetic:rho_ij_hat}) and using the orthogonality condition~(\ref{aniso_orthogonality}) we obtain
\begin{equation}
\hat{\mathcal{F}}_{\mp ij,nm}^{\left( \ell \right) } =\frac{\ell !}{\left(2\ell \right) !!}
\int \mathrm{d}K E_{\mathbf{k}u}^{\pm i}E_{\mathbf{k}l}^{j}\left(
\Delta ^{\alpha \beta }k_{\alpha }k_{\beta }\right) ^{\ell }
\hat{\mathcal{H}}_{\mathbf{k}nm}^{\left( \ell \right) }
\hat{f}_{0\mathbf{k}}\left( 1-a\hat{f}_{0\mathbf{k}}\right) \,.  
\label{F_ijnm}
\end{equation}%
The corresponding generalized auxiliary thermodynamic integrals are defined similarly to Eq.~(\ref{J_nq}), 
\begin{equation}
\hat{J}_{nrq} \left(\hat{\alpha}, \hat{\beta}_u , \hat{\beta}_l \right)
\equiv \left( \frac{\partial \hat{I}_{nrq}}{\partial\hat{\alpha}}\right)_{\hat{\beta}_{u},\hat{\beta}_{l}} 
=\frac{\left( -1\right)^{q}} {\left(2q\right) !!} \int \mathrm{d}K E_{\mathbf{k}u}^{n-r-2q} E_{\mathbf{k}l}^{r} \left(\Xi ^{\mu \nu }k_{\mu }k_{\nu }\right) ^{q} \hat{f}_{0\mathbf{k}}
\left( 1-a\hat{f}_{0\mathbf{k}}\right) \,.  
\label{J_nrq}
\end{equation}

\newpage
\section{Term by term evaluation of Eq.~(\protect\ref{Main_DM_i_mu1_mun})}
\label{Appendix:DM_i}

Inserting Eq.~(\ref{D_BTE_F_munu}) into Eq.~(\ref{def_D_M_i}) we obtain the following equation of tensor-rank $\ell$,
\begin{align}
D\mathcal{M}_{i}^{\left\langle \mu _{1}\cdots \mu _{\ell }\right\rangle }& =%
\mathcal{C}_{i-1}^{\mu _{1}\cdots \mu _{\ell }}+\Delta _{\nu _{1}\cdots \nu_{\ell }}^{\mu _{1} 
\cdots \mu _{\ell }} \int \mathrm{d}K D\left( E_{\mathbf{k}u}^{i}\right) 
k^{\left\langle \nu _{1}\right. }\cdots k^{\left. \nu _{\ell}\right\rangle } 
f_{\mathbf{k}}+\Delta _{\nu _{1}\cdots \nu _{\ell }}^{\mu_{1}\cdots \mu _{\ell }}
\int \mathrm{d}K E_{\mathbf{k}u}^{i}D\left(k^{\left\langle \nu _{1}\right. }\cdots 
k^{\left. \nu _{\ell }\right\rangle}\right) f_{\mathbf{k}}  \notag \\
&- \Delta _{\nu _{1}\cdots \nu _{\ell }}^{\mu _{1}\cdots \mu _{\ell }}
\int \mathrm{d}K E_{\mathbf{k}u}^{i-1}k^{\left\langle \nu _{1}\right. }\cdots
k^{\left. \nu _{\ell }\right\rangle }k^{\left\langle \nu _{\ell+1}\right\rangle }
\nabla _{\nu _{\ell +1}} f_{\mathbf{k}} - \mathbf{q}F^{\mu\nu }
\Delta _{\nu _{1}\cdots \nu _{\ell }}^{\mu _{1}\cdots \mu _{\ell}}
\int \mathrm{d}K E_{\mathbf{k}u}^{i-1}k^{\left\langle \nu _{1}\right.}\cdots 
k^{\left. \nu _{\ell }\right\rangle }k_{\nu }\frac{\partial f_{\mathbf{k}}}{\partial k^{\mu }}\, .  \label{DM_i_mu1_mun}
\end{align}%
Each term will be evaluated separately in the following, with the abbreviation $\int_\mathbf{k} \equiv \int \mathrm{d}K$. 

The first term on the right-hand side with $E_{\mathbf{k}u}=k^{\mu }u_{\mu}$, $u^{\mu }Du_{\mu }=0$, and $DE_{\mathbf{k}u}^{i}=iE_{\mathbf{k} u}^{i-1}k^{\left\langle \alpha \right\rangle }Du_{\alpha }$ leads to 
\begin{eqnarray}
M1 &\equiv &\Delta _{\nu _{1}\cdots \nu _{\ell }}^{\mu _{1}\cdots \mu _{\ell
}}\int_{\mathbf{k}}D\left( E_{\mathbf{k}u}^{i}\right) k^{\left\langle \nu
_{1}\right. }\cdots k^{\left. \nu _{\ell }\right\rangle }f_{\mathbf{k}%
}=iDu_{\nu _{\ell +1}}\Delta _{\nu _{1}\cdots \nu _{\ell }}^{\mu _{1}\cdots
\mu _{\ell }}\int_{\mathbf{k}}E_{\mathbf{k}u}^{i-1}k^{\left\langle \nu
_{1}\right. }\cdots k^{\left. \nu _{\ell }\right\rangle }k^{\left\langle \nu
_{\ell +1}\right\rangle }f_{\mathbf{k}}  \notag \\
&=&i\mathcal{M}_{i-1}^{\mu _{1}\cdots \mu _{\ell +1}}Du_{\mu _{\ell +1}}+i%
\frac{\ell }{2\ell +1}\left( m_{0}^{2}\mathcal{M}_{i-1}^{\left\langle \mu
_{1}\right. \cdots \mu _{\ell -1}}-\mathcal{M}_{i+1}^{\left\langle \mu
_{1}\right. \cdots \mu _{\ell -1}}\right) Du^{\left. \mu _{\ell}\right\rangle }\, ,  
\label{M1_term}
\end{eqnarray}%
where we used Eq.~(\ref{Main2_iso}) and replaced $\Delta_{\alpha \beta }k^{\alpha }k^{\beta } = m_{0}^{2}-E_{\mathbf{k}u}^{2}$.

The second term is 
\begin{eqnarray}
M2 &\equiv &\Delta _{\nu _{1}\cdots \nu _{\ell }}^{\mu _{1}\cdots \mu _{\ell
}}\int_{\mathbf{k}}E_{\mathbf{k}u}^{i}D\left( k^{\left\langle \nu
_{1}\right. }\cdots k^{\left. \nu _{\ell }\right\rangle }\right) f_{\mathbf{k%
}}=-\ell \Delta _{\nu _{1}\cdots \nu _{\ell }}^{\mu _{1}\cdots \mu _{\ell
}}\left( \mathbf{M}_{i+1}^{\nu _{1}\cdots \nu _{\ell -1}}Du^{\nu _{\ell
}}\right)  \notag \\
&=&-\ell \mathcal{M}_{i+1}^{\left\langle \mu _{1}\right. \cdots \mu _{\ell
-1}}Du^{\left. \mu _{\ell }\right\rangle }\, ,  \label{M2_term}
\end{eqnarray}%
where the proper-time derivative of the traceless projection is 
$\Delta_{\nu _{1}\cdots \nu _{\ell }}^{\mu _{1}\cdots \mu _{\ell }}D\left( \Delta_{\lambda _{1}\cdots \lambda _{\ell }}^{\nu_{1}\cdots \nu _{\ell }}\right) =-\ell \Delta _{\nu _{1}\cdots \nu _{\ell }}^{\mu _{1}\cdots \mu _{\ell}}\left( u_{\lambda _{1}}Du^{\nu _{1}}\right) \Delta _{\lambda _{2}}^{\nu_{2}}\cdots \Delta _{\lambda _{\ell }}^{\nu _{\ell }}$.

The third term leads to the following three integrals 
\begin{align}
M3& \equiv -\Delta _{\nu _{1}\cdots \nu _{\ell }}^{\mu _{1}\cdots \mu _{\ell
}}\int_{\mathbf{k}}E_{\mathbf{k}u}^{i-1}k^{\left\langle \nu _{1}\right.
}\cdots k^{\left. \nu _{\ell }\right\rangle }k^{\left\langle \mu
\right\rangle }\nabla _{\mu }f_{\mathbf{k}}=-\Delta _{\nu _{1}\cdots \nu
_{\ell }}^{\mu _{1}\cdots \mu _{\ell }}\nabla _{\mu }\left( \int_{\mathbf{k}%
}E_{\mathbf{k}u}^{i-1}k^{\left\langle \nu _{1}\right. }\cdots k^{\left. \nu
_{\ell }\right\rangle }k^{\left\langle \mu \right\rangle }f_{\mathbf{k}%
}\right)  \notag \\
& +\Delta _{\nu _{1}\cdots \nu _{\ell }}^{\mu _{1}\cdots \mu _{\ell }}\int_{%
\mathbf{k}}\left( \nabla _{\mu }E_{\mathbf{k}u}^{i-1}\right) k^{\left\langle
\nu _{1}\right. }\cdots k^{\left. \nu _{\ell }\right\rangle }k^{\left\langle
\mu \right\rangle }f_{\mathbf{k}}+\Delta _{\nu _{1}\cdots \nu _{\ell }}^{\mu
_{1}\cdots \mu _{\ell }}\int_{\mathbf{k}}E_{\mathbf{k}u}^{i-1}\nabla _{\mu
}\left( k^{\left\langle \nu _{1}\right. }\cdots k^{\left. \nu _{\ell
}\right\rangle }k^{\left\langle \mu \right\rangle }\right) f_{\mathbf{k}}\, .
\label{M3_term_def}
\end{align}%
Now making use of Eq.~(\ref{Main2_iso}) the first integral leads to%
\begin{align}
M3a& \equiv -\Delta _{\nu _{1}\cdots \nu _{\ell }}^{\mu _{1}\cdots \mu
_{\ell }}\nabla _{\mu }\left( \int_{\mathbf{k}}E_{\mathbf{k}%
u}^{i-1}k^{\left\langle \nu _{1}\right. }\cdots k^{\left. \nu _{\ell
}\right\rangle }k^{\left\langle \mu \right\rangle }f_{\mathbf{k}}\right) 
\notag \\
& =-\Delta _{\nu _{1}\cdots \nu _{\ell }}^{\mu _{1}\cdots \mu _{\ell
}}\left( \nabla _{\nu _{\ell +1}}\mathcal{M}_{i-1}^{\nu _{1}\cdots \nu
_{\ell +1}}\right) -\frac{\ell }{2\ell +1}\nabla ^{\left\langle \mu
_{1}\right. }\left( m_{0}^{2}\mathcal{M}_{i-1}^{\mu _{2}\cdots \left. \mu
_{\ell }\right\rangle }-\mathcal{M}_{i+1}^{\mu _{2}\cdots \left. \mu _{\ell
}\right\rangle }\right) \, ,  \label{M3a_term}
\end{align}%
where we used that $\Delta _{\nu _{1}\cdots \nu _{\ell }}^{\mu _{1}\cdots\mu _{\ell }}\left[ \nabla ^{\lambda _{\ell }}\left( \Delta _{\lambda_{1}\cdots \lambda _{\ell }}^{\nu _{1}\cdots \nu _{\ell }}\right) \right] =0$. 
The second integral from Eq.~(\ref{M3_term_def}) evaluates to 
\begin{align}
M3b& \equiv \Delta _{\nu _{1}\cdots \nu _{\ell }}^{\mu _{1}\cdots \mu _{\ell
}}\int_{\mathbf{k}}\left( \nabla _{\mu }E_{\mathbf{k}u}^{i-1}\right)
k^{\left\langle \nu _{1}\right. }\cdots k^{\left. \nu _{\ell }\right\rangle
}k^{\left\langle \mu \right\rangle }f_{\mathbf{k}}=\left( i-1\right) \Delta
_{\nu _{1}\cdots \nu _{\ell }}^{\mu _{1}\cdots \mu _{\ell }}\left( \nabla
_{\mu }u_{\nu }\right) \int_{\mathbf{k}}E_{\mathbf{k}u}^{i-2}k^{\left\langle
\nu _{1}\right. }\cdots k^{\left. \nu _{\ell }\right\rangle }k^{\left\langle
\mu \right\rangle }k^{\left\langle \nu \right\rangle }f_{\mathbf{k}}  \notag
\\
& =\frac{ i-1 }{3}\theta \left( m_{0}^{2}\mathcal{M}_{i-2}^{\mu
_{1}\cdots \mu _{\ell }}-\mathcal{M}_{i}^{\mu _{1}\cdots \mu _{\ell
}}\right) +\left( i-1\right) \mathcal{M}_{i-2}^{\mu _{1}\cdots \mu _{\ell
+2}}\sigma _{\mu _{\ell +1}\mu _{\ell +2}}  \notag \\
& +\left( i-1\right) \frac{2\ell }{2\ell +3}\left( m_{0}^{2}\mathcal{M}%
_{i-2}^{\nu \left\langle \mu _{1}\right. \cdots \mu _{\ell -1}}-\mathcal{M}%
_{i}^{\nu \left\langle \mu _{1}\right. \cdots \mu _{\ell -1}}\right) \sigma
_{\nu }^{\left. \mu _{\ell }\right\rangle }  \notag \\
& +\left( i-1\right) \frac{\ell \left( \ell -1\right) }{4\ell ^{2}-1}\left(
m_{0}^{4}\mathcal{M}_{i-2}^{\left\langle \mu _{1}\right. \cdots \mu _{\ell
-2}}-2m_{0}^{2}\mathcal{M}_{i}^{\left\langle \mu _{1}\right. \cdots \mu
_{\ell -2}}+\mathcal{M}_{i+2}^{\left\langle \mu _{1}\right. \cdots \mu
_{\ell -2}}\right) \sigma ^{\mu _{\ell -1}\left. \mu _{\ell }\right\rangle}\, ,  
\label{M3b_term}
\end{align}%
where we used that $\nabla _{\mu }E_{\mathbf{k}u}^{i-1}=\left( i-1\right) E_{\mathbf{k}u}^{i-2}k^{\left\langle \nu \right\rangle }\left( \nabla _{\mu}u_{\nu }\right) $. 
Furthermore, we also replaced 
$\nabla _{\mu }u_{\nu }=\frac{1}{3}\theta \Delta _{\mu \nu }+\sigma _{\mu \nu}+\omega _{\mu \nu }$ 
and applied Eq.~(\ref{Main22_iso})  
The last integral from Eq.~(\ref{M3_term_def}) is 
\begin{align}
M3c& \equiv \Delta _{\nu _{1}\cdots \nu _{\ell }}^{\mu _{1}\cdots \mu _{\ell
}}\int_{\mathbf{k}}E_{\mathbf{k}u}^{i-1}\nabla _{\mu }\left( k^{\left\langle
\nu _{1}\right. }\cdots k^{\left. \nu _{\ell }\right\rangle }k^{\left\langle
\mu \right\rangle }\right) f_{\mathbf{k}}=\Delta _{\nu _{1}\cdots \nu _{\ell
}}^{\mu _{1}\cdots \mu _{\ell }}\nabla _{\nu _{\ell +1}}\left( \Delta
_{\lambda _{1}\cdots \lambda _{\ell }}^{\nu _{1}\cdots \nu _{\ell }}\Delta
_{\lambda _{\ell +1}}^{\nu _{\ell +1}}\right) M_{i-1}^{\lambda _{1}\cdots
\lambda _{\ell +1}}  \notag \\
& =-\frac{\ell +3}{3} \theta \mathcal{M}_{i}^{\mu _{1}\cdots
\mu _{\ell }}-\frac{\ell \left( \ell -1\right) }{2\ell -1}\left( m_{0}^{2}%
\mathcal{M}_{i}^{\left\langle \mu _{1}\right. \cdots \mu _{\ell -2}}-%
\mathcal{M}_{i+2}^{\left\langle \mu _{1}\right. \cdots \mu _{\ell
-2}}\right) \sigma ^{\mu _{\ell -1}\left. \mu _{\ell }\right\rangle }  \notag
\\
& -\ell \mathcal{M}_{i}^{\nu \left\langle \mu _{1}\right. \cdots \mu _{\ell
-1}}\sigma _{\nu }^{\left. \mu _{\ell }\right\rangle }+\ell \mathcal{M}%
_{i}^{\nu \left\langle \mu _{1}\right. \cdots \mu _{\ell -1}}\omega _{\left.
{}\right. \nu }^{\left. \mu _{\ell }\right\rangle }\, .  \label{M3c_term}
\end{align}%
Adding the last three results leads to the expression for Eq.~(\ref{M3_term_def}) 
in the following form, 
\begin{align}
M3& \equiv -\Delta _{\nu _{1}\cdots \nu _{\ell }}^{\mu _{1}\cdots \mu _{\ell
}}\left( \nabla _{\nu _{\ell +1}}\mathcal{M}_{i-1}^{\nu _{1}\cdots \nu
_{\ell +1}}\right) -\frac{\ell }{\left( 2\ell +1\right) }\nabla
^{\left\langle \mu _{1}\right. }\left( m_{0}^{2}\mathcal{M}_{i-1}^{\mu
_{2}\cdots \left. \mu _{\ell }\right\rangle }-\mathcal{M}_{i+1}^{\mu
_{2}\cdots \left. \mu _{\ell }\right\rangle }\right)  \notag \\
& +\frac{1}{3}\theta \left[ m_{0}^{2}\left( i-1\right) \mathcal{M}%
_{i-2}^{\mu _{1}\cdots \mu _{\ell }}-\left( i+\ell +2\right) \mathcal{M}%
_{i}^{\mu _{1}\cdots \mu _{\ell }}\right] +\left( i-1\right) \mathcal{M}%
_{i-2}^{\mu _{1}\cdots \mu _{\ell +2}}\sigma _{\mu _{\ell +1}\mu _{\ell +2}}
\notag \\
& +\frac{\ell }{2\ell +3}\left[ m_{0}^{2}\left( 2i-2\right) \mathcal{M}%
_{i-2}^{\nu \left\langle \mu _{1}\right. \cdots \mu _{\ell -1}}-\left(
2i+2\ell +1\right) \mathcal{M}_{i}^{\nu \left\langle \mu _{1}\right. \cdots
\mu _{\ell -1}}\right] \sigma _{\nu }^{\left. \mu _{\ell }\right\rangle
}+\ell \mathcal{M}_{i}^{\nu \left\langle \mu _{1}\right. \cdots \mu _{\ell
-1}}\omega _{\left. {}\right. \nu }^{\left. \mu _{\ell }\right\rangle } 
\notag \\
& +\frac{\ell \left( \ell -1\right) }{4\ell ^{2}-1}\left[ m_{0}^{4}\left(
i-1\right) \mathcal{M}_{i-2}^{\left\langle \mu _{1}\right. \cdots \mu _{\ell
-2}}-m_{0}^{2}\left( 2i+2\ell -1\right) \mathcal{M}_{i}^{\left\langle \mu
_{1}\right. \cdots \mu _{\ell -2}}+\left( i+2\ell \right) \mathcal{M}%
_{i+2}^{\left\langle \mu _{1}\right. \cdots \mu _{\ell -2}}\right] \sigma
^{\mu _{\ell -1}\left. \mu _{\ell }\right\rangle }\, .  \label{M3_term}
\end{align}%
These results were also obtained previously by de Brito and Denicol, see Appendix A of Ref.~\cite{deBrito:2024vhm}.

Finally, the coupling to the electromagnetic field is given by the following term,
\begin{align}
M4& \equiv -\mathbf{q}F^{\mu \nu }\Delta _{\nu _{1}\cdots \nu _{\ell }}^{\mu
_{1}\cdots \mu _{\ell }}\int_{\mathbf{k}}E_{\mathbf{k}u}^{i-1}k^{\left%
\langle \nu _{1}\right. }\cdots k^{\left. \nu _{\ell }\right\rangle }k_{\nu }%
\frac{\partial f_{\mathbf{k}}}{\partial k^{\mu }}  \notag \\
& =\left( i-1\right) \mathbf{q}F^{\mu \nu }u_{\mu }g_{\nu \lambda _{\ell
+1}}\Delta _{\lambda _{1}\cdots \lambda _{\ell }}^{\mu _{1}\cdots \mu _{\ell
}}\int_{\mathbf{k}}E_{\mathbf{k}u}^{i-2}k^{\lambda _{1}}\cdots k^{\lambda
_{\ell }}k^{\lambda _{\ell +1}}f_{\mathbf{k}}+\mathbf{q}F^{\mu \nu }\Delta
_{\lambda _{1}\cdots \lambda _{\ell }}^{\mu _{1}\cdots \mu _{\ell }}\int_{%
\mathbf{k}}E_{\mathbf{k}u}^{i-1}\frac{\partial }{\partial k^{\mu }}\left(
k^{\lambda _{1}}\cdots k^{\lambda _{\ell }}k_{\nu }\right) f_{\mathbf{k}}\, ,
\end{align}%
where $g_{\nu \mu} = \partial k_\nu / \partial k^\mu$ and 
$\delta^{\nu}_{\mu} = \partial k^\nu / \partial k^\mu$, while in flat Minkowski space-time 
all Kronecker deltas can be replaced by
the mixed contravariant and covariant metric tensor, $\delta^{\mu}_{\nu} \equiv g^{\mu}_{\nu}$.

Here the first integral leads to
\begin{align}
M4a& \equiv \left( i-1\right) \mathbf{q}F^{\mu \nu }u_{\mu }g_{\nu \lambda
_{\ell +1}}\Delta _{\lambda _{1}\cdots \lambda _{\ell }}^{\mu _{1}\cdots \mu
_{\ell }}M_{i-2}^{\lambda _{1}\cdots \lambda _{\ell }\lambda _{\ell
+1}}=\left( i-1\right) \Delta _{\lambda _{1}\cdots \lambda _{\ell }}^{\mu
_{1}\cdots \mu _{\ell }}\mathcal{M}_{i-2}^{\lambda _{1}\cdots \lambda _{\ell
+1}}\left( \mathbf{q}F^{\mu \nu }u_{\mu }g_{\nu \lambda _{\ell +1}}\right) 
\notag \\
& +\left( i-1\right) \frac{\ell }{2\ell +1}\Delta _{\lambda _{1}\cdots
\lambda _{\ell }}^{\mu _{1}\cdots \mu _{\ell }}\left( m_{0}^{2}\mathcal{M}%
_{i-2}^{\lambda _{1}\cdots \lambda _{\ell -1}}-\mathcal{M}_{i}^{\lambda
_{1}\cdots \lambda _{\ell -1}}\right) \left( \mathbf{q}F^{\mu \nu }u_{\mu
}\Delta _{\nu }^{\lambda _{\ell }}\right) \, ,  \label{M4a_term}
\end{align}%
where we used Eq.~(\ref{eq:M_vecM_relation}) to write,
$\Delta _{\lambda _{1}\cdots \lambda _{\ell }}^{\mu _{1}\cdots \mu_{\ell }} 
M_{i-2}^{\lambda _{1}\cdots \lambda _{\ell }\lambda _{\ell+1}} 
= \Delta _{\lambda _{1}\cdots \lambda _{\ell }}^{\mu _{1}\cdots \mu_{\ell }} 
\left[\mathbf{M}_{i-2}^{\lambda _{1}\cdots \lambda _{\ell }\lambda _{\ell+1}} 
+ \left(\ell +1 \right)\mathbf{M}_{i-1}^{\left( \lambda _{1}\cdots \lambda _{\ell } \right.} 
u^{\left. \lambda _{\ell+1} \right)} \right]$ 
and then we used Eqs.~(\ref{Main1_iso}), (\ref{Main2_iso}) to express the orthogonally 
projected moments through IIMs.

The second integral is rewritten in a similar fashion,
\begin{align}
M4b& \equiv \mathbf{q}F^{\mu \nu }g_{\nu \lambda _{\ell +1}}\Delta _{\lambda
_{1}\cdots \lambda _{\ell }}^{\mu _{1}\cdots \mu _{\ell }}\int_{\mathbf{k}%
}E_{\mathbf{k}u}^{i-1}k^{\lambda _{\ell +1}}\frac{\partial }{\partial k^{\mu
}}\left( k^{\lambda _{1}}\cdots k^{\lambda _{\ell }}\right) f_{\mathbf{k}}
=\mathbf{q}F^{\mu \nu }g_{\nu \lambda _{\ell +1}}\Delta _{\lambda _{1}\cdots
\lambda _{\ell }}^{\mu _{1}\cdots \mu _{\ell }}\sum_{n=1}^{\ell }\delta
_{\mu }^{\lambda _{n}}M_{i-1}^{\lambda _{1}\cdots \lambda _{n-1}
\lambda _{n +1}}  \notag \\
& =\ell \mathcal{M}_{i}^{\left\langle \mu _{1}\right. \cdots \mu _{\ell -1}}%
\mathbf{q}F^{\left. \mu _{\ell }\right\rangle \nu }u_{\nu }+\ell \Delta
_{\lambda _{1}\cdots \lambda _{\ell }}^{\mu _{1}\cdots \mu _{\ell }}\mathcal{%
M}_{i-1}^{\lambda _{1}\cdots \lambda _{\ell -1}\lambda _{\ell +1}}\left( 
\mathbf{q}F^{\lambda _{\ell }\nu }g_{\nu \lambda _{\ell +1}}\right)  \notag
\\
& +\frac{\ell \left( \ell -1\right) }{2\ell -1}\Delta _{\lambda _{1}\cdots
\lambda _{\ell }}^{\mu _{1}\cdots \mu _{\ell }}\left( m_{0}^{2}\mathcal{M}%
_{i-1}^{\lambda _{1}\cdots \lambda _{\ell -2}}-\mathcal{M}_{i+1}^{\lambda
_{1}\cdots \lambda _{\ell -2}}\right) \left( \mathbf{q}F^{\lambda _{\ell
}\nu }\Delta _{\nu }^{\lambda _{\ell -1}}\right) \, ,
\end{align}%
where we used Eq.~(\ref{eq:M_vecM_relation}) to give $M_{i-1}^{\lambda _{1}\cdots \lambda _{n-1}\lambda _{n +1}} = \mathbf{M}_{i-1}^{\lambda _{1}\cdots \lambda _{n-1}\lambda _{n +1}} + n \mathbf{M}_{i}^{\left(\lambda _{1}\cdots \lambda _{n-1} \right.} u^{\left. \lambda_{n+1} \right)} + \cdots$, and then we used Eqs.~(\ref{Main1_iso}), (\ref{Main2_iso}) to express the orthogonally projected moments through IIMs.

Adding these terms together and using that $\Delta _{\lambda _{1}\cdots
\lambda _{\ell }}^{\mu _{1}\cdots \mu _{\ell }}F^{\lambda _{\ell }\nu
}\Delta _{\nu }^{\lambda _{\ell -1}}\equiv \Delta _{\lambda _{1}\cdots
\lambda _{\ell }}^{\mu _{1}\cdots \mu _{\ell }}F^{\lambda _{\ell }\lambda
_{\ell -1}}=0$, leads to%
\begin{align}
M4& \equiv \left( i-1\right) \Delta _{\lambda _{1}\cdots \lambda _{\ell
}}^{\mu _{1}\cdots \mu _{\ell }}\mathcal{M}_{i-2}^{\lambda _{1}\cdots
\lambda _{\ell }\lambda _{\ell +1}}\left( \mathbf{q}F^{\mu \nu }u_{\mu
}g_{\nu \lambda _{\ell +1}}\right) +\ell \Delta _{\lambda _{1}\cdots \lambda
_{\ell }}^{\mu _{1}\cdots \mu _{\ell }}\mathcal{M}_{i-1}^{\lambda _{1}\cdots
\lambda _{\ell -1}\lambda _{\ell +1}}\left( \mathbf{q}F^{\lambda _{\ell }\nu
}g_{\nu \lambda _{\ell +1}}\right) \   \notag \\
& -\frac{\ell }{2\ell +1}\left[ m_{0}^{2}\left( i-1\right) \mathcal{M}%
_{i-2}^{\left\langle \mu _{1}\right. \cdots \mu _{\ell -1}}-\left( i+2\ell
\right) \mathcal{M}_{i}^{\left\langle \mu _{1}\right. \cdots \mu _{\ell -1}}%
\right] \mathbf{q}F^{\left. \mu _{\ell }\right\rangle \nu }u_{\nu }\, .
\end{align}%
Now expressing the Faraday tensor in terms of the
electric field and the magnetic tensor, 
$F^{\mu \nu }\equiv E^{\mu }u^{\nu}-E^{\nu }u^{\mu }+B^{\mu \nu }$, we obtain 
\begin{align}
M4& \equiv -\left( i-1\right) \mathcal{M}_{i-2}^{\left\langle \mu
_{1}\right. \cdots \left. \mu _{\ell }\right\rangle \mu _{\ell +1}}\mathbf{q}%
E_{\mu _{\ell +1}}-\frac{\ell }{2\ell +1}\left[ m_{0}^{2}\left( i-1\right) 
\mathcal{M}_{i-2}^{\left\langle \mu _{1}\right. \cdots \mu _{\ell
-1}}-\left( i+2\ell \right) \mathcal{M}_{i}^{\left\langle \mu _{1}\right.
\cdots \mu _{\ell -1}}\right] \mathbf{q}E^{\left. \mu _{\ell }\right\rangle }
\notag \\
& +\ell \Delta _{\nu _{1}\cdots \nu _{\ell }}^{\mu _{1}\cdots \mu _{\ell }}%
\mathcal{M}_{i-1}^{\nu _{1}\cdots \nu _{\ell -1}\nu _{\ell +1}}\mathbf{q}%
B^{\nu _{\ell }\nu }g_{\nu \nu _{\ell +1}}\, .  \label{M4_term}
\end{align}%
Finally, combining all these results we obtain the equation of motion for
the IIMs (\ref{Main_DM_i_mu1_mun}).

\section{Term by term evaluation of Eq.~(\protect\ref{Main_DA_ij_mu1_mun})}
\label{Appendix:DA_ij}

Similarly to the previous section we start from the main equation of motion, 
\begin{align}
D\mathcal{A}_{ij}^{\left\{ \mu _{1}\cdots \mu _{\ell }\right\} } 
& =\mathcal{C}_{i-1,j}^{\mu _{1}\cdots \mu _{\ell }}+\Xi _{\nu _{1}
\cdots \nu _{\ell}}^{\mu _{1}\cdots \mu _{\ell }}
\int \mathrm{d}K D\left( E_{\mathbf{k}u}^{i}E_{\mathbf{k}l}^{j}\right) 
k^{\left\{ \nu _{1}\right. }\cdots k^{\left. \nu _{\ell }\right\} }
f_{\mathbf{k}}+\Xi _{\nu _{1}\cdots \nu_{\ell }}^{\mu _{1}\cdots \mu _{\ell }} 
\int \mathrm{d}K E_{\mathbf{k}u}^{i}E_{\mathbf{k}l}^{j}D\left( k^{\left\{ \nu _{1}\right. }\cdots
k^{\left. \nu _{\ell }\right\} }\right) f_{\mathbf{k}}  \notag \\
& -\Xi _{\nu _{1}\cdots \nu _{\ell }}^{\mu _{1}\cdots \mu _{\ell }}
\int \mathrm{d}K E_{\mathbf{k}u}^{i-1}E_{\mathbf{k}l}^{j}k^{\left\{ \nu_{1}\right. }
\cdots k^{\left. \nu _{\ell }\right\} }k^{\left\{ \nu _{\ell+1}\right\} }
\tilde{\nabla}_{\nu _{\ell +1}}f_{\mathbf{k}} 
+\Xi _{\nu_{1}\cdots \nu _{\ell }}^{\mu _{1}\cdots \mu _{\ell }} 
\int \mathrm{d}K E_{\mathbf{k}u}^{i-1}E_{\mathbf{k}l}^{j+1}k^{\left\{ \nu _{1}\right. }
\cdots k^{\left. \nu _{\ell }\right\} }D_{l}f_{\mathbf{k}}  \notag \\
& -\mathbf{q}F^{\mu \nu }\Xi _{\nu _{1}\cdots \nu _{\ell }}^{\mu _{1}\cdots \mu _{\ell }} 
\int \mathrm{d}K E_{\mathbf{k}u}^{i-1}E_{\mathbf{k}l}^{j}k^{\left\{ \nu _{1}\right. }
\cdots k^{\left. \nu _{\ell }\right\} }k_{\nu }\frac{\partial f_{\mathbf{k}}}{\partial k^{\mu }}\,.
\label{DA_ij_mu1_mun}
\end{align}%
The first term follows similarly to Eq.~(\ref{M1_term}), where using that 
$E_{\mathbf{k}u}=k^{\mu }u_{\mu }$, $E_{\mathbf{k}l}=-k^{\mu }l_{\mu }$ and $u^{\mu }Du_{\mu }=l^{\mu }Dl_{\mu }=0$, while $DE_{\mathbf{k}u}^{i} = iE_{\mathbf{k}u}^{i-1}\left( E_{\mathbf{k}l}l^{\alpha }+k^{\left\{ \alpha \right\} }\right) Du_{\alpha }$ and $DE_{\mathbf{k}l}^{j}=-jE_{\mathbf{k}l}^{j-1}\left( E_{\mathbf{k}u}u^{\alpha }+k^{\left\{ \alpha \right\}}\right) Dl_{\alpha }$, leads to 
\begin{align}
A1& \equiv \Xi _{\nu _{1}\cdots \nu _{\ell }}^{\mu _{1}\cdots \mu _{\ell
}}\int_{\mathbf{k}}D\left( E_{\mathbf{k}u}^{i}E_{\mathbf{k}l}^{j}\right)
k^{\left\{ \nu _{1}\right. }\cdots k^{\left. \nu _{\ell }\right\} }f_{%
\mathbf{k}}  \notag \\
& =il^{\nu _{\ell +1}}Du_{\nu _{\ell +1}}\Xi _{\nu _{1}\cdots \nu _{\ell
}}^{\mu _{1}\cdots \mu _{\ell }}\int_{\mathbf{k}}E_{\mathbf{k}u}^{i-1}E_{%
\mathbf{k}l}^{j+1}k^{\left\{ \nu _{1}\right. }\cdots k^{\left. \nu _{\ell
}\right\} }f_{\mathbf{k}}+iDu_{\nu _{\ell +1}}\Xi _{\nu _{1}\cdots \nu
_{\ell }}^{\mu _{1}\cdots \mu _{\ell }}\int_{\mathbf{k}}E_{\mathbf{k}%
u}^{i-1}E_{\mathbf{k}l}^{j}k^{\left\{ \nu _{1}\right. }\cdots k^{\left. \nu
_{\ell }\right\} }k^{\left\{ \nu _{\ell +1}\right\} }f_{\mathbf{k}}  \notag
\\
& +jl^{\nu _{\ell +1}}Du_{\nu _{\ell +1}}\Xi _{\nu _{1}\cdots \nu _{\ell
}}^{\mu _{1}\cdots \mu _{\ell }}\int_{\mathbf{k}}E_{\mathbf{k}u}^{i+1}E_{%
\mathbf{k}l}^{j-1}k^{\left\{ \nu _{1}\right. }\cdots k^{\left. \nu _{\ell
}\right\} }f_{\mathbf{k}}-jDl_{\nu _{\ell +1}}\Xi _{\nu _{1}\cdots \nu
_{\ell }}^{\mu _{1}\cdots \mu _{\ell }}\int_{\mathbf{k}}E_{\mathbf{k}%
u}^{i}E_{\mathbf{k}l}^{j-1}k^{\left\{ \nu _{1}\right. }\cdots k^{\left. \nu
_{\ell }\right\} }k^{\left\{ \nu _{\ell +1}\right\} }f_{\mathbf{k}}  \notag
\\
& =i\mathcal{A}_{i-1,j+1}^{\mu _{1}\cdots \mu _{\ell }}l_{\mu _{\ell
+1}}Du^{\mu _{\ell +1}}+i\mathcal{A}_{i-1,j}^{\mu _{1}\cdots \mu _{\ell
+1}}Du_{\mu _{\ell +1}}+\frac{i}{2}\left( m_{0}^{2}\mathcal{A}%
_{i-1,j}^{\left\{ \mu _{1}\right. \cdots \mu _{\ell -1}}-\mathcal{A}%
_{i+1,j}^{\left\{ \mu _{1}\right. \cdots \mu _{\ell -1}}+\mathcal{A}%
_{i-1,j+2}^{\left\{ \mu _{1}\right. \cdots \mu _{\ell -1}}\right) Du^{\left.
\mu _{\ell }\right\} }  \notag \\
& +j\mathcal{A}_{i+1,j-1}^{\mu _{1}\cdots \mu _{\ell }}l_{\mu _{\ell
+1}}Du^{\mu _{\ell +1}}-j\mathcal{A}_{i,j-1}^{\mu _{1}\cdots \mu _{\ell
+1}}Dl_{\mu _{\ell +1}}-\frac{j}{2}\left( m_{0}^{2}\mathcal{A}%
_{i,j-1}^{\left\{ \mu _{1}\right. \cdots \mu _{\ell -1}}-\mathcal{A}%
_{i+2,j-1}^{\left\{ \mu _{1}\right. \cdots \mu _{\ell -1}}+\mathcal{A}%
_{i,j+1}^{\left\{ \mu _{1}\right. \cdots \mu _{\ell -1}}\right) Dl^{\left.
\mu _{\ell }\right\} }\, ,  \label{A1_term}
\end{align}%
where we used Eq.~(\ref{Main2_aniso}) and replaced $\Xi _{\alpha \beta
}k^{\alpha }k^{\beta }=m_{0}^{2}-E_{\mathbf{k}u}^{2}+E_{\mathbf{k}l}^{2}$.

The second term is similar to Eq.~(\ref{M2_term}), 
\begin{eqnarray}
A2 &\equiv &\Xi _{\nu _{1}\cdots \nu _{\ell }}^{\mu _{1}\cdots \mu _{\ell
}}\int_{\mathbf{k}}E_{\mathbf{k}u}^{i}E_{\mathbf{k}l}^{j}D\left( k^{\left\{
\nu _{1}\right. }\cdots k^{\left. \nu _{\ell }\right\} }\right) f_{\mathbf{k}%
}=-\ell \Xi _{\nu _{1}\cdots \nu _{\ell }}^{\mu _{1}\cdots \mu _{\ell
}}\left( \mathbf{A}_{i+1,j}^{\nu _{1}\cdots \nu _{\ell -1}}Du^{\nu _{\ell }}-%
\mathbf{A}_{i,j+1}^{\nu _{1}\cdots \nu _{\ell -1}}Dl^{\nu _{\ell }}\right) 
\notag \\
&=&-\ell \mathcal{A}_{i+1,j}^{\left\{ \mu _{1}\right. \cdots \mu _{\ell
-1}}Du^{\left. \mu _{\ell }\right\} }-\ell \mathcal{A}_{i,j+1}^{\left\{ \mu
_{1}\right. \cdots \mu _{\ell -1}}Dl^{\left. \mu _{\ell }\right\} }\, ,
\label{A2_term}
\end{eqnarray}%
where we used that $\Xi _{\nu _{1}\cdots \nu _{\ell }}^{\mu _{1}\cdots \mu
_{\ell }}D\left( \Xi _{\lambda _{1}\cdots \lambda _{\ell }}^{\nu _{1}\cdots
\nu _{\ell }}\right) = -\ell \Xi _{\nu _{1}\cdots \nu _{\ell }}^{\mu
_{1}\cdots \mu _{\ell }}\left( u_{\lambda _{1}}Du^{\nu _{1}}-l_{\lambda
_{1}}Dl^{\nu _{1}}\right) \Xi _{\lambda _{2}}^{\nu _{2}}\cdots \Xi _{\lambda
_{\ell }}^{\nu _{\ell }}$.

The third term is analogous to Eq.~(\ref{M3_term_def}) and leads to the following three integrals
\begin{align}
A3& \equiv -\Xi _{\nu _{1}\cdots \nu _{\ell }}^{\mu _{1}\cdots \mu _{\ell
}}\int_{\mathbf{k}}E_{\mathbf{k}u}^{i-1}E_{\mathbf{k}l}^{j}k^{\left\{ \nu
_{1}\right. }\cdots k^{\left. \nu _{\ell }\right\} }k^{\left\{ \mu \right\} }%
\tilde{\nabla}_{\mu }f_{\mathbf{k}}=-\Xi _{\nu _{1}\cdots \nu _{\ell }}^{\mu
_{1}\cdots \mu _{\ell }}\tilde{\nabla}_{\mu }\left( \int_{\mathbf{k}}E_{%
\mathbf{k}u}^{i-1}E_{\mathbf{k}l}^{j}k^{\left\{ \nu _{1}\right. }\cdots
k^{\left. \nu _{\ell }\right\} }k^{\left\{ \mu \right\} }f_{\mathbf{k}%
}\right)  \notag \\
& +\Xi _{\nu _{1}\cdots \nu _{\ell }}^{\mu _{1}\cdots \mu _{\ell }}\int_{%
\mathbf{k}}\tilde{\nabla}_{\mu }\left( E_{\mathbf{k}u}^{i-1}E_{\mathbf{k}%
l}^{j}\right) k^{\left\{ \nu _{1}\right. }\cdots k^{\left. \nu _{\ell
}\right\} }k^{\left\{ \mu \right\} }f_{\mathbf{k}}+\Xi _{\nu _{1}\cdots \nu
_{\ell }}^{\mu _{1}\cdots \mu _{\ell }}\int_{\mathbf{k}}E_{\mathbf{k}%
u}^{i-1}E_{\mathbf{k}l}^{j}\tilde{\nabla}_{\mu }\left( k^{\left\{ \nu
_{1}\right. }\cdots k^{\left. \nu _{\ell }\right\} }k^{\left\{ \mu \right\}
}\right) f_{\mathbf{k}}\, .  \label{A3_term0}
\end{align}%
Making use of Eq.~(\ref{Main2_aniso}) the first integral leads to, 
\begin{align}
A3a& \equiv -\Xi _{\nu _{1}\cdots \nu _{\ell }}^{\mu _{1}\cdots \mu _{\ell }}%
\tilde{\nabla}_{\mu }\left( \int_{\mathbf{k}}E_{\mathbf{k}u}^{i-1}E_{\mathbf{%
k}l}^{j}k^{\left\{ \nu _{1}\right. }\cdots k^{\left. \nu _{\ell }\right\}
}k^{\left\{ \mu \right\} }f_{\mathbf{k}}\right)  \notag \\
& =-\Xi _{\nu _{1}\cdots \nu _{\ell }}^{\mu _{1}\cdots \mu _{\ell }}\left( 
\tilde{\nabla}_{\nu _{\ell +1}}\mathcal{A}_{i-1,j}^{\nu _{1}\cdots \nu
_{\ell +1}}\right) -\frac{1}{2}\tilde{\nabla}^{\left\{ \mu _{1}\right.
}\left( m_{0}^{2}\mathcal{A}_{i-1,j}^{\mu _{2}\cdots \left. \mu _{\ell
}\right\} }-\mathcal{A}_{i+1,j}^{\mu _{2}\cdots \left. \mu _{\ell }\right\}
}+\mathcal{A}_{i-1,j+2}^{\mu _{2}\cdots \left. \mu _{\ell }\right\} }\right)\, ,  
\label{A31_term}
\end{align}%
where we used that $\Xi _{\nu _{1}\cdots \nu _{\ell }}^{\mu _{1}\cdots \mu
_{\ell }}\left[ \tilde{\nabla}^{\lambda _{\ell }}\left( \Xi _{\lambda
_{1}\cdots \lambda _{\ell }}^{\nu _{1}\cdots \nu _{\ell }}\right) \right] =0$. 

The next integral is similar to Eq.~(\ref{M3b_term}) and leads to%
\begin{align}
A3b& \equiv \Xi _{\nu _{1}\cdots \nu _{\ell }}^{\mu _{1}\cdots \mu _{\ell
}}\int_{\mathbf{k}}\tilde{\nabla}_{\mu }\left( E_{\mathbf{k}u}^{i-1}E_{%
\mathbf{k}l}^{j}\right) k^{\left\{ \nu _{1}\right. }\cdots k^{\left. \nu
_{\ell }\right\} }k^{\left\{ \mu \right\} }f_{\mathbf{k}} \notag \\
&=\left( i-1\right)
\Xi _{\nu _{1}\cdots \nu _{\ell }}^{\mu _{1}\cdots \mu _{\ell }}\left( 
\tilde{\nabla}_{\mu }u_{\nu }\right) \int_{\mathbf{k}}E_{\mathbf{k}%
u}^{i-2}E_{\mathbf{k}l}^{j}k^{\left\{ \nu _{1}\right. }\cdots k^{\left. \nu
_{\ell }\right\} }k^{\left\{ \mu \right\} }k^{\left\{ \nu \right\} }f_{%
\mathbf{k}}  \notag \\
& +\left( i-1\right) \Xi _{\nu _{1}\cdots \nu _{\ell }}^{\mu _{1}\cdots \mu
_{\ell }}\left( l^{\nu }\tilde{\nabla}_{\mu }u_{\nu }\right) \int_{\mathbf{k}%
}E_{\mathbf{k}u}^{i-2}E_{\mathbf{k}l}^{j+1}k^{\left\{ \nu _{1}\right.
}\cdots k^{\left. \nu _{\ell }\right\} }k^{\left\{ \mu \right\} }f_{\mathbf{k%
}}  \notag \\
& - j\Xi _{\nu _{1}\cdots \nu _{\ell }}^{\mu _{1}\cdots \mu _{\ell }}\left( 
\tilde{\nabla}_{\mu }l_{\nu }\right) \int_{\mathbf{k}}E_{\mathbf{k}%
u}^{i-1}E_{\mathbf{k}l}^{j-1}k^{\left\{ \nu _{1}\right. }\cdots k^{\left.
\nu _{\ell }\right\} }k^{\left\{ \mu \right\} }k^{\left\{ \nu \right\} }f_{\mathbf{k}} \notag \\
&+j\Xi _{\nu _{1}\cdots \nu _{\ell }}^{\mu _{1}\cdots \mu _{\ell
}}\left( l^{\nu }\tilde{\nabla}_{\mu }u_{\nu }\right) \int_{\mathbf{k}}E_{%
\mathbf{k}u}^{i}E_{\mathbf{k}l}^{j-1}k^{\left\{ \nu _{1}\right. }\cdots
k^{\left. \nu _{\ell }\right\} }k^{\left\{ \mu \right\} }f_{\mathbf{k}} 
\notag \\
& =\frac{ i-1 }{2}\tilde{\theta}\left( m_{0}^{2}\mathcal{A}%
_{i-2,j}^{\mu _{1}\cdots \mu _{\ell }}-\mathcal{A}_{i,j}^{\mu _{1}\cdots \mu
_{\ell }}+\mathcal{A}_{i-2,j+2}^{\mu _{1}\cdots \mu _{\ell }}\right) +\left(
i-1\right) \mathcal{A}_{i-2,j}^{\mu _{1}\cdots \mu _{\ell +2}}\tilde{\sigma}%
_{\mu _{\ell +1}\mu _{\ell +2}}  \notag \\
& +\left( i-1\right) \frac{\ell }{\ell +1}\left( m_{0}^{2}\mathcal{A}%
_{i-2,j}^{\nu \left\{ \mu _{1}\right. \cdots \mu _{\ell -1}}-\mathcal{A}%
_{i,j}^{\nu \left\{ \mu _{1}\right. \cdots \mu _{\ell -1}}+\mathcal{A}%
_{i-2,j+2}^{\nu \left\{ \mu _{1}\right. \cdots \mu _{\ell -1}}\right) \tilde{%
\sigma}_{\nu }^{\left. \mu _{\ell }\right\} }  \notag \\
& +\frac{ i-1 }{4}\left( m_{0}^{4}\mathcal{A}_{i-2,j}^{\left\{
\mu _{1}\right. \cdots \mu _{\ell -2}}-2m_{0}^{2}\mathcal{A}_{i,j}^{\left\{
\mu _{1}\right. \cdots \mu _{\ell -2}}+2m_{0}^{2}\mathcal{A}%
_{i-2,j+2}^{\left\{ \mu _{1}\right. \cdots \mu _{\ell -2}}-2\mathcal{A}%
_{i,j+2}^{\left\{ \mu _{1}\right. \cdots \mu _{\ell -2}}+\mathcal{A}%
_{i+2,j}^{\left\{ \mu _{1}\right. \cdots \mu _{\ell -2}}+\mathcal{A}%
_{i-2,j+4}^{\left\{ \mu _{1}\right. \cdots \mu _{\ell -2}}\right) \tilde{%
\sigma}^{\mu _{\ell -1}\left. \mu _{\ell }\right\} }  \notag \\
& +\left( i-1\right) \Xi _{\nu _{1}\cdots \nu _{\ell }}^{\mu _{1}\cdots \mu
_{\ell }}\mathcal{A}_{i-2,j+1}^{\nu _{1}\cdots \nu _{\ell +1}}\left(
l_{\lambda }\tilde{\nabla}_{\nu _{\ell +1}}u^{\lambda }\right) +\frac{
i-1 }{2}\Xi _{\nu _{1}\cdots \nu _{\ell }}^{\mu _{1}\cdots \mu _{\ell
}}\left( m_{0}^{2}\mathcal{A}_{i-2,j+1}^{\nu _{1}\cdots \nu _{\ell -1}}-%
\mathcal{A}_{i,j+1}^{\nu _{1}\cdots \nu _{\ell -1}}+\mathcal{A}%
_{i-2,j+3}^{\nu _{1}\cdots \nu _{\ell -1}}\right) l_{\lambda }\tilde{\nabla}%
^{\nu _{\ell }}u^{\lambda }  \notag \\
& -\frac{j}{2}\tilde{\theta}_{l}\left( m_{0}^{2}\mathcal{A}_{i-1,j-1}^{\mu
_{1}\cdots \mu _{\ell }}-\mathcal{A}_{i+1,j-1}^{\mu _{1}\cdots \mu _{\ell }}+%
\mathcal{A}_{i-1,j+1}^{\mu _{1}\cdots \mu _{\ell }}\right) -j\mathcal{A}%
_{i-1,j-1}^{\mu _{1}\cdots \mu _{\ell +2}}\tilde{\sigma}_{l,\mu _{\ell
+1}\mu _{\ell +2}}  \notag \\
& -j\frac{\ell }{\ell +1}\left( m_{0}^{2}\mathcal{A}_{i-1,j-1}^{\nu \left\{
\mu _{1}\right. \cdots \mu _{\ell -1}}-\mathcal{A}_{i+1,j-1}^{\nu \left\{
\mu _{1}\right. \cdots \mu _{\ell -1}}+\mathcal{A}_{i-1,j+1}^{\nu \left\{
\mu _{1}\right. \cdots \mu _{\ell -1}}\right) \tilde{\sigma}_{l,\nu
}^{\left. \mu _{\ell }\right\} }  \notag \\
& -\frac{j}{4}\left( m_{0}^{4}\mathcal{A}_{i-1,j-1}^{\left\{ \mu _{1}\right.
\cdots \mu _{\ell -2}}-2m_{0}^{2}\mathcal{A}_{i+1,j-1}^{\left\{ \mu
_{1}\right. \cdots \mu _{\ell -2}}+2m_{0}^{2}\mathcal{A}_{i-1,j+1}^{\left\{
\mu _{1}\right. \cdots \mu _{\ell -2}}-2\mathcal{A}_{i+1,j+1}^{\left\{ \mu
_{1}\right. \cdots \mu _{\ell -2}}+\mathcal{A}_{i+3,j-1}^{\left\{ \mu
_{1}\right. \cdots \mu _{\ell -2}}+\mathcal{A}_{i-1,j+3}^{\left\{ \mu
_{1}\right. \cdots \mu _{\ell -2}}\right) \tilde{\sigma}_{l}^{\mu _{\ell
-1}\left. \mu _{\ell }\right\} }  \notag \\
& +j\Xi _{\nu _{1}\cdots \nu _{\ell }}^{\mu _{1}\cdots \mu _{\ell }}\mathcal{%
A}_{i,j-1}^{\nu _{1}\cdots \nu _{\ell +1}}\left( l_{\lambda }\tilde{\nabla}%
_{\nu _{\ell +1}}u^{\lambda }\right) +\frac{j}{2}\Xi _{\nu _{1}\cdots \nu
_{\ell }}^{\mu _{1}\cdots \mu _{\ell }}\left( m_{0}^{2}\mathcal{A}%
_{i,j-1}^{\nu _{1}\cdots \nu _{\ell -1}}-\mathcal{A}_{i+2,j-1}^{\nu
_{1}\cdots \nu _{\ell -1}}+\mathcal{A}_{i,j+1}^{\nu _{1}\cdots \nu _{\ell
-1}}\right) l_{\lambda }\tilde{\nabla}^{\nu _{\ell }}u^{\lambda }\, ,
\end{align}%
where we used that $\tilde{\nabla}_{\mu }E_{\mathbf{k}u}^{i-1}=\left(
i-1\right) E_{\mathbf{k}u}^{i-2}\left( E_{\mathbf{k}l}l^{\nu }+k^{\left\{
\nu \right\} }\right) \left( \tilde{\nabla}_{\mu }u_{\nu }\right) $ and $%
\tilde{\nabla}_{\mu }E_{\mathbf{k}l}^{j}=-jE_{\mathbf{k}l}^{j-1}\left( E_{%
\mathbf{k}u}u^{\nu }+k^{\left\{ \nu \right\} }\right) \left( \tilde{\nabla}%
_{\mu }l_{\nu }\right) $ and we have also replaced $\tilde{\nabla}_{\mu }u_{\nu
}\equiv \frac{1}{2}\tilde{\theta}\Xi _{\mu \nu }-l_{\beta }l_{\nu }\tilde{%
\nabla}_{\mu }u^{\beta }+\tilde{\sigma}_{\mu \nu }+\tilde{\omega}_{\mu \nu }$
and $\tilde{\nabla}_{\mu }l_{\nu }\equiv \frac{1}{2}\tilde{\theta}_{l}\Xi
_{\mu \nu }+u_{\beta }u_{\nu }\tilde{\nabla}_{\mu }l^{\beta }+\tilde{\sigma}%
_{l,\mu \nu }+\tilde{\omega}_{l,\mu \nu }$.

The last integral is analogous to Eq.~(\ref{M3c_term}) and leads to%
\begin{align}
A3c& \equiv \Xi _{\nu _{1}\cdots \nu _{\ell }}^{\mu _{1}\cdots \mu _{\ell
}}\int_{\mathbf{k}}E_{\mathbf{k}u}^{i-1}E_{\mathbf{k}l}^{j}\tilde{\nabla}%
_{\mu }\left( k^{\left\{ \nu _{1}\right. }\cdots k^{\left. \nu _{\ell
}\right\} }k^{\left\{ \mu \right\} }\right) f_{\mathbf{k}}=\Xi _{\nu
_{1}\cdots \nu _{\ell }}^{\mu _{1}\cdots \mu _{\ell }}\tilde{\nabla}_{\nu
_{\ell +1}}\left( \Xi _{\lambda _{1}\cdots \lambda _{\ell }}^{\nu _{1}\cdots
\nu _{\ell }}\Xi _{\lambda _{\ell +1}}^{\nu _{\ell +1}}\right)
A_{i-1,j}^{\lambda _{1}\cdots \lambda _{\ell +1}}  \notag \\
& =- \frac{\ell +2}{2} \tilde{\theta}\mathcal{A}_{i,j}^{\mu
_{1}\cdots \mu _{\ell }}-\frac{\ell }{2}\left( m_{0}^{2}\mathcal{A}%
_{i,j}^{\left\{ \mu _{1}\right. \cdots \mu _{\ell -2}}-\mathcal{A}%
_{i+2,j}^{\left\{ \mu _{1}\right. \cdots \mu _{\ell -2}}+\mathcal{A}%
_{i,j+2}^{\left\{ \mu _{1}\right. \cdots \mu _{\ell -2}}\right) \tilde{\sigma%
}^{\mu _{\ell -1}\left. \mu _{\ell }\right\} }  \notag \\
& - \frac{\ell +2}{2} \tilde{\theta}_{l}\mathcal{A}%
_{i-1,j+1}^{\mu _{1}\cdots \mu _{\ell }}-\frac{\ell }{2}\left( m_{0}^{2}%
\mathcal{A}_{i-1,j+1}^{\left\{ \mu _{1}\right. \cdots \mu _{\ell -2}}-%
\mathcal{A}_{i+1,j+1}^{\left\{ \mu _{1}\right. \cdots \mu _{\ell -2}}+%
\mathcal{A}_{i-1,j+3}^{\left\{ \mu _{1}\right. \cdots \mu _{\ell -2}}\right) 
\tilde{\sigma}_{l}^{\mu _{\ell -1}\left. \mu _{\ell }\right\} }  \notag \\
& -\ell \mathcal{A}_{i,j}^{\nu \left\{ \mu _{1}\right. \cdots \mu _{\ell -1}}%
\tilde{\sigma}_{\nu }^{\left. \mu _{\ell }\right\} }+\ell \mathcal{A}%
_{i,j}^{\nu \left\{ \mu _{1}\right. \cdots \mu _{\ell -1}}\tilde{\omega}%
_{\left. {}\right. \nu }^{\left. \mu _{\ell }\right\} }-\ell \mathcal{A}%
_{i-1,j+1}^{\nu \left\{ \mu _{1}\right. \cdots \mu _{\ell -1}}\tilde{\sigma}%
_{l,\nu }^{\left. \mu _{\ell }\right\} }+\ell \mathcal{A}_{i-1,j+1}^{\nu
\left\{ \mu _{1}\right. \cdots \mu _{\ell -1}}\tilde{\omega}_{l,\left.
{}\right. \nu }^{\left. \mu _{\ell }\right\} }\, ,
\end{align}%
where for the vorticity tensors we applied the following notation convention regarding the placement of contra/co-variant or upper/lower indices: $\omega^{\mu}_{\   \nu} = \omega^{\mu \alpha} g_{\alpha \nu}$ and  $\omega^{\ \mu}_{\nu} = \omega^{\alpha \mu} g_{\alpha \nu}$, which is such that the antisymmetric nature of the tensors is preserved $\omega^{\mu}_{\   \nu} + \omega^{\ \mu}_{\nu} =0$.

Adding all up we obtain,%
\begin{align}
A3& \equiv -\Xi _{\nu _{1}\cdots \nu _{\ell }}^{\mu _{1}\cdots \mu _{\ell
}}\left( \tilde{\nabla}_{\nu _{\ell +1}}\mathcal{A}_{i-1,j}^{\nu _{1}\cdots
\nu _{\ell +1}}\right) -\frac{1}{2}\tilde{\nabla}^{\left\{ \mu _{1}\right.
}\left( m_{0}^{2}\mathcal{A}_{i-1,j}^{\mu _{2}\cdots \left. \mu _{\ell
}\right\} }-\mathcal{A}_{i+1,j}^{\mu _{2}\cdots \left. \mu _{\ell }\right\}
}+\mathcal{A}_{i-1,j+2}^{\mu _{2}\cdots \left. \mu _{\ell }\right\} }\right)
\notag \\
& +\frac{1}{2}\tilde{\theta}\left[ m_{0}^{2}\left( i-1\right) \mathcal{A}%
_{i-2,j}^{\mu _{1}\cdots \mu _{\ell }}-\left( i+\ell +1\right) \mathcal{A}%
_{i,j}^{\mu _{1}\cdots \mu _{\ell }}+\left( i-1\right) \mathcal{A}%
_{i-2,j+2}^{\mu _{1}\cdots \mu _{\ell }}\right] +\left( i-1\right) \mathcal{A%
}_{i-2,j}^{\mu _{1}\cdots \mu _{\ell +2}}\tilde{\sigma}_{\mu _{\ell +1}\mu
_{\ell +2}}  \notag \\
& +\frac{\ell }{\ell +1}\left[ m_{0}^{2}\left( i-1\right) \mathcal{A}%
_{i-2,j}^{\nu \left\{ \mu _{1}\right. \cdots \mu _{\ell -1}}-\left( i+\ell
\right) \mathcal{A}_{i,j}^{\nu \left\{ \mu _{1}\right. \cdots \mu _{\ell
-1}}+\left( i-1\right) \mathcal{A}_{i-2,j+2}^{\nu \left\{ \mu _{1}\right.
\cdots \mu _{\ell -1}}\right] \tilde{\sigma}_{\nu }^{\left. \mu _{\ell
}\right\} }+\ell \mathcal{A}_{i,j}^{\nu \left\{ \mu _{1}\right. \cdots \mu
_{\ell -1}}\tilde{\omega}_{\left. {}\right. \nu }^{\left. \mu _{\ell
}\right\} }  \notag \\
& +\frac{1}{4}\left[ m_{0}^{4}\left( i-1\right) \mathcal{A}_{i-2,j}^{\left\{
\mu _{1}\right. \cdots \mu _{\ell -2}}-2m_{0}^{2}\left( i+\ell -1\right) 
\mathcal{A}_{i,j}^{\left\{ \mu _{1}\right. \cdots \mu _{\ell -2}}-2\left(
i+\ell -1\right) \mathcal{A}_{i,j+2}^{\left\{ \mu _{1}\right. \cdots \mu
_{\ell -2}}\right] \tilde{\sigma}^{\mu _{\ell -1}\left. \mu _{\ell }\right\}
}  \notag \\
& +\frac{1}{4}\left[ 2m_{0}^{2}\left( i-1\right) \mathcal{A}%
_{i-2,j+2}^{\left\{ \mu _{1}\right. \cdots \mu _{\ell -2}}+\left( i+2\ell
-1\right) \mathcal{A}_{i+2,j}^{\left\{ \mu _{1}\right. \cdots \mu _{\ell
-2}}+\left( i-1\right) \mathcal{A}_{i-2,j+4}^{\left\{ \mu _{1}\right. \cdots
\mu _{\ell -2}}\right] \tilde{\sigma}^{\mu _{\ell -1}\left. \mu _{\ell
}\right\} }  \notag \\
& +\left( i-1\right) \Xi _{\nu _{1}\cdots \nu _{\ell }}^{\mu _{1}\cdots \mu
_{\ell }}\mathcal{A}_{i-2,j+1}^{\nu _{1}\cdots \nu _{\ell +1}}\left(
l_{\lambda }\tilde{\nabla}_{\nu _{\ell +1}}u^{\lambda }\right) +\frac{
i-1 }{2}\Xi _{\nu _{1}\cdots \nu _{\ell }}^{\mu _{1}\cdots \mu _{\ell
}}\left( m_{0}^{2}\mathcal{A}_{i-2,j+1}^{\nu _{1}\cdots \nu _{\ell -1}}-%
\mathcal{A}_{i,j+1}^{\nu _{1}\cdots \nu _{\ell -1}}+\mathcal{A}%
_{i-2,j+3}^{\nu _{1}\cdots \nu _{\ell -1}}\right) l_{\lambda }\tilde{\nabla}%
^{\nu _{\ell }}u^{\lambda }  \notag \\
& -\frac{1}{2}\tilde{\theta}_{l}\left[ m_{0}^{2}j\mathcal{A}_{i-1,j-1}^{\mu
_{1}\cdots \mu _{\ell }}-j\mathcal{A}_{i+1,j-1}^{\mu _{1}\cdots \mu _{\ell
}}+\left( j+\ell +2\right) \mathcal{A}_{i-1,j+1}^{\mu _{1}\cdots \mu _{\ell
}}\right] -j\mathcal{A}_{i-1,j-1}^{\mu _{1}\cdots \mu _{\ell +2}}\tilde{%
\sigma}_{l,\mu _{\ell +1}\mu _{\ell +2}}  \notag \\
& -\frac{\ell }{\ell +1}\left[ m_{0}^{2}j\mathcal{A}_{i-1,j-1}^{\nu \left\{
\mu _{1}\right. \cdots \mu _{\ell -1}}-j\mathcal{A}_{i+1,j-1}^{\nu \left\{
\mu _{1}\right. \cdots \mu _{\ell -1}}+\left( j+\ell +1\right) \mathcal{A}%
_{i-1,j+1}^{\nu \left\{ \mu _{1}\right. \cdots \mu _{\ell -1}}\right] \tilde{%
\sigma}_{l,\nu }^{\left. \mu _{\ell }\right\} }+\ell \mathcal{A}%
_{i-1,j+1}^{\nu \left\{ \mu _{1}\right. \cdots \mu _{\ell -1}}\tilde{\omega}%
_{l,\left. {}\right. \nu }^{\left. \mu _{\ell }\right\} }  \notag \\
& -\frac{1}{4}\left[ m_{0}^{4}j\mathcal{A}_{i-1,j-1}^{\left\{ \mu
_{1}\right. \cdots \mu _{\ell -2}}-2m_{0}^{2}j\mathcal{A}_{i+1,j-1}^{\left\{
\mu _{1}\right. \cdots \mu _{\ell -2}}-2\left( j+\ell \right) \mathcal{A}%
_{i+1,j+1}^{\left\{ \mu _{1}\right. \cdots \mu _{\ell -2}}\right] \tilde{%
\sigma}_{l}^{\mu _{\ell -1}\left. \mu _{\ell }\right\} }  \notag \\
& -\frac{1}{4}\left[ 2m_{0}^{2}\left( j+\ell \right) \mathcal{A}%
_{i-1,j+1}^{\left\{ \mu _{1}\right. \cdots \mu _{\ell -2}}+j\mathcal{A}%
_{i+3,j-1}^{\left\{ \mu _{1}\right. \cdots \mu _{\ell -2}}+\left( j+2\ell
\right) \mathcal{A}_{i-1,j+3}^{\left\{ \mu _{1}\right. \cdots \mu _{\ell -2}}%
\right] \tilde{\sigma}_{l}^{\mu _{\ell -1}\left. \mu _{\ell }\right\} } 
\notag \\
& +j\Xi _{\nu _{1}\cdots \nu _{\ell }}^{\mu _{1}\cdots \mu _{\ell }}\mathcal{%
A}_{i,j-1}^{\nu _{1}\cdots \nu _{\ell +1}}\left( l_{\lambda }\tilde{\nabla}%
_{\nu _{\ell +1}}u^{\lambda }\right) +\frac{j}{2}\Xi _{\nu _{1}\cdots \nu
_{\ell }}^{\mu _{1}\cdots \mu _{\ell }}\left( m_{0}^{2}\mathcal{A}%
_{i,j-1}^{\nu _{1}\cdots \nu _{\ell -1}}-\mathcal{A}_{i+2,j-1}^{\nu
_{1}\cdots \nu _{\ell -1}}+\mathcal{A}_{i,j+1}^{\nu _{1}\cdots \nu _{\ell
-1}}\right) l_{\lambda }\tilde{\nabla}^{\nu _{\ell }}u^{\lambda }\, .
\label{A3_term}
\end{align}

The next term is 
\begin{align}
A4& \equiv \Xi _{\nu _{1}\cdots \nu _{\ell }}^{\mu _{1}\cdots \mu _{\ell
}}\int_{\mathbf{k}}E_{\mathbf{k}u}^{i-1}E_{\mathbf{k}l}^{j+1}k^{\left\{ \nu
_{1}\right. }\cdots k^{\left. \nu _{\ell }\right\} }D_{l}f_{\mathbf{k}}=\Xi
_{\nu _{1}\cdots \nu _{\ell }}^{\mu _{1}\cdots \mu _{\ell }}D_{l}\left(
\int_{\mathbf{k}}E_{\mathbf{k}u}^{i-1}E_{\mathbf{k}l}^{j+1}k^{\left\{ \nu
_{1}\right. }\cdots k^{\left. \nu _{\ell }\right\} }f_{\mathbf{k}}\right) 
\notag \\
& -\Xi _{\nu _{1}\cdots \nu _{\ell }}^{\mu _{1}\cdots \mu _{\ell }}\int_{%
\mathbf{k}}D_{l}\left( E_{\mathbf{k}u}^{i-1}E_{\mathbf{k}l}^{j+1}\right)
k^{\left\{ \nu _{1}\right. }\cdots k^{\left. \nu _{\ell }\right\} }f_{%
\mathbf{k}}-\Xi _{\nu _{1}\cdots \nu _{\ell }}^{\mu _{1}\cdots \mu _{\ell
}}\int_{\mathbf{k}}E_{\mathbf{k}u}^{i-1}E_{\mathbf{k}l}^{j+1}D_{l}\left(
k^{\left\{ \nu _{1}\right. }\cdots k^{\left. \nu _{\ell }\right\} }\right)
f_{\mathbf{k}}\, .  \label{A4_term0}
\end{align}%
The evaluation follows in a similar fashion as for the $A3$ terms, now replacing $\tilde{\nabla}_{\mu }$ with $D_{l}$, hence we get,
\begin{equation}
A4a\equiv \Xi _{\nu _{1}\cdots \nu _{\ell }}^{\mu _{1}\cdots \mu _{\ell
}}D_{l}\left( \int_{\mathbf{k}}E_{\mathbf{k}u}^{i-1}E_{\mathbf{k}%
l}^{j+1}k^{\left\{ \nu _{1}\right. }\cdots k^{\left. \nu _{\ell }\right\}
}f_{\mathbf{k}}\right) =\Xi _{\nu _{1}\cdots \nu _{\ell }}^{\mu _{1}\cdots
\mu _{\ell }}\left( D_{l}\mathcal{A}_{i-1,j+1}^{\nu _{1}\cdots \nu _{\ell
}}\right) \, .  \label{A41_term}
\end{equation}%
The second integral is%
\begin{align}
A4b& \equiv -\Xi _{\nu _{1}\cdots \nu _{\ell }}^{\mu _{1}\cdots \mu _{\ell
}}\int_{\mathbf{k}}D_{l}\left( E_{\mathbf{k}u}^{i-1}E_{\mathbf{k}%
l}^{j+1}\right) k^{\left\{ \nu _{1}\right. }\cdots k^{\left. \nu _{\ell
}\right\} }f_{\mathbf{k}}  \notag \\
& =-\left[ \left( i-1\right) \mathcal{A}_{i-2,j+2}^{\mu _{1}\cdots \mu
_{\ell }}+\left( j+1\right) \mathcal{A}_{i,j}^{\mu _{1}\cdots \mu _{\ell }}%
\right] l_{\nu }D_{l}u^{\nu }-\left( i-1\right) \mathcal{A}_{i-2,j+1}^{\mu
_{1}\cdots \mu _{\ell +1}}D_{l}u_{\mu _{\ell +1}}+\left( j+1\right) \mathcal{%
A}_{i-1,j}^{\mu _{1}\cdots \mu _{\ell +1}}D_{l}l_{\mu _{\ell +1}}  \notag \\
& -\frac{ i-1 }{2}\left( m_{0}^{2}\mathcal{A}%
_{i-2,j+1}^{\left\{ \mu _{1}\right. \cdots \mu _{\ell -1}}-\mathcal{A}%
_{i,j+1}^{\left\{ \mu _{1}\right. \cdots \mu _{\ell -1}}+\mathcal{A}%
_{i-2,j+3}^{\left\{ \mu _{1}\right. \cdots \mu _{\ell -1}}\right)
D_{l}u^{\left. \mu _{\ell }\right\} }  \notag \\
& +\frac{ j+1 }{2}\left( m_{0}^{2}\mathcal{A}_{i-1,j}^{\left\{
\mu _{1}\right. \cdots \mu _{\ell -1}}-\mathcal{A}_{i+1,j}^{\left\{ \mu
_{1}\right. \cdots \mu _{\ell -1}}+\mathcal{A}_{i-1,j+2}^{\left\{ \mu
_{1}\right. \cdots \mu _{\ell -1}}\right) D_{l}l^{\left. \mu _{\ell
}\right\} }\, ,
\end{align}%
while the third integral is%
\begin{eqnarray}
A4c &\equiv &-\Xi _{\nu _{1}\cdots \nu _{\ell }}^{\mu _{1}\cdots \mu _{\ell
}}\int_{\mathbf{k}}E_{\mathbf{k}u}^{i-1}E_{\mathbf{k}l}^{j+1}D_{l}\left(
k^{\left\{ \nu _{1}\right. }\cdots k^{\left. \nu _{\ell }\right\} }\right)
f_{\mathbf{k}}=-\Xi _{\nu _{1}\cdots \nu _{\ell }}^{\mu _{1}\cdots \mu
_{\ell }}D_{l}\left( \Xi _{\lambda _{1}\cdots \lambda _{\ell }}^{\nu
_{1}\cdots \nu _{\ell }}\right) A_{i-1,j+1}^{\lambda _{1}\cdots \lambda
_{\ell }}  \notag \\
&=&\ell \mathcal{A}_{i,j+1}^{\left\{ \mu _{1}\right. \cdots \mu _{\ell
-1}}D_{l}u^{\left. \mu _{\ell }\right\} }+\ell \mathcal{A}%
_{i-1,j+2}^{\left\{ \mu _{1}\right. \cdots \mu _{\ell -1}}D_{l}l^{\left. \mu
_{\ell }\right\} }\, .
\end{eqnarray}%
Adding these terms together leads to 
\begin{align}
A4& \equiv \Xi _{\nu _{1}\cdots \nu _{\ell }}^{\mu _{1}\cdots \mu _{\ell
}}\left( D_{l}\mathcal{A}_{i-1,j+1}^{\nu _{1}\cdots \nu _{\ell }}\right) -%
\left[ \left( i-1\right) \mathcal{A}_{i-2,j+2}^{\mu _{1}\cdots \mu _{\ell
}}+\left( j+1\right) \mathcal{A}_{i,j}^{\mu _{1}\cdots \mu _{\ell }}\right]
l_{\mu _{\ell +1}}D_{l}u^{\mu _{\ell +1}}  \notag \\
& -\left( i-1\right) \mathcal{A}_{i-2,j+1}^{\mu _{1}\cdots \mu _{\ell
+1}}D_{l}u_{\mu _{\ell +1}}+\left( j+1\right) \mathcal{A}_{i-1,j}^{\mu
_{1}\cdots \mu _{\ell +1}}D_{l}l_{\mu _{\ell +1}}  \notag \\
& -\frac{1}{2}\left[ m_{0}^{2}\left( i-1\right) \mathcal{A}%
_{i-2,j+1}^{\left\{ \mu _{1}\right. \cdots \mu _{\ell -1}}-\left( i+2\ell
-1\right) \mathcal{A}_{i,j+1}^{\left\{ \mu _{1}\right. \cdots \mu _{\ell
-1}}+\left( i-1\right) \mathcal{A}_{i-2,j+3}^{\left\{ \mu _{1}\right. \cdots
\mu _{\ell -1}}\right] D_{l}u^{\left. \mu _{\ell }\right\} }  \notag \\
& +\frac{1}{2}\left[ m_{0}^{2}\left( j+1\right) \mathcal{A}_{i-1,j}^{\left\{
\mu _{1}\right. \cdots \mu _{\ell -1}}-\left( j+1\right) \mathcal{A}%
_{i+1,j}^{\left\{ \mu _{1}\right. \cdots \mu _{\ell -1}}+\left( j+2\ell
+1\right) \mathcal{A}_{i-1,j+2}^{\left\{ \mu _{1}\right. \cdots \mu _{\ell
-1}}\right] D_{l}l^{\left. \mu _{\ell }\right\} }\, .  \label{A4_term}
\end{align}

The electromagnetic term is 
\begin{align}
A5& \equiv -\mathbf{q}F^{\mu \nu }\Xi _{\nu _{1}\cdots \nu _{\ell }}^{\mu
_{1}\cdots \mu _{\ell }}\int_{\mathbf{k}}E_{\mathbf{k}u}^{i-1}E_{\mathbf{k}%
l}^{j}k^{\left\{ \nu _{1}\right. }\cdots k^{\left. \nu _{\ell }\right\}
}k_{\nu }\frac{\partial f_{\mathbf{k}}}{\partial k^{\mu }}=\left( i-1\right) 
\mathbf{q}F^{\mu \nu }u_{\mu }g_{\nu \lambda _{\ell +1}}\Xi _{\lambda
_{1}\cdots \lambda _{\ell }}^{\mu _{1}\cdots \mu _{\ell }}\int_{\mathbf{k}%
}E_{\mathbf{k}u}^{i-2}E_{\mathbf{k}l}^{j}k^{\lambda _{1}}\cdots k^{\lambda
_{\ell }}k^{\lambda _{\ell +1}}f_{\mathbf{k}}  \notag \\
& -j\mathbf{q}F^{\mu \nu }l_{\mu }g_{\nu \lambda _{\ell +1}}\Xi _{\lambda
_{1}\cdots \lambda _{\ell }}^{\mu _{1}\cdots \mu _{\ell }}\int_{\mathbf{k}%
}E_{\mathbf{k}u}^{i-1}E_{\mathbf{k}l}^{j-1}k^{\lambda _{1}}\cdots k^{\lambda
_{\ell }}k^{\lambda _{\ell +1}}f_{\mathbf{k}}+\mathbf{q}F^{\mu \nu }\Xi
_{\lambda _{1}\cdots \lambda _{\ell }}^{\mu _{1}\cdots \mu _{\ell }}\int_{%
\mathbf{k}}E_{\mathbf{k}u}^{i-1}E_{\mathbf{k}l}^{j}\frac{\partial }{\partial
k^{\mu }}\left( k^{\lambda _{1}}\cdots k^{\lambda _{\ell }}k_{\nu }\right)
f_{\mathbf{k}}\, .
\end{align}
Here the first and second integrals are analogous to Eq.~(\ref{M4a_term}) and lead to 
\begin{align}
A5a& \equiv \left( i-1\right) \mathbf{q}F^{\mu \nu }u_{\mu }g_{\nu \lambda
_{\ell +1}}\Xi _{\lambda _{1}\cdots \lambda _{\ell }}^{\mu _{1}\cdots \mu
_{\ell }}A_{i-2,j}^{\lambda _{1}\cdots \lambda _{\ell }\lambda _{\ell +1}} 
\notag \\
& =\left( i-1\right) \mathcal{A}_{i-2,j+1}^{\mu _{1}\cdots \mu _{\ell
}}\left( \mathbf{q}F^{\mu \nu }u_{\mu }l_{\nu }\right) +\left( i-1\right)
\Xi _{\lambda _{1}\cdots \lambda _{\ell }}^{\mu _{1}\cdots \mu _{\ell }}%
\mathcal{A}_{i-2,j}^{\lambda _{1}\cdots \lambda _{\ell }\lambda _{\ell
+1}}\left( \mathbf{q}F^{\mu \nu }u_{\mu }g_{\nu \lambda _{\ell +1}}\right)  
\notag \\
& +\frac{ i-1 }{2}\Xi _{\lambda _{1}\cdots \lambda _{\ell
}}^{\mu _{1}\cdots \mu _{\ell }}\left( m_{0}^{2}\mathcal{A}_{i-2,j}^{\lambda
_{1}\cdots \lambda _{\ell -1}}-\mathcal{A}_{i,j}^{\lambda _{1}\cdots \lambda
_{\ell -1}}+\mathcal{A}_{i-2,j+2}^{\lambda _{1}\cdots \lambda _{\ell
-1}}\right) \left( \mathbf{q}F^{\mu \nu }u_{\mu }\Xi _{\nu }^{\lambda _{\ell
}}\right) \, ,
\end{align}
and
\begin{align}
A5b& \equiv -j\mathbf{q}F^{\mu \nu }l_{\mu }g_{\nu \lambda _{\ell +1}}\Xi
_{\lambda _{1}\cdots \lambda _{\ell }}^{\mu _{1}\cdots \mu _{\ell
}}A_{i-1,j-1}^{\lambda _{1}\cdots \lambda _{\ell }\lambda _{\ell +1}}  \notag
\\
& =-j\mathcal{A}_{i,j-1}^{\mu _{1}\cdots \mu _{\ell }}\left( \mathbf{q}%
F^{\mu \nu }l_{\mu }u_{\nu }\right) -j\Xi _{\lambda _{1}\cdots \lambda
_{\ell }}^{\mu _{1}\cdots \mu _{\ell }}\mathcal{A}_{i-1,j-1}^{\lambda
_{1}\cdots \lambda _{\ell }\lambda _{\ell +1}}\left( \mathbf{q}F^{\mu \nu
}l_{\mu }g_{\nu \lambda _{\ell +1}}\right)   \notag \\
& -\frac{j}{2}\Xi _{\lambda _{1}\cdots \lambda _{\ell }}^{\mu _{1}\cdots \mu
_{\ell }}\left( m_{0}^{2}\mathcal{A}_{i-1,j-1}^{\lambda _{1}\cdots \lambda
_{\ell -1}}-\mathcal{A}_{i+1,j-1}^{\lambda _{1}\cdots \lambda _{\ell -1}}+%
\mathcal{A}_{i-1,j+1}^{\lambda _{1}\cdots \lambda _{\ell -1}}\right) \left( 
\mathbf{q}F^{\mu \nu }l_{\mu }\Xi _{\nu }^{\lambda _{\ell }}\right) \, ,
\end{align}
where we have used Eq.~(\ref{eq:A_vecA_relation}) together with Eq.~(\ref{Main1_aniso}),  (\ref{Main2_aniso}). 
The last integral evaluates to 
\begin{align}
A5c& \equiv \mathbf{q}F^{\mu \nu }g_{\nu \lambda _{\ell +1}}\Xi _{\lambda
_{1}\cdots \lambda _{\ell }}^{\mu _{1}\cdots \mu _{\ell }}\int_{\mathbf{k}%
}E_{\mathbf{k}u}^{i-1}E_{\mathbf{k}l}^{j}k^{\lambda _{\ell +1}}\frac{%
\partial }{\partial k^{\mu }}\left( k^{\lambda _{1}}\cdots k^{\lambda _{\ell
}}\right) f_{\mathbf{k}}=\mathbf{q}F^{\mu \nu }g_{\nu \lambda _{\ell +1}}\Xi
_{\lambda _{1}\cdots \lambda _{\ell }}^{\mu _{1}\cdots \mu _{\ell
}}\sum_{n=1}^{\ell }\delta _{\mu }^{\lambda _{n}}A_{i-1,j}^{\lambda
_{1}\cdots \lambda _{n-1}\lambda _{n +1}}  \notag \\
& =\ell \mathcal{A}_{i,j}^{\left\{ \mu _{1}\right. \cdots \mu _{\ell -1}}%
\mathbf{q}F^{\left. \mu _{\ell }\right\} \nu }u_{\nu }+\ell \mathcal{A}%
_{i-1,j+1}^{\left\{ \mu _{1}\right. \cdots \mu _{\ell -1}}\mathbf{q}%
F^{\left. \mu _{\ell }\right\} \nu }l_{\nu }+\ell \Xi _{\lambda _{1}\cdots
\lambda _{\ell }}^{\mu _{1}\cdots \mu _{\ell }}\mathcal{A}_{i-1,j}^{\lambda
_{1}\cdots \lambda _{\ell -1}\lambda _{\ell +1}}\left( \mathbf{q}F^{\lambda
_{\ell }\nu }g_{\nu \lambda _{\ell +1}}\right)   \notag \\
& +\frac{\ell }{2}\Xi _{\lambda _{1}\cdots \lambda _{\ell }}^{\mu _{1}\cdots
\mu _{\ell }}\left( m_{0}^{2}\mathcal{A}_{i-1,j}^{\lambda _{1}\cdots \lambda
_{\ell -2}}-\mathcal{A}_{i+1,j}^{\lambda _{1}\cdots \lambda _{\ell -2}}+%
\mathcal{A}_{i-1,j+2}^{\lambda _{1}\cdots \lambda _{\ell -2}}\right) \left( 
\mathbf{q}F^{\lambda _{\ell }\nu }\Xi _{\nu }^{\lambda _{\ell -1}}\right) \, .
\end{align}%
Adding these terms together\ and noting that $\Xi _{\lambda _{1}\cdots
\lambda _{\ell }}^{\mu _{1}\cdots \mu _{\ell }}F^{\lambda _{\ell }\nu }\Xi
_{\nu }^{\lambda _{\ell -1}}=0$, we get
\begin{align}
A5& \equiv \left( i-1\right) \mathcal{A}_{i-2,j+1}^{\mu _{1}\cdots \mu
_{\ell }}\left( \mathbf{q}F^{\mu \nu }u_{\mu }l_{\nu }\right) -j\mathcal{A}%
_{i,j-1}^{\mu _{1}\cdots \mu _{\ell }}\left( \mathbf{q}F^{\mu \nu }l_{\mu
}u_{\nu }\right) +\left( i-1\right) \Xi _{\lambda _{1}\cdots \lambda _{\ell
}}^{\mu _{1}\cdots \mu _{\ell }}\mathcal{A}_{i-2,j}^{\lambda _{1}\cdots
\lambda _{\ell }\lambda _{\ell +1}}\left( \mathbf{q}F^{\mu \nu }u_{\mu
}g_{\nu \lambda _{\ell +1}}\right)   \notag \\
& -\frac{1}{2}\left[ m_{0}^{2}\left( i-1\right) \mathcal{A}_{i-2,j}^{\left\{
\mu _{1}\right. \cdots \mu _{\ell -1}}-\left( i+2\ell -1\right) \mathcal{A}%
_{i,j}^{\left\{ \mu _{1}\right. \cdots \mu _{\ell -1}}+\left( i-1\right) 
\mathcal{A}_{i-2,j+2}^{\left\{ \mu _{1}\right. \cdots \mu _{\ell -1}}\right] 
\mathbf{q}F^{\left. \mu _{\ell }\right\} \nu }u_{\nu }  \notag \\
& -j\Xi _{\lambda _{1}\cdots \lambda _{\ell }}^{\mu _{1}\cdots \mu _{\ell }}%
\mathcal{A}_{i-1,j-1}^{\lambda _{1}\cdots \lambda _{\ell }\lambda _{\ell
+1}}\left( \mathbf{q}F^{\mu \nu }l_{\mu }g_{\nu \lambda _{\ell +1}}\right)
+\ell \Xi _{\lambda _{1}\cdots \lambda _{\ell }}^{\mu _{1}\cdots \mu _{\ell
}}\mathcal{A}_{i-1,j}^{\lambda _{1}\cdots \lambda _{\ell -1}\lambda _{\ell
+1}}\left( \mathbf{q}F^{\lambda _{\ell }\nu }g_{\nu \lambda _{\ell
+1}}\right)   \notag \\
& +\frac{1}{2}\left[ m_{0}^{2}j\mathcal{A}_{i-1,j-1}^{\left\{ \mu
_{1}\right. \cdots \mu _{\ell -1}}-j\mathcal{A}_{i+1,j-1}^{\left\{ \mu
_{1}\right. \cdots \mu _{\ell -1}}+\left( j+2\ell \right) \mathcal{A}%
_{i-1,j+1}^{\left\{ \mu _{1}\right. \cdots \mu _{\ell -1}}\right] \mathbf{q}%
F^{\left. \mu _{\ell }\right\} \nu }l_{\nu }\, .
\end{align}%

Now, replacing the Faraday tensor from Eq.~(\ref{F_munu}) we obtain%
\begin{align}
A5& \equiv -\left[ \left( i-1\right) \mathcal{A}_{i-2,j+1}^{\mu _{1}\cdots
\mu _{\ell }}+j\mathcal{A}_{i,j-1}^{\mu _{1}\cdots \mu _{\ell }}\right] 
\mathbf{q}E^{\nu }l_{\nu }-\left( i-1\right) \Xi _{\nu _{1}\cdots \nu _{\ell
}}^{\mu _{1}\cdots \mu _{\ell }}\mathcal{A}_{i-2,j}^{\nu _{1}\cdots \nu
_{\ell +1}}\mathbf{q}E_{\nu _{\ell +1}}  \notag \\
& -\frac{1}{2}\left[ m_{0}^{2}\left( i-1\right) \mathcal{A}_{i-2,j}^{\left\{
\mu _{1}\right. \cdots \mu _{\ell -1}}-\left( i+2\ell -1\right) \mathcal{A}%
_{i,j}^{\left\{ \mu _{1}\right. \cdots \mu _{\ell -1}}+\left( i-1\right) 
\mathcal{A}_{i-2,j+2}^{\left\{ \mu _{1}\right. \cdots \mu _{\ell -1}}\right] 
\mathbf{q}E^{\left. \mu _{\ell }\right\} }  \notag \\
& -j\Xi _{\nu _{1}\cdots \nu _{\ell }}^{\mu _{1}\cdots \mu _{\ell }}\mathcal{%
A}_{i-1,j-1}^{\nu _{1}\cdots \nu _{\ell +1}}\mathbf{q}B^{\mu \nu }l_{\mu
}g_{\nu \nu _{\ell +1}}+\ell \Xi _{\nu _{1}\cdots \nu _{\ell }}^{\mu
_{1}\cdots \mu _{\ell }}\mathcal{A}_{i-1,j}^{\nu _{1}\cdots \nu _{\ell
-1}\nu _{\ell +1}}\mathbf{q}B^{\nu _{\ell }\nu }g_{\nu \nu _{\ell +1}} 
\notag \\
& +\frac{1}{2}\left[ m_{0}^{2}j\mathcal{A}_{i-1,j-1}^{\left\{ \mu
_{1}\right. \cdots \mu _{\ell -1}}-j\mathcal{A}_{i+1,j-1}^{\left\{ \mu
_{1}\right. \cdots \mu _{\ell -1}}+\left( j+2\ell \right) \mathcal{A}%
_{i-1,j+1}^{\left\{ \mu _{1}\right. \cdots \mu _{\ell -1}}\right] \mathbf{q}%
B^{\left. \mu _{\ell }\right\} \nu }l_{\nu }\, .
\end{align}%
Finally, combining all these results we obtain the equation of motion for
the AIMs~(\ref{Main_DA_ij_mu1_mun}).

\newpage
\section{Equations of motion of resistive and dissipative MHD}
\label{app:resistiveMHD}

The equations of motion for resistive and dissipative MHD follow from Eq.~(\ref{Main_DM_i_mu1_mun}) 
with $\mathcal{M}_{r}^{\mu_{1}\cdots \mu _{\ell }}
=\mathcal{I}_{r}^{\mu _{1}\cdots \mu _{\ell }}+\rho_{r}^{\mu _{1}\cdots \mu _{\ell }}$. 
The scalar equation for tensor-rank $\ell =0$ reads
\begin{align}
D\rho _{i} &= \mathcal{C}_{i-1}-D\mathcal{I}_{i}+\frac{1}{3}\theta 
\left[ m_{0}^{2}\left( i-1\right) \mathcal{I}_{i-2}-\left( i+2\right) \mathcal{I}%
_{i}\right] +\frac{1}{3}\theta \left[ m_{0}^{2}\left( i-1\right) 
\rho_{i-2}-\left( i+2\right) \rho _{i}\right]  \notag \\
&+  i\rho _{i-1}^{\mu _{1}}Du_{\mu _{1}}-\nabla _{\nu _{1}}\rho _{i-1}^{\nu
_{1}}+\left( i-1\right) \rho _{i-2}^{\mu _{1}\mu _{2}}\sigma _{\mu _{1}\mu
_{2}}-\left( i-1\right) \rho _{i-2}^{\mu _{1}}\mathbf{q}E_{\mu _{1}}\, ,
\label{Drho_i}
\end{align}%
where the moments of the collision integral vanish only for the particle 
number or charge and energy conservation equations, see Eqs.~(\ref{C_0_C_1_C_0_mu}).
The vector equation follows from Eq.~(\ref{Main_DM_i_mu1_mun}) for $\ell =1$,
\begin{align}
D\rho _{i}^{\left\langle \mu _{1}\right\rangle }& =\mathcal{C}_{i-1}^{\mu
_{1}}+\frac{1}{3}\theta \left[ m_{0}^{2}\left( i-1\right) \rho _{i-2}^{\mu
_{1}}-\left( i+3\right) \rho _{i}^{\mu _{1}}\right] +i\rho _{i-1}^{\mu
_{1}\mu _{2}}Du_{\mu _{2}}  \notag \\
& -\Delta _{\nu _{1}}^{\mu _{1}}\nabla _{\nu _{2}}\rho _{i-1}^{\nu _{1}\nu
_{2}}+\frac{1}{3}\left[ m_{0}^{2}i\mathcal{I}_{i-1}-\left( i+3\right) 
\mathcal{I}_{i+1}\right] Du^{\mu _{1}}+\frac{1}{3}\left[ m_{0}^{2}i\rho
_{i-1}-\left( i+3\right) \rho _{i+1}\right] Du^{\mu _{1}}  \notag \\
& -\frac{1}{3}\nabla ^{\mu _{1}}\left( m_{0}^{2}\mathcal{I}_{i-1}-\mathcal{I}%
_{i+1}\right) -\frac{1}{3}\nabla ^{\mu _{1}}\left( m_{0}^{2}\rho _{i-1}-\rho
_{i+1}\right) +\left( i-1\right) \rho _{i-2}^{\mu _{1}\mu _{2}\mu
_{3}}\sigma _{\mu _{2}\mu _{3}}  \notag \\
& +\frac{1}{5}\left[ m_{0}^{2}\left( 2i-2\right) \rho _{i-2}^{\lambda
}-\left( 2i+3\right) \rho _{i}^{\lambda }\right] \sigma _{\lambda }^{\mu
_{1}}+\rho _{i}^{\lambda }\omega _{\left. {}\right. \lambda }^{\mu _{1}} 
\notag \\
& -\left( i-1\right) \mathcal{I}_{i-2}^{\left\langle \mu _{1}\right\rangle
\mu _{2}}\mathbf{q}E_{\mu _{2}}-\left( i-1\right) \rho _{i-2}^{\left\langle
\mu _{1}\right\rangle \mu _{2}}\mathbf{q}E_{\mu _{2}}-\frac{1}{3}\left[
m_{0}^{2}\left( i-1\right) \mathcal{I}_{i-2}-\left( i+2\right) \mathcal{I}%
_{i}\right] \mathbf{q}E^{\mu _{1}}  \notag \\
& +\frac{1}{3}\left[ m_{0}^{2}\left( i-1\right) \rho _{i-2}-\left(
i+2\right) \rho _{i}\right] \mathbf{q}E^{\mu _{1}}+\Delta _{\nu _{1}}^{\mu
_{1}}\rho _{i-1}^{\nu _{2}}\mathbf{q}B^{\nu _{1}\nu }g_{\nu \nu _{2}}\, ,
\label{Drho_i_mu1}
\end{align}%
where the collision integral vanishes only for $\mathcal{C}_{0}^{\mu _{1}}=0$.

The equation of motion for tensor-rank $\ell =2$ follows similarly, 
\begin{align}
D\rho _{i}^{\left\langle \mu _{1}\mu _{2}\right\rangle }& =\mathcal{C}%
_{i-1}^{\mu _{1}\mu _{2}}+\frac{1}{3}\theta \left[ m_{0}^{2}\left(
i-1\right) \rho _{i-2}^{\mu _{1}\mu _{2}}-\left( i+4\right) \rho _{i}^{\mu
_{1}\mu _{2}}\right] +i\rho _{i-1}^{\mu _{1}\mu _{2}\mu _{3}}Du_{\mu _{3}} 
\notag \\
& -\Delta _{\nu _{1}\nu _{2}}^{\mu _{1}\mu _{2}}\nabla _{\nu _{3}}\rho
_{i-1}^{\nu _{1}\nu _{2}\nu _{3}}+\frac{2}{5}\left[ m_{0}^{2}i\rho
_{i-1}^{\left\langle \mu _{1}\right. }-\left( i+5\right) \rho
_{i+1}^{\left\langle \mu _{1}\right. }\right] Du^{\left. \mu
_{2}\right\rangle }-\frac{2}{5}\nabla ^{\left\langle \mu _{1}\right. }\left(
m_{0}^{2}\rho _{i-1}^{\left. \mu _{2}\right\rangle }-\rho _{i+1}^{\left. \mu
_{2}\right\rangle }\right)  \notag \\
& +\left( i-1\right) \rho _{i-2}^{\mu _{1}\mu _{2}\mu _{3}\mu _{4}}\sigma
_{\mu _{3}\mu _{4}}+\frac{2}{7}\left[ m_{0}^{2}\left( 2i-2\right) \rho
_{i-2}^{\lambda \left\langle \mu _{1}\right. }-\left( 2i+5\right) \rho
_{i}^{\lambda \left\langle \mu _{1}\right. }\right] \sigma _{\lambda
}^{\left. \mu _{2}\right\rangle }+2\rho _{i}^{\lambda \left\langle \mu
_{1}\right. }\omega _{\left. {}\right. \lambda }^{\left. \mu
_{2}\right\rangle }  \notag \\
& +\frac{2}{15}\left[ m_{0}^{4}\left( i-1\right) \mathcal{I}%
_{i-2}-m_{0}^{2}\left( 2i+3\right) \mathcal{I}_{i}+\left( i+4\right) 
\mathcal{I}_{i+2}\right] \sigma ^{\mu _{1}\left. \mu _{2}\right\rangle } 
\notag \\
& +\frac{2}{15}\left[ m_{0}^{4}\left( i-1\right) \rho _{i-2}-m_{0}^{2}\left(
2i+3\right) \rho _{i}+\left( i+4\right) \rho _{i+2}\right] \sigma ^{\mu
_{1}\left. \mu _{2}\right\rangle }  \notag \\
& -\left( i-1\right) \rho _{i-2}^{\left\langle \mu _{1}\mu _{2}\right\rangle
\mu _{3}}\mathbf{q}E_{\mu _{3}}-\frac{2}{5}\left[ m_{0}^{2}\left(
i-1\right) \rho _{i-2}^{\left\langle \mu _{1}\right. }-\left( i+4\right)
\rho _{i}^{\left\langle \mu _{1}\right. }\right] \mathbf{q}E^{\left. \mu
_{2}\right\rangle } 
+ 2\Delta _{\nu _{1}\nu _{2}}^{\mu _{1}\mu _{2}}\rho _{i-1}^{\nu _{1}\nu
_{3}}\mathbf{q}B^{\nu _{2}\nu }g_{\nu \nu _{3}}\, .  
\label{Drho_i_mu1_mu2}
\end{align}%
Finally, the equation of motion for tensor-rank $\ell =3$ reads
\begin{align}
D\rho _{i}^{\left\langle \mu _{1}\mu _{2}\mu _{3}\right\rangle }& =\mathcal{C%
}_{i-1}^{\mu _{1}\mu _{2}\mu _{3}}+\frac{1}{3}\theta \left[ m_{0}^{2}\left(
i-1\right) \rho _{i-2}^{\mu _{1}\mu _{2}\mu _{3}}-\left( i+5\right) \rho
_{i}^{\mu _{1}\mu _{2}\mu _{3}}\right] +i\rho _{i-1}^{\mu _{1}\mu _{2}\mu
_{3}\mu _{4}}Du_{\mu _{4}}  \notag \\
& -\Delta _{\nu _{1}\nu _{2}\nu _{3}}^{\mu _{1}\mu _{2}\mu _{3}}\nabla _{\nu
_{4}}\rho _{i-1}^{\nu _{1}\nu _{2}\nu _{3}\nu _{4}}+\frac{3}{7}\left[
m_{0}^{2}i\rho _{i-1}^{\left\langle \mu _{1}\right. \mu _{2}}-\left(
i+7\right) \rho _{i+1}^{\left\langle \mu _{1}\right. \mu _{2}}\right]
Du^{\left. \mu _{3}\right\rangle } 
-\frac{3}{7}\nabla ^{\left\langle \mu _{1}\right. }\left( m_{0}^{2}\rho
_{i-1}^{\mu _{2}\left. \mu _{3}\right\rangle } 
-\rho _{i+1}^{\mu _{2}\left.\mu _{3}\right\rangle }\right) \notag \\
& +\left( i-1\right) \rho _{i-2}^{\mu _{1}\mu_{2}\mu _{3}\mu _{4}\mu _{5}}\sigma_{\mu _{4}\mu _{5}} +\frac{1}{3}\left[ m_{0}^{2}\left( 2i-2\right) \rho _{i-2}^{\lambda
\left\langle \mu _{1}\right. \mu _{2}}-\left( 2i+7\right) \rho _{i}^{\lambda
\left\langle \mu _{1}\right. \mu _{2}}\right] \sigma _{\lambda }^{\left. \mu
_{3}\right\rangle }+3\rho _{i}^{\lambda \left\langle \mu _{1}\right. \mu
_{2}}\omega _{\left. {}\right. \lambda }^{\left. \mu _{3}\right\rangle } 
\notag \\
& +\frac{6}{35}\left[ m_{0}^{4}\left( i-1\right) \rho _{i-2}^{\left\langle
\mu _{1}\right. }-m_{0}^{2}\left( 2i+5\right) \rho _{i}^{\left\langle \mu
_{1}\right. }+\left( i+6\right) \rho _{i+2}^{\left\langle \mu _{1}\right. }%
\right] \sigma ^{\mu _{2}\left. \mu _{3}\right\rangle }  \notag \\
& -\left( i-1\right) \rho _{i-2}^{\left\langle \mu _{1}\right. \mu
_{2}\left. \mu _{3}\right\rangle \mu _{4}}\mathbf{q}E_{\mu _{4}}-\frac{3}{7}%
\left[ m_{0}^{2}\left( i-1\right) \rho _{i-2}^{\left\langle \mu _{1}\right.
\mu _{2}}-\left( i+6\right) \rho _{i}^{\left\langle \mu _{1}\right. \mu _{2}}%
\right] \mathbf{q}E^{\left. \mu _{3}\right\rangle } 
+3\Delta _{\nu _{1}\nu _{2}\nu _{3}}^{\mu _{1}\mu _{2}\mu _{3}}\rho
_{i-1}^{\nu _{1}\nu _{2}\nu _{4}}\mathbf{q}B^{\nu _{3}\nu }g_{\nu \nu _{4}}\, .  
\label{Drho_i_mu1_mu2_m3}
\end{align}
Higher-rank tensor equations follow in a similar fashion from Eq.~(\ref{Main_DM_i_mu1_mun}).

\section{Resistive anisotropic MHD}
\label{app:aniso_resistiveMHD}

The scalar equation of motion from Eq.~(\ref{Main_DA_ij_mu1_mun}) for tensor-rank $\ell =0$ reads
\begin{align}
	D\hat{\rho}_{ij}& =\mathcal{C}_{i-1,j}-D\hat{\mathcal{I}}_{ij}+\left[
	i\left( \hat{\mathcal{I}}_{i-1,j+1}+\hat{\rho}_{i-1,j+1}\right) +j\left( 
	\hat{\mathcal{I}}_{i+1,j-1}+\hat{\rho}_{i+1,j-1}\right) \right] l_{\lambda
	}Du^{\lambda }  \notag \\
	& -\left[ \left( i-1\right) \left( \hat{\mathcal{I}}_{i-2,j+2}+\hat{\rho}%
	_{i-2,j+2}\right) +\left( j+1\right) \left( \hat{\mathcal{I}}_{i,j}+\hat{\rho%
	}_{i,j}\right) \right] l_{\lambda }D_{l}u^{\lambda }  \notag \\
	& +i\hat{\rho}_{i-1,j}^{\mu _{1}}Du_{\mu _{1}}-\left( i-1\right) \hat{\rho}%
	_{i-2,j+1}^{\mu _{1}}D_{l}u_{\mu _{1}}-j\hat{\rho}_{i,j-1}^{\mu _{1}}Dl_{\mu
		_{1}}+\left( j+1\right) \hat{\rho}_{i-1,j}^{\mu _{1}}D_{l}l_{\mu _{1}} 
	\notag \\
	& +D_{l}\hat{\mathcal{I}}_{i-1,j+1}+D_{l}\hat{\rho}_{i-1,j+1}-\tilde{\nabla}%
	_{\nu _{1}}\hat{\rho}_{i-1,j}^{\nu _{1}}+\left( i-1\right) \hat{\rho}%
	_{i-2,j+1}^{\nu _{1}}l_{\lambda }\tilde{\nabla}_{\nu _{1}}u^{\lambda }+j\hat{%
		\rho}_{i,j-1}^{\nu _{1}}l_{\lambda }\tilde{\nabla}_{\nu _{1}}u^{\lambda } 
	\notag \\
	& +\frac{1}{2}\tilde{\theta}\left[ m_{0}^{2}\left( i-1\right) \left( \hat{%
		\mathcal{I}}_{i-2,j}+\hat{\rho}_{i-2,j}\right) -\left( i+1\right) \left( 
	\hat{\mathcal{I}}_{i,j}+\hat{\rho}_{i,j}\right) +\left( i-1\right) \left( 
	\hat{\mathcal{I}}_{i-2,j+2}+\hat{\rho}_{i-2,j+2}\right) \right]   \notag \\
	& -\frac{1}{2}\tilde{\theta}_{l}\left[ m_{0}^{2}j\left( \hat{\mathcal{I}}%
	_{i-1,j-1}+\hat{\rho}_{i-1,j-1}\right) -j\left( \hat{\mathcal{I}}_{i+1,j-1}+%
	\hat{\rho}_{i+1,j-1}\right) +\left( j+2\right) \left( \hat{\mathcal{I}}%
	_{i-1,j+1}+\hat{\rho}_{i-1,j+1}\right) \right]   \notag \\
	& +\left( i-1\right) \hat{\rho}_{i-2,j}^{\mu _{1}\mu _{2}}\tilde{\sigma}%
	_{\mu _{1}\mu _{2}}-j\hat{\rho}_{i-1,j-1}^{\mu _{1}\mu _{2}}\tilde{\sigma}%
	_{l,\mu _{1}\mu _{2}}  \notag \\
	& -\left[ \left( i-1\right) \left( \hat{\mathcal{I}}_{i-2,j+1}+\hat{\rho}%
	_{i-2,j+1}\right) +j\left( \hat{\mathcal{I}}_{i,j-1}+\hat{\rho}%
	_{i,j-1}\right) \right] \mathbf{q}E^{\nu }l_{\nu }  \notag \\
	& -\left( i-1\right) \hat{\rho}_{i-2,j}^{\nu _{1}}\mathbf{q}E_{\nu _{1}}-j%
	\hat{\rho}_{i-1,j-1}^{\nu _{1}}\mathbf{q}B^{\mu \nu }l_{\mu }g_{\nu \nu_{1}}\, .  
	\label{Drho_ij}
\end{align}%
The vector equation follows from Eq.~(\ref{Main_DA_ij_mu1_mun}) for tensor-rank $\ell =1$,
\begin{align}
	D\hat{\rho}_{ij}^{\left\{ \mu _{1}\right\} }& =\mathcal{C}_{i-1,j}^{\mu
		_{1}}+\left[ i\hat{\rho}_{i-1,j+1}^{\mu _{1}}+j\hat{\rho}_{i+1,j-1}^{\mu
		_{1}}\right] l_{\lambda }Du^{\lambda }-\left[ \left( i-1\right) \hat{\rho}%
	_{i-2,j+2}^{\mu _{1}}+\left( j+1\right) \hat{\rho}_{i,j}^{\mu _{1}}\right]
	l_{\lambda }D_{l}u^{\lambda }  \notag \\
	& +i\hat{\rho}_{i-1,j}^{\mu _{1}\mu _{2}}Du_{\mu _{2}}-\left( i-1\right) 
	\hat{\rho}_{i-2,j+1}^{\mu _{1}\mu _{2}}D_{l}u_{\mu _{2}}-j\hat{\rho}%
	_{i,j-1}^{\mu _{1}\mu _{2}}Dl_{\mu _{2}}+\left( j+1\right) \hat{\rho}%
	_{i-1,j}^{\mu _{1}\mu _{2}}D_{l}l_{\mu _{2}}  \notag \\
	& -\frac{1}{2}\tilde{\nabla}^{\mu _{1}}\left[
	m_{0}^{2}\left( \hat{\mathcal{I}}_{i-1,j}+\hat{\rho}_{i-1,j}\right) -\left( 
	\hat{\mathcal{I}}_{i+1,j}+\hat{\rho}_{i+1,j}\right) +\left( \hat{\mathcal{I}}%
	_{i-1,j+2}+\hat{\rho}_{i-1,j+2}\right) \right]   \notag \\
	& +\Xi _{\nu _{1}}^{\mu _{1}}D_{l}\hat{\rho}_{i-1,j+1}^{\nu _{1}}-\Xi _{\nu
		_{1}}^{\mu _{1}}\left( \tilde{\nabla}_{\nu _{2}}\hat{\rho}_{i-1,j}^{\nu
		_{1}\nu _{2}}\right) 
	+\left( i-1\right) \Xi _{\nu _{1}}^{\mu _{1}}\hat{\rho}_{i-2,j+1}^{\nu
		_{1}\nu _{2}}l_{\lambda }\tilde{\nabla}_{\nu _{2}}u^{\lambda }+j\Xi _{\nu
		_{1}}^{\mu _{1}}\hat{\rho}_{i,j-1}^{\nu _{1}\nu _{2}}l_{\lambda }\tilde{%
		\nabla}_{\nu _{2}}u^{\lambda }  \notag \\
	& +\frac{ i-1 }{2}\left[ m_{0}^{2}\left( \hat{\mathcal{I}}%
	_{i-2,j+1}+\hat{\rho}_{i-2,j+1}\right) -\left( \hat{\mathcal{I}}_{i,j+1}+%
	\hat{\rho}_{i,j+1}\right) +\left( \hat{\mathcal{I}}_{i-2,j+3}+\hat{\rho}%
	_{i-2,j+3}\right) \right] l_{\lambda }\tilde{\nabla}^{\mu _{1}}u^{\lambda } 
	\notag \\
	& +\frac{j}{2}\left[ m_{0}^{2}\left( \hat{\mathcal{I}}_{i,j-1}+\hat{\rho}%
	_{i,j-1}\right) -\left( \hat{\mathcal{I}}_{i+2,j-1}+\hat{\rho}%
	_{i+2,j-1}\right) +\left( \hat{\mathcal{I}}_{i,j+1}+\hat{\rho}%
	_{i,j+1}\right) \right] l_{\lambda }\tilde{\nabla}^{\mu _{1}}u^{\lambda } 
	\notag \\
	& +\frac{1}{2}\left[ m_{0}^{2}i\left( \hat{\mathcal{I}}_{i-1,j}+\hat{\rho}%
	_{i-1,j}\right) -\left( i+2\right) \left( \hat{\mathcal{I}}_{i+1,j}+\hat{\rho%
	}_{i+1,j}\right) +i\left( \hat{\mathcal{I}}_{i-1,j+2}+\hat{\rho}%
	_{i-1,j+2}\right) \right] \Xi _{\lambda }^{\mu _{1}}Du^{\lambda }  \notag \\
	& -\frac{1}{2}\left[ m_{0}^{2}\left( i-1\right) \left( \hat{\mathcal{I}}%
	_{i-2,j+1}+\hat{\rho}_{i-2,j+1}\right) -\left( i+1\right) \left( \hat{%
		\mathcal{I}}_{i,j+1}+\hat{\rho}_{i,j+1}\right) +\left( i-1\right) \left( 
	\hat{\mathcal{I}}_{i-2,j+3}+\hat{\rho}_{i-2,j+3}\right) \right] \Xi
	_{\lambda }^{\mu _{1}}D_{l}u^{\lambda }  \notag \\
	& -\frac{1}{2}\left[ m_{0}^{2}j\left( \hat{\mathcal{I}}_{i,j-1}+\hat{\rho}%
	_{i,j-1}\right) -j\left( \hat{\mathcal{I}}_{i+2,j-1}+\hat{\rho}%
	_{i+2,j-1}\right) +\left( j+2\right) \left( \hat{\mathcal{I}}_{i,j+1}+\hat{%
		\rho}_{i,j+1}\right) \right] \Xi _{\lambda }^{\mu _{1}}Dl^{\lambda }  \notag
	\\
	& +\frac{1}{2}\left[ m_{0}^{2}\left( j+1\right) \left( \hat{\mathcal{I}}%
	_{i-1,j}+\hat{\rho}_{i-1,j}\right) -\left( j+1\right) \left( \hat{\mathcal{I}%
	}_{i+1,j}+\hat{\rho}_{i+1,j}\right) +\left( j+3\right) \left( \hat{\mathcal{I%
	}}_{i-1,j+2}+\hat{\rho}_{i-1,j+2}\right) \right] \Xi _{\lambda }^{\mu
		_{1}}D_{l}l^{\lambda }  \notag \\
	& +\frac{1}{2}\tilde{\theta}\left[ m_{0}^{2}\left( i-1\right) \hat{\rho}%
	_{i-2,j}^{\mu _{1}}-\left( i+2\right) \hat{\rho}_{i,j}^{\mu _{1}}+\left(
	i-1\right) \hat{\rho}_{i-2,j+2}^{\mu _{1}}\right] +\left( i-1\right) \hat{%
		\rho}_{i-2,j}^{\mu _{1}\mu _{2}\mu _{3}}\tilde{\sigma}_{\mu _{2}\mu _{3}} 
	\notag \\
	& +\frac{1}{2}\left[ m_{0}^{2}\left( i-1\right) \hat{\rho}_{i-2,j}^{\lambda
	}-\left( i+1\right) \hat{\rho}_{i,j}^{\lambda }+\left( i-1\right) \hat{\rho}%
	_{i-2,j+2}^{\lambda }\right] \tilde{\sigma}_{\lambda }^{\mu _{1}}+\hat{\rho}%
	_{i,j}^{\lambda }\tilde{\omega}_{\left. {}\right. \lambda }^{\mu _{1}} 
	\notag \\
	& -\frac{1}{2}\tilde{\theta}_{l}\left[ m_{0}^{2}j\hat{\rho}_{i-1,j-1}^{\mu
		_{1}}-j\hat{\rho}_{i+1,j-1}^{\mu _{1}}+\left( j+3\right) \hat{\rho}%
	_{i-1,j+1}^{\mu _{1}}\right] -j\hat{\rho}_{i-1,j-1}^{\mu _{1}\mu _{2}\mu
		_{3}}\tilde{\sigma}_{l,\mu _{2}\mu _{3}}  \notag \\
	& -\frac{1}{2}\left[ m_{0}^{2}j\hat{\rho}_{i-1,j-1}^{\lambda }-j\hat{\rho}%
	_{i+1,j-1}^{\lambda }+\left( j+2\right) \hat{\rho}_{i-1,j+1}^{\lambda }%
	\right] \tilde{\sigma}_{l,\lambda }^{\mu _{1}}+\hat{\rho}_{i-1,j+1}^{\lambda
	}\tilde{\omega}_{l,\left. {}\right. \lambda }^{\mu _{1}}  \notag \\
	& -\left[ \left( i-1\right) \hat{\rho}_{i-2,j+1}^{\mu _{1}}+j\hat{\rho}%
	_{i,j-1}^{\mu _{1}}\right] \mathbf{q}E^{\nu }l_{\nu }-\left( i-1\right) \Xi
	_{\nu _{1}}^{\mu _{1}}\hat{\rho}_{i-2,j}^{\nu _{1}\nu _{2}}\mathbf{q}E_{\nu
		_{2}}  \notag \\
	& -\frac{1}{2}\left[ m_{0}^{2}\left( i-1\right) \left( \hat{\mathcal{I}}%
	_{i-2,j}+\hat{\rho}_{i-2,j}\right) -\left( i+1\right) \left( \hat{\mathcal{I}%
	}_{i,j}+\hat{\rho}_{i,j}\right) +\left( i-1\right) \left( \hat{\mathcal{I}}%
	_{i-2,j+2}+\hat{\rho}_{i-2,j+2}\right) \right] \Xi _{\nu _{1}}^{\mu _{1}}%
	\mathbf{q}E^{\nu _{1}}  \notag \\
	& -j\Xi _{\nu _{1}}^{\mu _{1}}\hat{\rho}_{i-1,j-1}^{\nu _{1}\nu _{2}}\mathbf{%
		q}B^{\mu \nu }l_{\mu }g_{\nu \nu _{2}}+\Xi _{\nu _{1}}^{\mu _{1}}\hat{\rho}%
	_{i-1,j}^{\nu _{2}}\mathbf{q}B^{\nu _{1}\nu }g_{\nu \nu _{2}}  \notag \\
	& +\frac{1}{2}\left[ m_{0}^{2}j\left( \hat{\mathcal{I}}_{i-1,j-1}+\hat{\rho}%
	_{i-1,j-1}\right) -j\left( \hat{\mathcal{I}}_{i+1,j-1}+\hat{\rho}%
	_{i+1,j-1}\right) +\left( j+2\right) \left( \hat{\mathcal{I}}_{i-1,j+1}+\hat{%
		\rho}_{i-1,j+1}\right) \right] \Xi _{\nu _{1}}^{\mu _{1}}\mathbf{q}B^{\nu
		_{1}\nu }l_{\nu }\, .  \label{Drho_ij_mu1}
\end{align}%
The equation of motion for tensor-rank $\ell =2$ is 
\begin{align}
	D\hat{\rho}_{ij}^{\left\{ \mu _{1}\mu _{2}\right\} }& =\mathcal{C}%
	_{i-1,j}^{\mu _{1}\mu _{2}}+\left[ i\hat{\rho}_{i-1,j+1}^{\mu _{1}\mu _{2}}+j%
	\hat{\rho}_{i+1,j-1}^{\mu _{1}\mu _{2}}\right] l_{\lambda }Du^{\lambda }-%
	\left[ \left( i-1\right) \hat{\rho}_{i-2,j+2}^{\mu _{1}\mu _{2}}+\left(
	j+1\right) \hat{\rho}_{i,j}^{\mu _{1}\mu _{2}}\right] l_{\lambda
	}D_{l}u^{\lambda }  \notag \\
	& +i\hat{\rho}_{i-1,j}^{\mu _{1}\mu _{2}\mu _{3}}Du_{\mu _{3}}-\left(
	i-1\right) \hat{\rho}_{i-2,j+1}^{\mu _{1}\mu _{2}\mu _{3}}D_{l}u_{\mu _{3}}-j%
	\hat{\rho}_{i,j-1}^{\mu _{1}\mu _{2}\mu _{3}}Dl_{\mu _{3}}+\left( j+1\right) 
	\hat{\rho}_{i-1,j}^{\mu _{1}\mu _{2}\mu _{3}}D_{l}l_{\mu _{3}}  \notag \\
	& +\Xi _{\nu _{1}\nu _{2}}^{\mu _{1}\mu _{2}}D_{l}\hat{\rho}_{i-1,j+1}^{\nu
		_{1}\nu _{2}}-\Xi _{\nu _{1}\nu _{2}}^{\mu _{1}\mu _{2}}\left( \tilde{\nabla}%
	_{\nu _{3}}\hat{\rho}_{i-1,j}^{\nu _{1}\nu _{2}\nu _{3}}\right) -\frac{1}{2}%
	\tilde{\nabla}^{\left\{ \mu _{1}\right. }\left( m_{0}^{2}\hat{\rho}%
	_{i-1,j}^{\left. \mu _{2}\right\} }-\hat{\rho}_{i+1,j}^{\left. \mu
		_{2}\right\} }+\hat{\rho}_{i-1,j+2}^{\left. \mu _{2}\right\} }\right)  
	\notag \\
	& +\left( i-1\right) \Xi _{\nu _{1}\nu _{2}}^{\mu _{1}\mu _{2}}\hat{\rho}%
	_{i-2,j+1}^{\nu _{1}\nu _{2}\nu _{3}}l_{\lambda }\tilde{\nabla}_{\nu
		_{3}}u^{\lambda }+\frac{ i-1 }{2}\Xi _{\nu _{1}\nu _{2}}^{\mu
		_{1}\mu _{2}}\left( m_{0}^{2}\hat{\rho}_{i-2,j+1}^{\nu _{1}}-\hat{\rho}%
	_{i,j+1}^{\nu _{1}}+\hat{\rho}_{i-2,j+3}^{\nu _{1}}\right) l_{\lambda }%
	\tilde{\nabla}^{\nu _{2}}u^{\lambda }  \notag \\
	& +j\Xi _{\nu _{1}\nu _{2}}^{\mu _{1}\mu _{2}}\hat{\rho}_{i,j-1}^{\nu
		_{1}\nu _{2}\nu _{3}}l_{\lambda }\tilde{\nabla}_{\nu _{3}}u^{\lambda }+\frac{%
		j}{2}\Xi _{\nu _{1}\nu _{2}}^{\mu _{1}\mu _{2}}\left( m_{0}^{2}\hat{\rho}%
	_{i,j-1}^{\nu _{1}}-\hat{\rho}_{i+2,j-1}^{\nu _{1}}+\hat{\rho}_{i,j+1}^{\nu
		_{1}}\right) l_{\lambda }\tilde{\nabla}^{\nu _{2}}u^{\lambda }  \notag \\
	& +\frac{1}{2}\left[ m_{0}^{2}i\hat{\rho}_{i-1,j}^{\left\{ \mu _{1}\right.
	}-\left( i+4\right) \hat{\rho}_{i+1,j}^{\left\{ \mu _{1}\right. }+i\hat{\rho}%
	_{i-1,j+2}^{\left\{ \mu _{1}\right. }\right] Du^{\left. \mu _{2}\right\} }-%
	\frac{1}{2}\left[ m_{0}^{2}j\hat{\rho}_{i,j-1}^{\left\{ \mu _{1}\right. }-j%
	\hat{\rho}_{i+2,j-1}^{\left\{ \mu _{1}\right. }+\left( j+4\right) \hat{\rho}%
	_{i,j+1}^{\left\{ \mu _{1}\right. }\right] Dl^{\left. \mu _{2}\right\} } 
	\notag \\
	& -\frac{1}{2}\left[ m_{0}^{2}\left( i-1\right) \hat{\rho}%
	_{i-2,j+1}^{\left\{ \mu _{1}\right. }-\left( i+3\right) \hat{\rho}%
	_{i,j+1}^{\left\{ \mu _{1}\right. }+\left( i-1\right) \hat{\rho}%
	_{i-2,j+3}^{\left\{ \mu _{1}\right. }\right] D_{l}u^{\left. \mu _{2}\right\}
	}  \notag \\
	& +\frac{1}{2}\left[ m_{0}^{2}\left( j+1\right) \hat{\rho}_{i-1,j}^{\left\{
		\mu _{1}\right. }-\left( j+1\right) \hat{\rho}_{i+1,j}^{\left\{ \mu
		_{1}\right. }+\left( j+5\right) \hat{\rho}_{i-1,j+2}^{\left\{ \mu
		_{1}\right. }\right] D_{l}l^{\left. \mu _{2}\right\} }  \notag \\
	& +\frac{1}{2}\tilde{\theta}\left[ m_{0}^{2}\left( i-1\right) \hat{\rho}%
	_{i-2,j}^{\mu _{1}\mu _{2}}-\left( i+3\right) \hat{\rho}_{i,j}^{\mu _{1}\mu
		_{2}}+\left( i-1\right) \hat{\rho}_{i-2,j+2}^{\mu _{1}\mu _{2}}\right]
	+\left( i-1\right) \hat{\rho}_{i-2,j}^{\mu _{1}\mu _{2}\mu _{3}\mu _{4}}%
	\tilde{\sigma}_{\mu _{3}\mu _{4}}  \notag \\
	& +\frac{2}{3}\left[ m_{0}^{2}\left( i-1\right) \hat{\rho}_{i-2,j}^{\lambda
		\left\{ \mu _{1}\right. }-\left( i+2\right) \hat{\rho}_{i,j}^{\lambda
		\left\{ \mu _{1}\right. }+\left( i-1\right) \hat{\rho}_{i-2,j+2}^{\lambda
		\left\{ \mu _{1}\right. }\right] \tilde{\sigma}_{\lambda }^{\left. \mu
		_{2}\right\} }+2\hat{\rho}_{i,j}^{\lambda \left\{ \mu _{1}\right. }\tilde{%
		\omega}_{\left. {}\right. \lambda }^{\left. \mu _{2}\right\} }  \notag \\
	& +\frac{1}{4}\left[ m_{0}^{4}\left( i-1\right) \left( \hat{\mathcal{I}}%
	_{i-2,j}+\hat{\rho}_{i-2,j}\right) -2m_{0}^{2}\left( i+1\right) \left( \hat{%
		\mathcal{I}}_{i,j}+\hat{\rho}_{i,j}\right) -2\left( i+1\right) \left( \hat{%
		\mathcal{I}}_{i,j+2}+\hat{\rho}_{i,j+2}\right) \right] \tilde{\sigma}^{\mu
		_{1}\mu _{2}}  \notag \\
	& +\frac{1}{4}\left[ 2m_{0}^{2}\left( i-1\right) \left( \hat{\mathcal{I}}%
	_{i-2,j+2}+\hat{\rho}_{i-2,j+2}\right) +\left( i+3\right) \left( \hat{%
		\mathcal{I}}_{i+2,j}+\hat{\rho}_{i+2,j}\right) +\left( i-1\right) \left( 
	\hat{\mathcal{I}}_{i-2,j+4}+\hat{\rho}_{i-2,j+4}\right) \right] \tilde{\sigma%
	}^{\mu _{1}\mu _{2}}  \notag \\
	& -\frac{1}{2}\tilde{\theta}_{l}\left[ m_{0}^{2}j\hat{\rho}_{i-1,j-1}^{\mu
		_{1}\mu _{2}}-j\hat{\rho}_{i+1,j-1}^{\mu _{1}\mu _{2}}+\left( j+4\right) 
	\hat{\rho}_{i-1,j+1}^{\mu _{1}\mu _{2}}\right] -j\hat{\rho}_{i-1,j-1}^{\mu
		_{1}\mu _{2}\mu _{3}\mu _{4}}\tilde{\sigma}_{l,\mu _{3}\mu _{4}}  \notag \\
	& -\frac{2}{3}\left[ m_{0}^{2}j\hat{\rho}_{i-1,j-1}^{\lambda \left\{ \mu
		_{1}\right. }-j\hat{\rho}_{i+1,j-1}^{\lambda \left\{ \mu _{1}\right.
	}+\left( j+3\right) \hat{\rho}_{i-1,j+1}^{\lambda \left\{ \mu _{1}\right. }%
	\right] \tilde{\sigma}_{l,\lambda }^{\left. \mu _{2}\right\} }+2\hat{\rho}%
	_{i-1,j+1}^{\lambda \left\{ \mu _{1}\right. }\tilde{\omega}_{l,\left.
		{}\right. \lambda }^{\left. \mu _{2}\right\} }  \notag \\
	& -\frac{1}{4}\left[ m_{0}^{4}j\left( \hat{\mathcal{I}}_{i-1,j-1}+\hat{\rho}%
	_{i-1,j-1}\right) -2m_{0}^{2}j\left( \hat{\mathcal{I}}_{i+1,j-1}+\hat{\rho}%
	_{i+1,j-1}\right) -2\left( j+2\right) \left( \hat{\mathcal{I}}_{i+1,j+1}+%
	\hat{\rho}_{i+1,j+1}\right) \right] \tilde{\sigma}_{l}^{\mu _{1}\mu _{2}} 
	\notag \\
	& -\frac{1}{4}\left[ 2m_{0}^{2}\left( j+2\right) \left( \hat{\mathcal{I}}%
	_{i-1,j+1}+\hat{\rho}_{i-1,j+1}\right) +j\left( \hat{\mathcal{I}}_{i+3,j-1}+%
	\hat{\rho}_{i+3,j-1}\right) +\left( j+4\right) \left( \hat{\mathcal{I}}%
	_{i-1,j+3}+\hat{\rho}_{i-1,j+3}\right) \right] \tilde{\sigma}_{l}^{\mu
		_{1}\mu _{2}}  \notag \\
	& -\left[ \left( i-1\right) \hat{\rho}_{i-2,j+1}^{\mu _{1}\mu _{2}}+j\hat{%
		\rho}_{i,j-1}^{\mu _{1}\mu _{2}}\right] \mathbf{q}E^{\nu }l_{\nu }-\left(
	i-1\right) \Xi _{\nu _{1}\nu _{2}}^{\mu _{1}\mu _{2}}\hat{\rho}_{i-2,j}^{\nu
		_{1}\nu _{2}\nu _{3}}\mathbf{q}E_{\nu _{3}}  \notag \\
	& -\frac{1}{2}\left[ m_{0}^{2}\left( i-1\right) \hat{\rho}_{i-2,j}^{\left\{
		\mu _{1}\right. }-\left( i+3\right) \hat{\rho}_{i,j}^{\left\{ \mu
		_{1}\right. }+\left( i-1\right) \hat{\rho}_{i-2,j+2}^{\left\{ \mu
		_{1}\right. }\right] \mathbf{q}E^{\left. \mu _{2}\right\} }  \notag \\
	& -j\Xi _{\nu _{1}\nu _{2}}^{\mu _{1}\mu _{2}}\hat{\rho}_{i-1,j-1}^{\nu
		_{1}\nu _{2}\nu _{3}}\mathbf{q}B^{\mu \nu }l_{\mu }g_{\nu \nu _{3}}+2\Xi
	_{\nu _{1}\nu _{2}}^{\mu _{1}\mu _{2}}\hat{\rho}_{i-1,j}^{\nu _{1}\nu _{3}}%
	\mathbf{q}B^{\nu _{2}\nu }g_{\nu \nu _{3}}  \notag \\
	& +\frac{1}{2}\left[ m_{0}^{2}j\hat{\rho}_{i-1,j-1}^{\left\{ \mu _{1}\right.
	}-j\hat{\rho}_{i+1,j-1}^{\left\{ \mu _{1}\right. }+\left( j+4\right) \hat{%
		\rho}_{i-1,j+1}^{\left\{ \mu _{1}\right. }\right] \mathbf{q}B^{\left. \mu
		_{2}\right\} \nu }l_{\nu }\, .  \label{Drho_ij_mu1_mu2}
\end{align}


\end{document}